\documentclass{aa}  

\usepackage{graphicx}
\usepackage{txfonts}
\usepackage{natbib}
\usepackage{xcolor}
\usepackage{caption}
\usepackage{subfigure}
\usepackage{booktabs}

\usepackage[colorlinks=true,citecolor=blue,linkcolor=magenta,urlcolor=blue]{hyperref}
\newcommand{\msun}{$\text{M}_\odot$}

\newcommand{\teff}{$T_{\text{eff}}$}

\newcommand{\heh}{log\,$N$(He)/$N$(H)}
\newcommand{\logg}{\(\log g\)}
\newcommand{\het}{$^3$He}

\begin{document}

   \title{The Arizona-Montréal spectroscopic survey of hot subluminous stars \thanks{Dedicated to the memory of Prof. Gilles Fontaine}
   }

   \author{M. Latour\inst{1},
            E. M. Green\inst{2},
            M. Dorsch\inst{3}, 
            V. Van Grootel\inst{4},
            P. Chayer\inst{5},
            S. Charpinet\inst{6},
            U. Heber\inst{7},
            S.K. Randall\inst{8},\\ and
            X.-Y. Ma \inst{4}
          }

   \institute{Institut für Astrophysik und Geophysik, Georg-August-Universität Göttingen, Friedrich-Hund-Platz 1, 37077 Göttingen, Germany
\\ \email{marilyn.latour@uni-goettingen.de}
         \and Steward Observatory, University of Arizona, 933 N. Cherry Avenue, Tucson, AZ 85721, USA
         \and Institut f\"ur Physik und Astronomie, Universit\"at Potsdam, Haus 28, Karl-Liebknecht-Str. 24/25, 14476 Potsdam, Germany
             \and Space sciences, Technologies and Astrophysics Research (STAR) Institute, Universit\' e
  de Li\` ege, 19C All\'ee du 6 Ao\^ ut, B-4000 Li\` ege, Belgium
         \and Space Telescope Science Institute, 3700 San Martin Drive, 
Baltimore, MD 21218, USA
         \and Institut de Recherche en Astrophysique et Plan\'etologie, CNRS, Universit\'e de Toulouse, CNES, 14 Avenue Edouard Belin, 31400 Toulouse, France
         \and Dr. Karl Remeis-Observatory and Erlangen Centre for Astroparticle Physics, Friedrich-Alexander-Universit\"at Erlangen-N\"urnberg, Sternwartstr. 7, 96049 Bamberg, Germany
         \and ESO, Karl-Schwarzschild-Str. 2, 85748 Garching bei M\"unchen, Germany
             }

   \date{Received 19/09/2025; accepted 3/11/2025}

  \abstract
   {
   Hot subdwarf B (sdB) and O (sdO) type stars are evolved helium-burning objects that lost their hydrogen envelope before the helium flash when their progenitors were close to the tip of the red giant branch. They populate the extreme horizontal branch (EHB) in the Hertzsprung-Russel diagram (HRD). The mass distribution of canonical hot subdwarfs is expected to peak at the core mass required for helium ignition under degenerate conditions, in the 0.45 to 0.5 \msun\ range. However, non-degenerate helium ignition from intermediate-mass progenitor and non-canonical pathways, such as the merger of helium white dwarfs and delayed helium flashes, are also expected to contribute to the hot subdwarf population.
   }
   {Using the high-quality, homogeneous spectra of 336 hot subluminous star candidates from the Arizona-Montréal Spectroscopic Survey, we aim to improve our understanding of the atmospheric and stellar properties of hot subdwarf stars. Our focus is on the mass distribution of the different types of hot subdwarfs and their connections to the various formation scenarios. 
   }
   { We used large grids of model atmospheres to fit the observed spectra and derived their atmospheric parameters: effective temperature (\teff), surface gravity, and helium abundance. The model grids were further utilized to fit the spectral energy distribution of each star and the $Gaia$ parallax was used to compute the stellar parameters radius, luminosity, and mass.
   }
   {Our spectroscopic sample mostly consists of H-rich sdBs and sdOs, but also contains 41 He-rich sdOs. Additionally, the sample includes 11 intermediate-helium stars and 19 horizontal branch objects with \teff\ $\gtrsim$ 14 kK. 
   We detected the presence of helium stratification in six sdB stars with \teff\ around 30 kK, making them good candidates for also showing \het\ enrichment in their atmospheres.
Our sdB distribution along the EHB shows a gap near 33 kK, visible in both the Kiel (\logg - \teff) diagram and HRD, corroborating previous observations and predictions.
   The mass distributions of H-rich sdBs and sdOs are similar and centered around 0.47~\msun, consistent with the canonical formation scenario of helium ignition under degenerate conditions. Among the H-rich hot subdwarfs, we found no difference between the mass distributions of close binaries and apparently single stars. The He-sdOs have a significantly wider mass distribution than their H-rich counterparts, with an average mass of about 0.78~\msun. In the HRD, the He-sdOs lie on the theoretical helium main sequence for masses between 0.6 and 1~\msun. This strongly favors a merger origin for these He-rich objects. We identified a small number of candidate low-mass ($<$0.45~\msun) sdBs located below the EHB that might have originated from more massive progenitors. These low-mass sdBs preferentially show low helium abundances. Finally, we identified more than 80 pulsating stars in our sample and find these to fall into well-defined $p$- and $g$-mode instability regions. }
   {}

\authorrunning{Latour et al.}
\titlerunning{The Arizona-Montréal survey of hot subdwarf stars}
   \maketitle
%
%-------------------------------------------------------------------

\section{Introduction}

The majority of hot subdwarf B and O type stars (sdBs and sdOs) are evolved helium-burning objects that have been stripped of their hydrogen envelope before the helium flash, when their low-mass ($\lesssim 2.3$ \msun) progenitors were at the tip of the red giant branch (RGB, \citealt{2009ARA&A..47..211H,2016PASP..128h2001H,2024arXiv241011663H}). This is what is referred to as the canonical formation scenario \citep{1987ApJS...65...95S,1993ApJ...419..596D}. Hot subdwarfs essentially sit at the very hot end of the horizontal branch and are also referred to as extreme horizontal-branch stars (EHB).
The mass distribution of the canonical hot subdwarfs is expected to peak at the core mass required for helium ignition under degenerate conditions, which lies in the 0.45$-$0.5 \msun\ range \citep{1967ApJ...147..624I,1987ApJS...65...95S,SalarisCassisi2005}. The evolution of intermediate-mass progenitors ($\gtrsim$ 2.3 \msun), igniting core-He burning under non-degenerate conditions, is expected to produce lower core and total masses, down to $\sim$0.33 \msun\ \citep{2013EPJWC..4303002M,10.1088/2514-3433/adcf15}. The part of the Hertzsprung-Russell diagram (HRD) where hot subdwarfs are located can also be crossed by stars in other evolutionary phases, such as low-mass post asymptotic giant branch (post-AGB, \citealt{1991IAUS..145..363H}) stars and stars cooling down as He-core white dwarfs \citep[WDs,][]{1998A&A...339..123D,2016A&A...595A..35I}. More exotic stellar products, resulting from the merger of, among other possibilities, two He-core WDs, can also end up as core-He burning objects and be found close to the EHB and the adjacent helium main sequence in the HRD \citep{1984ApJ...277..355W,2002MNRAS.333..121S}. 

One way to disentangle the various evolutionary phases populating the HRD in the hot subdwarf region is via the masses of the stars. 
However, precise stellar mass measurements independent of evolutionary tracks can readily be obtained only for certain special cases, such as eclipsing binary systems. For hot subdwarfs, it has also been possible to derive accurate masses from asteroseismology for pulsating sdBs \citep{2002ApJS..139..487C,2005A&A...437..575C} and through light-curve modeling coupled with spectroscopic observations for close binaries exhibiting eclipses, reflection effects, or ellipsoidal deformations \citep{2007A&A...471..605V,2010ApJ...708..253F}. A compilation of literature masses derived for 22 hot subdwarf pulsators and binaries led to a first empirical mass distribution for hot subdwarfs, peaking at 0.47 \msun\ with a narrow range of 0.44$-$0.50~\msun\ containing 68\% of the stars \citep{2012A&A...539A..12F}. This aligned very well with the expectations from canonical stellar evolution theory. Moreover, the seismic models obtained for the 14 pulsators in the sample yielded predicted distances in good agreement with the parallaxes from the second \textit{Gaia} Data Release \citep{2019ApJ...880...79F}. 

The availability of accurate parallaxes, and thus distances, for most of the nearby ($\lesssim$3 kpc) hot subdwarfs from the \textit{Gaia} Data Release (EDR3, \citealt{gaia_edr3}) finally permits the mass to be estimated for large samples of hot subdwarfs based on the fit of their spectral energy distribution (SED). However, this requires a few additional key ingredients: observed magnitudes in different filters, theoretical model grids to compute the expected flux in these filters, and reliable estimates of the atmospheric parameters of the star, most notably the effective temperature and surface gravity \citep{2018OAst...27...35H}. 
Mass distributions obtained from a combination of spectroscopic analysis, parallaxes and SED fits for a large sample of hot subdwarfs were first presented in \citet{Lei2023_mass} using LAMOST spectra. While the authors found an average mass ($\sim$0.47 \msun) in line with the theoretical expectations for their hydrogen-rich sdBs, their distribution is very broad, extending from 0.2 to 1.0 \msun. Their mass distribution for hydrogen-rich sdO masses revealed itself to be unrealistically flat, with a very low mean mass of 0.36~\msun. Similarly, the mass estimates of hot subdwarfs in globular clusters, whose distances are relatively well constrained by methods other than parallax measurements, remain systematically lower than expected, despite significant improvements in model atmospheres and data obtained over the years (see, e.g. \citealt{monibidin2011,shotglas1}).
Note that most of the dispersion seen in the mass distributions from parallaxes and SED fits in the literature is due to the uncertainty on the atmospheric parameters - especially the surface gravity, which dominates the error budget in the mass determination - rather than intrinsic to the stars themselves.

Atmospheric parameters derived from high-quality and homogeneous spectroscopic samples are crucial for minimizing the scatter in the masses derived from parallaxes and SED fits and yielding distributions of sufficient quality for a meaningful comparison to predictions from different formation scenarios and population synthesis models \citep{2002MNRAS.336..449H,2003MNRAS.341..669H}. This is what we aim to do with our samples of hot subdwarf spectra from the Arizona-Montréal spectroscopic program, whose early results were reported in conference proceedings \citep{2004Ap&SS.291..267G,2005ASPC..334..363G,2008ASPC..392...75G,2014ASPC..481...83F}. The spectroscopic data consists of two samples with different wavelength coverage and resolution. The larger sample, which we refer to as the Bok sample, comprises spectra for 337 stars obtained at the 2.3m Bok telescope at the Kitt Peak observatory. The second sample consists of spectra for a subset of 116 stars (all from the Bok sample), observed with the MMT telescope. This sample is referred to as the MMT sample. All of these high-quality Bok and MMT spectra were fit with synthetic model spectra to estimate their atmospheric parameters: effective temperature (\teff), surface gravity (\logg), and helium abundance (\heh). The atmospheric parameters were then used, along with the $Gaia$ parallaxes and magnitudes from various photometric surveys, to perform SED fits and derive the stellar parameters: radius ($R$), luminosity ($L$), and mass ($M$).  

We present the results of our work as follows. Sections~\ref{sec:data} and~\ref{sec:method} include the description of our observational material and analysis methods. Section~\ref{sec:res} gives the results of our spectroscopic and SED fits for the Bok and MMT samples separately.
In the discussion (Sect.~\ref{sec:diss}), we focus on the mass distributions (Sects.~\ref {sec:diss:mass} to ~\ref{sec:diss:mass:b_ehb}) and examine the properties of the pulsating hot subdwarfs included in our sample in Sect.~\ref{sec:diss:puls}.
Finally, a summary of our work and our main conclusions are presented in Sect.~\ref{sec:end}.
  
%--------------------------------------------------------------------
\section{Observational material}\label{sec:data}

\subsection{Low-resolution Bok spectra}

\begin{figure*}
\resizebox{\hsize}{!}{
   \includegraphics{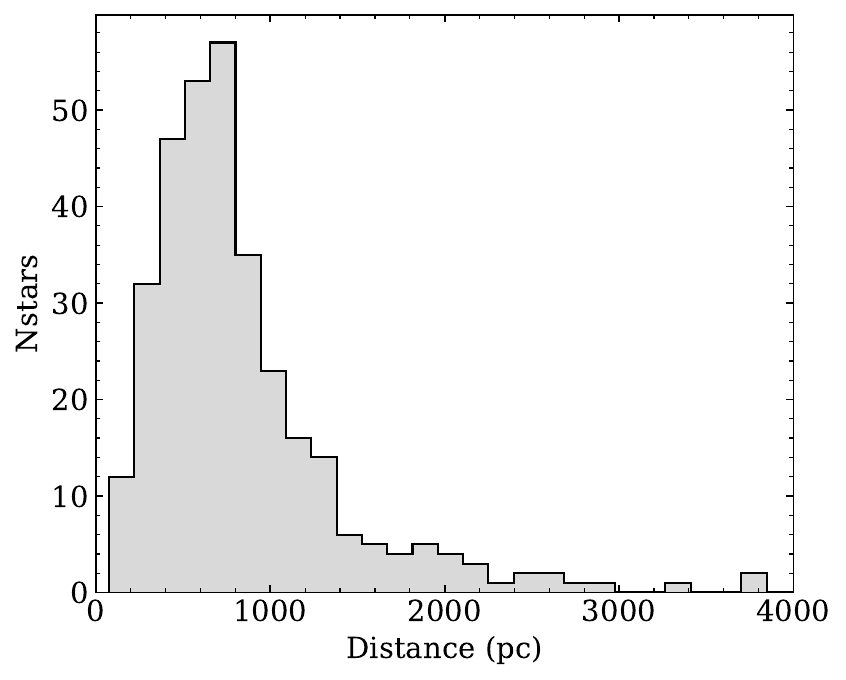}
   \includegraphics{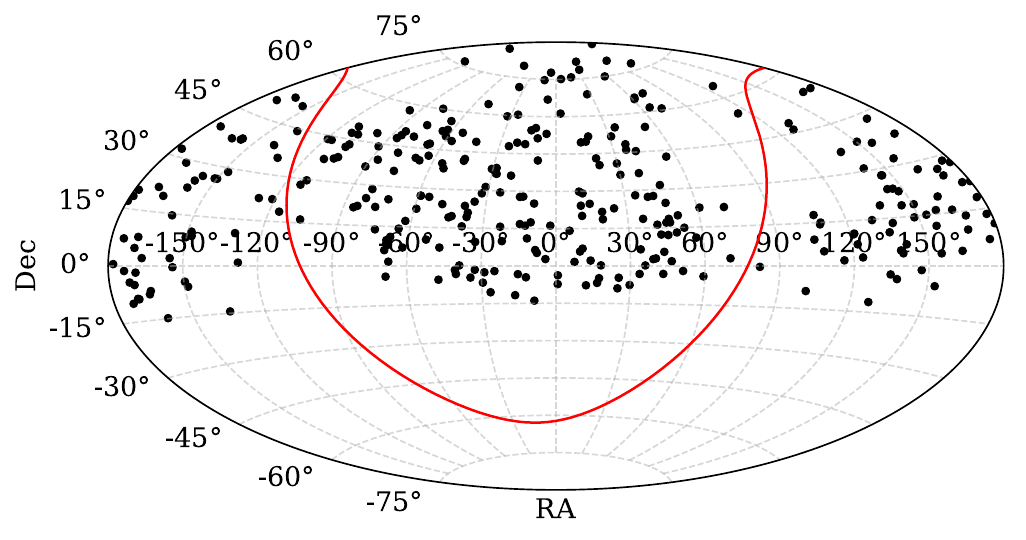}}\vspace{1pt}
     \caption{Distance and sky distribution of the stars in the Bok sample. Six objects are located beyond 4000 pc, but they are all most likely MS B-type stars (see Sect.~\ref{sec:res:bok:stellar}). The red line indicates the position of the galactic plane on the sky projection.
     }
     \label{fig:Sample}
\end{figure*}

Low-resolution spectra for 336 hot subluminous star candidates were obtained by one of us (E.M.G) as part of a project aiming to better characterize the hot subdwarf population \citep{2008ASPC..392...75G}. The stars were selected as relatively bright (V $\lesssim$ 14.5) blue objects visible from Arizona. They were mainly identified from early small-scale surveys of blue objects, such as Palomar-Green \citep{1986ApJS...61..305G}, Feige \citep{1958ApJ...128..267F}, and Kitt Peak Downes \citep{1986ApJS...61..569D}. All objects, except for the sdB binary \object{NGC 188 2019} \citep{1962ApJ...135..333S,2005ASPC..334..363G}, are members of the galactic field. The distances derived from the Gaia EDR3 parallaxes and the sky distribution of these 336 stars are shown in Fig.~\ref{fig:Sample}. 

The spectra were obtained at the Bok 2.3m telescope of the Steward Observatory on Kitt Peak, mainly between 1999 and 2004. The standard setup for the spectroscopic survey was to use the Bollers \& Chivens Cassegrain spectrograph with the 400 mm$^{-1}$ first-order grating with a 2.5\arcsec\ slit. This results in a typical resolution of 9~$\AA$ over the wavelength interval 3620$-$6900~\AA. Before each exposure, the instrument rotator was set to align with the parallactic angle. HeAr comparison spectra were taken immediately following the stellar exposures. 
The spectra were bias-subtracted, flat-fielded, background-subtracted, optimally extracted, wavelength-calibrated, and flux-calibrated using standard IRAF
\citep{1986SPIE..627..733T,1993ASPC...52..173T} tasks.
The flux calibration was performed using observations of the standard stars Feige\,34 or BD+28$\degr$4211. 
All individual spectra were also corrected for heliocentric velocities. The number of observations per object varies from star to star but is typically between one and ten. 
For stars with more than one observation, the radial velocity (RV) was measured via cross-correlation with the highest S/N spectrum for each given star. The spectra were then aligned in radial velocity prior to being combined into a single spectrum. We note that this way of extracting the RV is independent of synthetic spectra but provides only relative, rather than absolute, RVs. 
Except for a few faint objects, the signal-to-noise ratios ($S/N$) of the final spectra lie in the range 100$-$500. The median $S/N$ of the sample is 197. 

\subsection{Medium-resolution MMT spectra}

Additional medium-resolution spectra were obtained for a subset of 116 stars in the Bok sample by one of us (E.M.G.) with the blue spectrograph attached to the 6.5 m MMT telescope, between 1996 and 2003 as part of a radial velocity program \citep{2004Ap&SS.291..267G,2005ASPC..334..363G}. 
Throughout the observing seasons, the same experimental setup was consistently used: the 832 mm$^{-1}$ grating in second order and a 1\arcsec\ slit that provided a resolution $R$ of $\sim$4250 (1.0 \AA) over a wavelength range of 4000$-$4950 \AA. The spectra were reduced and combined in the same way as those obtained with the Bok telescope.

The MMT sample consists of hydrogen-rich sdBs and sdOs cooler than 45~kK. Again, a number (between 2 and 8) of individual spectra were combined. The $S/N$ for the combined spectra typically ranges from 100 to 200. Exposure times were mostly kept below 15 min and depended on the brightness of the star. Fig.~D.1 shows our MMT spectral atlas: the combined spectrum for each star, ordered by increasing effective temperature.

\subsection{Light curves}
\label{id_puls}
Most of the stars in the MMT sample were also observed as part of a photometric monitoring campaign (see \citealt{2003ApJ...583L..31G,2004Ap&SS.291..267G}) by one of us (E.M.G.) at the 1.6m Kuiper Telescope at Mount Bigelow (Steward Observatory) and at the 2.3m Telescope on Kitt Peak with conventional 2K CCDs binned 3 $\times$ 3 (0.45$\arcsec$ pixel $^{-1}$). Integration times were kept short, between 10 and 60 s, depending on the brightness of the target. The stars were typically observed for a total duration of 2 to 20 hours. 
Differential magnitudes were derived from aperture photometry of the sdB relative to reference stars of comparable magnitude within each frame.
As a result of these observations, we have information on the pulsation properties for all but six stars in the MMT sample, and can classify them as rapid ($p$-mode), slow ($g$-mode) or hybrid pulsators, or non-pulsating objects. 

Given that these ground-based observations date from earlier than 2005, many of the targets have since also been monitored for variability from space. We cross-checked our information from the ground-based data with the light curves obtained from the Transiting Exoplanet Survey Satellite (TESS, \citealt{Tess2014}). For all stars in our Bok sample, we retrieved and examined, when available, the Lomb-Scargle Periodograms of the TESS data, using our dedicated software FELIX (\citealt{2010A&A...516L...6C,2016A&A...585A..22Z}), to search for pulsation signals.
In addition, we searched for pulsators among the list of sdB pulsators from the Kepler extended K2 mission (W. Zong, priv. comm.)
and the literature compilation of pulsators made by \citet{2024yCat..36840118U}.
In total, we identified 22 $p$-mode pulsators, 53 $g$-mode pulsators, and 9 hybrid pulsators in our sample. The light-curves of 201 stars did not show any signal that could be associated with $p$- or $g$-modes pulsations and we classified them as non-pulsating. Finally, 44 stars do not have TESS light-curves and their pulsation status remains unknown.
The pulsating stars and their properties are further discussed in Sect.~\ref{sec:diss:puls}.

\section{Analysis Methods}\label{sec:method}

\subsection{Atmospheric parameters}\label{sec:method:spectro_fit}

Our samples include a large variety of hot subluminous stars, from cool blue horizontal branch (BHB) stars to extremely helium-enhanced sdO stars. In order to cover all the atmospheric parameter ranges of these stars, two different types of model atmospheres were used. 

For the spectral fit of the hydrogen-rich hot subdwarfs and BHB stars, we used the grids of hybrid LTE/NLTE models that have been used in similar studies (e.g., \citealt{2024A&A...690A.368G}, Heber et al., in prep., Dawson et al., in prep.). The model grid extends from 9~kK to 75~kK, from \logg\ = 3.0 to 7.0, and from \heh\ = $-$5.0 to 2.5. The lower limit on the surface gravity gradually increases to \logg\ = 5.25, for models at 75~kK, with increasing effective temperature, following the Eddington limit (see Heber et al., in prep. for a figure showing the grid coverage).
The models are constructed using the combination of \textsc{Atlas{\footnotesize12}/Detail/Surface} (ADS) codes. ADS is a hybrid LTE/NLTE method, which was first developed by \citet{przybilla2006} and \citet{nieva2007} and further improved by \citet{przybilla2011} and \citet{irrgang2014,irrgang2018,2021A&A...650A.102I,2022NatAs...6.1414I}. The improvements include a proper treatment of level dissolution for hydrogen \citep{1994A&A...282..151H} and the implementation of the Stark broadening line profiles of hydrogen and \ion{He}{i} \citep{2009ApJ...696.1755T,1997ApJS..108..559B}.
The synthetic spectra are obtained by running the three codes mentioned above in succession. First, a LTE line-blanketed, plane-parallel, homogeneous and hydrostatic model atmosphere is calculated using \textsc{Atlas{\footnotesize12}} \citep{Kurucz96}. The LTE atmospheric structure is then used by \textsc{Detail} \citep{giddings81,butler85} to calculate the population numbers of hydrogen and helium \citep{przybilla04,przybilla05} assuming NLTE and appropriate model atoms. The other chemical species are treated in LTE. 
In this work, we use metal abundances that reflect the typical abundance patterns of hot subdwarfs (see e.g. \citealt{2008ApJ...678.1329B,2011PhDT.......261P,2013MNRAS.434.1920N}). 
Most notably, iron-peak elements are enhanced by a factor of 10 with respect to solar, except iron itself which is kept at the solar value.  
The final synthetic spectra are computed with \textsc{Surface} \citep{giddings81, butler85} and include only lines of hydrogen and helium.

The second type of model atmospheres we used are tailored to the helium-enhanced objects. These models were used in \citet{2024PhDT........36D} and are described in their section 2.2. 
The model atmospheres and synthetic spectra were computed with \textsc{Tlusty}, version 205, and \textsc{Synspec}, version 51  (\citealt{2011ascl.soft09022H,2011ascl.soft09021H}, but see also \citealt{2017arXiv170601859H,2017arXiv170601935H,2017arXiv170601937H} for the specificity of the versions used here), respectively. 
They are NLTE models that include line-blanketing from C, N, O, Ne, Si, P, S, Fe, and Ni. The abundances are set to the solar values \citep{asplund09} except for C (2$\times$ solar), O (0.1$\times$ solar), Ne (2$\times$ solar), Fe (1.5$\times$ solar), and Ni (10$\times$ solar), which roughly represents the abundance patterns of most helium-enriched hot subdwarfs. 
Specific improvements were made in \textsc{Synspec} regarding the treatment of \ion{He}{i}, the most important being the consideration of Stark broadening following the line profiles calculated by \citet{1997ApJS..108..559B} and implemented in \textsc{Synspec} by \citet{2020ApJ...901...93B}. For the \ion{He}{i} lines at 4472~$\AA$ and 4922~$\AA$, the Stark profiles of \citet{stark4471} and \citet{stark4922} were used. The grid extends from \teff\ = 25 to 65 kK, from \logg\ = 3.0 to 6.5 and from \heh\ = $-$1.75 to $+$4.0. However, the lower limit on the surface gravity gradually reaches \logg\ = 4.5 (for models at 65 kK) as the effective temperature increases, following the Eddington limit.

The atmospheric parameters, namely \teff, \logg, and \heh, were derived for each stellar spectrum using a full spectral fit to one of the model grids described above. The fit to a given star is not restricted to one specific grid, instead, the best-fit solution can be found in either of the two grids, as long as the parameters do not lie on the grid's edge.
These fits were performed using the \textit{Interactive Spectral Interpretation System} (ISIS, \citealt{houck00}) with a modified version of the $\chi^2$-minimization method presented by \citet{irrgang2014}. 
The spectral continuum was modeled using a spline with anchor points spaced every 100\,\AA, but avoiding the hydrogen and helium lines. 
This continuum spline was fitted simultaneously with the atmospheric parameters, which is necessary to account for correlations.

\subsection{Spectral energy distribution and stellar parameters}\label{sec:method:SED}

To complement the spectral analysis, we also derive the stellar radius, luminosity, and mass for each star by performing a fit of its SED combined with the $Gaia$ parallax. The method was first described in \citet{2018OAst...27...35H} and is explained in more detail in section 2.3 of \citet{2024PhDT........36D}. 
Here we briefly summarize the main points. 
For each star, observed magnitudes are queried via Table Access Protocol (TAP) using the Astronomical Data Query Language (ADQL). This way, many photometric surveys can be queried through the VizieR database or other databases like the Astro Data Lab \citep{Fitzpatrick2014}. Some of the major surveys providing magnitudes for the stars in our sample are the Sloan Digital Sky Survey (SDSS, \citealt{sdss3}), the Two Micron All Sky Survey (2MASS, \citealt{2mass}), Pan-STARRS DR2 \citep{pan-starrsdr2}, GALEX \citep{galex}, and \textit{Gaia} \citep{gaia_edr3}. Also relevant here are the
\textit{Gaia} DR3 photometric low-resolution spectra \citep{gaia_spectra} that were used to construct a sequence of 14 box filters. An exhaustive list of the photometric catalogs queried is presented in \citet{2024A&A...685A.134C}.

\begin{figure*}
\sidecaption
   \includegraphics[width=12cm]{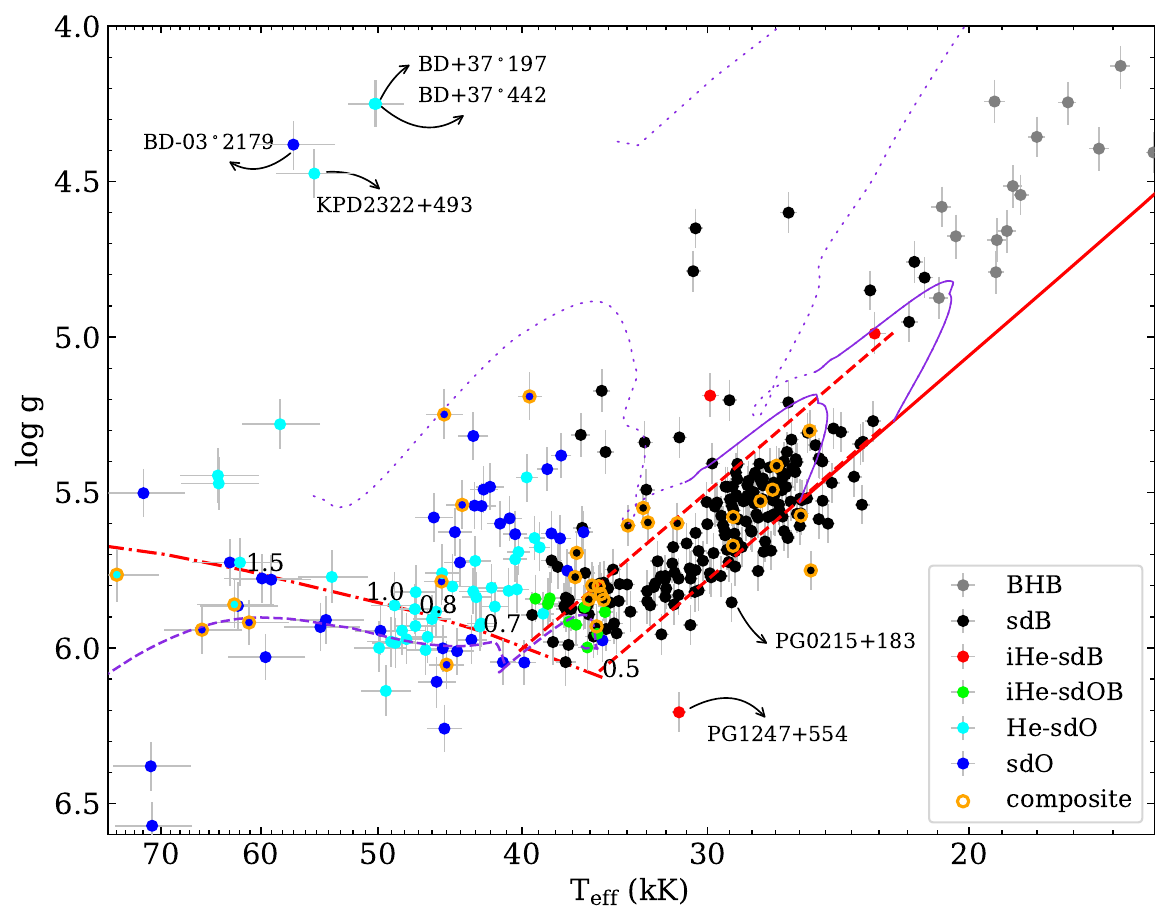}
     \caption{Surface gravity (log~$g$) as function of \teff\ (Kiel diagram) for the stars in the Bok sample. The six spectral groups are indicated with symbols of different colors. 
     Stars with IR excess are marked with an additional orange circle (see Sect.~\ref{sec:res:bok:stellar}). The ZAEHB and TAEHB computed with STELUM for a core mass of 0.47~\msun\ are shown with red dashed lines. The ZAEHB extension below 20 kK (solid red line) is from BaSTI models. Two BaSTI evolutionary tracks are shown in purple, the solid part represents the core-He burning phase while the dotted part is the post-EHB phase. The dashed purple track is from a late-flasher model. The dashed-dotted red line is the ZAHeMS with the stellar masses indicated along the line. References for theoretical models are listed in Sect.~\ref{sec:res:bok:atmo}.
     }\label{fig:Kiel}
\end{figure*}

\begin{figure}
\resizebox{\hsize}{!}{
   \includegraphics{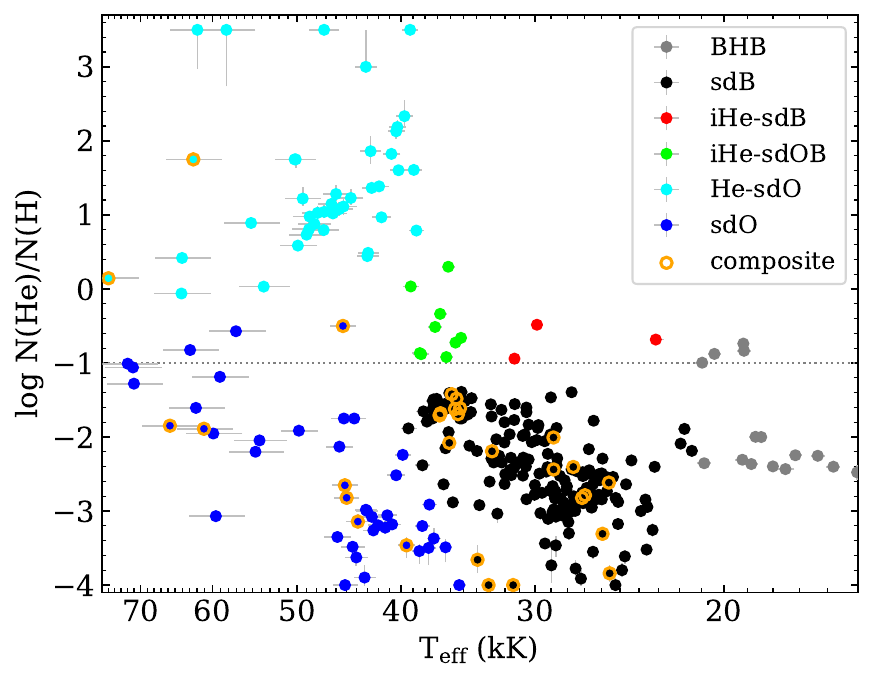}
}
     \caption{Helium abundance as a function of \teff\ for the stars in the Bok sample. The spectral groups are indicated following the same color scheme as in Fig.~\ref{fig:Kiel}. The solar helium abundance is indicated with the dotted line.
     }
     \label{fig:Teff_He}
\end{figure}

During the SED fit procedure, the distance to the star is kept fixed to the value derived from the \textit{Gaia} EDR3 parallax ($\varpi$). 
We correct the parallax for its zero-point offset following \citet{Lindegren2021} and inflate the corresponding uncertainty using the function suggested by \citet{El-Badry2021}.
The effective temperature, \logg, and helium abundance of the star are fixed to the value obtained from the spectroscopic fit of the Bok spectra. Thus the only parameters left to optimize are the angular diameter, $\Theta=2R/D$, or in terms of the parallax, $\Theta =2 \varpi R$, as well as the interstellar colour excess $E(44-55)$\footnote{$E$(44$-$55) is analogous to $E$($B$$-$$V$), but with the monochromatic measures of the extinction at 4400 and
5500 \AA\ substituting for measurements with the $B$ and $V$ filters. 
Conversion factors to the $UBV$ systems are given in table 4 of \citet{2019ApJ...886..108F}. They are close to 1 for hot stars.}. To account for interstellar extinction, we use the functions of \citet{2019ApJ...886..108F} and adopt the standard parameter $R(55)$=3.02. The model atmospheres used to compute the synthetic fluxes in the various photometric passbands are the same as used for the spectroscopic fits, i.e. the emergent flux from \textsc{Atlas{\footnotesize12}} and \textsc{Tlusty}, including metal lines. From the value of $\Theta$ we directly obtain the radius $R$ of the star and we compute the luminosity and the mass via the formulae 
\begin{equation*}
\centering
L=4\pi R^2 \sigma T^4 _{\rm eff}\qquad \text{and} \quad M=\frac{gR^2}{G}.
\end{equation*}
All uncertainties are propagated using the Monte Carlo method; the resulting best-fit values (for $R$, $L$, $M$) are stated as the median with the 68\% (1$\sigma$) uncertainties. 
For these computations,
we included systematic uncertainties, alongside the statistical uncertainties, for the spectroscopic \teff\ and \logg\ measurements, adding them in quadrature following the prescription detailed in Dawson et al. (in prep.). The authors used their spectroscopic observations of hot subdwarfs within 500~pc to characterize systematic uncertainties arising from three sources: variability in data reduction, offsets between different instruments, and the choice of metallicity in the model atmospheres used to fit the spectra. As a result, they provided third-order polynomial functions to evaluate systematic uncertainties as a function of \teff, for the three atmospheric parameters (\teff, \logg, and, \heh). The systematic uncertainties are about 0.07 dex for \logg\ and 1.5\,\% for \teff, although the uncertainty on \teff\ steadily increases to 5\% for stars hotter than 40 kK.
The surface gravity is typically the atmospheric parameter with the largest uncertainty and is also more affected by systematics (e.g., from fitting different spectra of the same star, see Dawson et al. in prep.). Because the stellar parameters $R$ and $L$ do not strongly depend on log~$g$, they can be relatively well constrained, provided that the $Gaia$ parallax is well determined. On the other hand, because $M$ depends linearly on the surface gravity, we are typically left with rather large uncertainties on the masses derived from the combination of spectroscopy, SED fits and parallaxes ($\sim$16-20\%, see Sect.~\ref{sec:diss}).

For some of our targets, the SED fit indicates an IR excess, attributed to the target being a binary system with a cool companion that is bright enough to emit more IR flux than the hot subdwarf itself (see App.~\ref{App:composites}). This is typically the case for main sequence companions earlier than spectral type M. In such cases, we fit the flux contribution of the companion with synthetic spectra for main sequence stars from the G\"{o}ttingen spectral library \citep{husser2013} by including two additional parameters: the surface ratio A$_\textnormal{MS}$/A$_\textnormal{sd}$ between the MS companion and the hot subdwarf and \teff\ of the companion. We assume the companion star to have a log~$g$ typical of MS stars (\logg\ = 4) and a metallicity of [Fe/H]~=~$-$0.3, which is a mean value for the F/G/K-type companions \citep{Vos2018}. 

\section{Results}\label{sec:res}

\subsection{The Bok sample}\label{sec:res:bok}

 \begin{figure*}
\resizebox{\hsize}{!}{
   \includegraphics{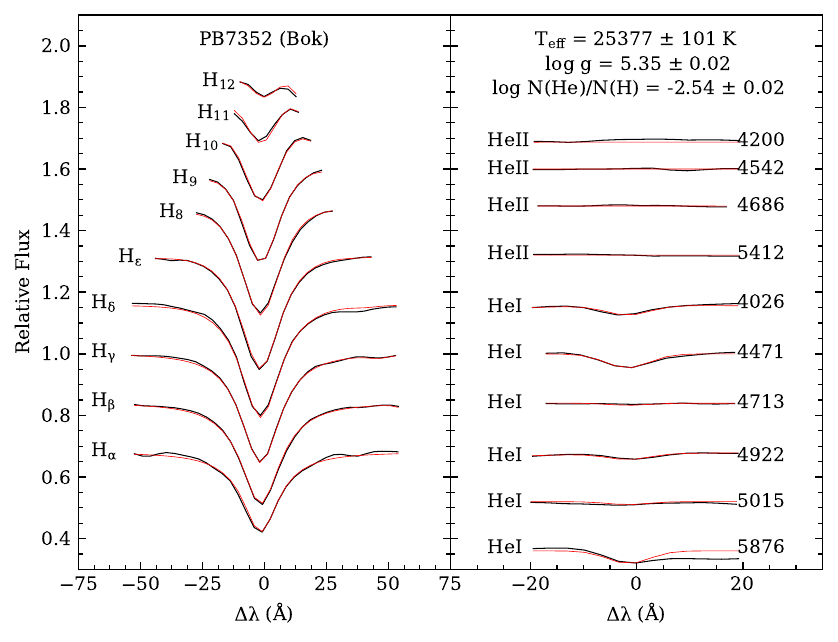}
   \includegraphics{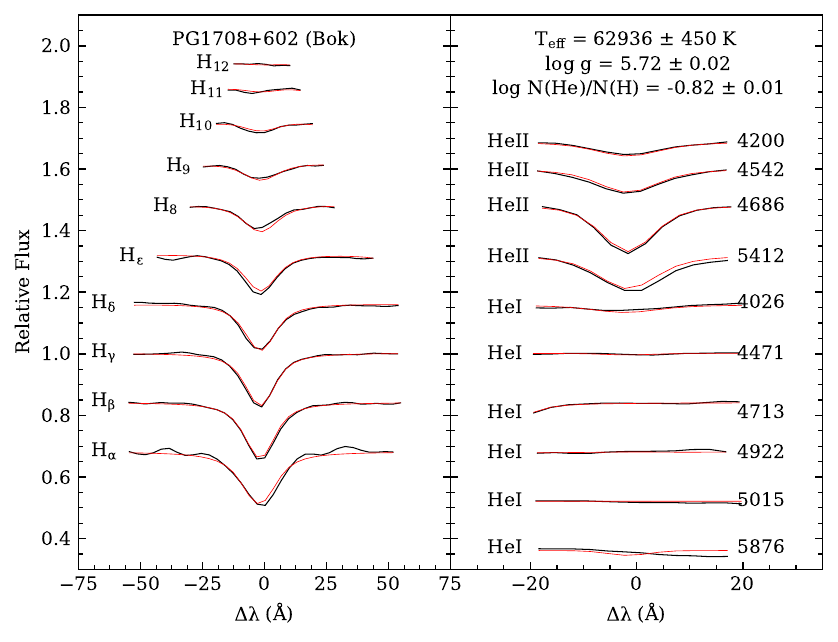}}\vspace{1pt}
\resizebox{\hsize}{!}{   
    \includegraphics{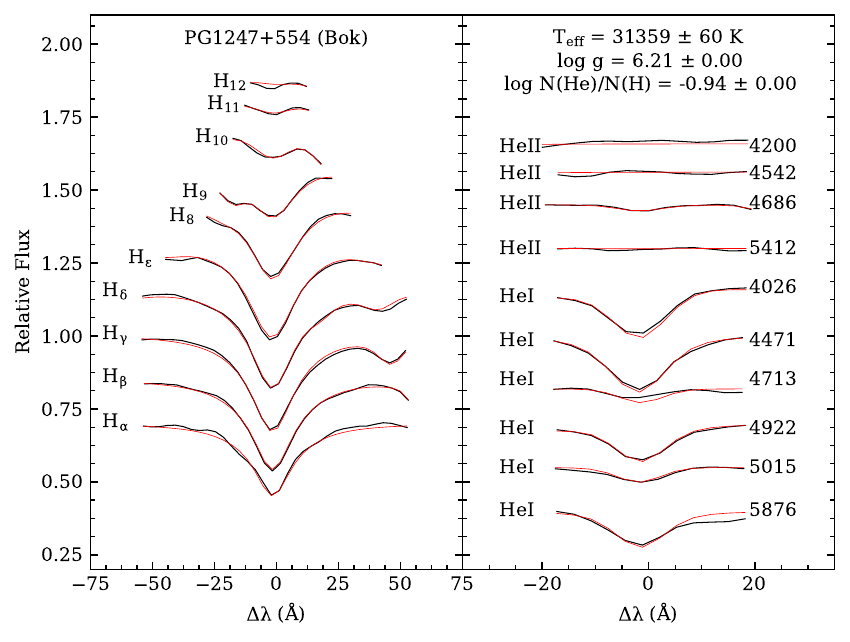}
   \includegraphics{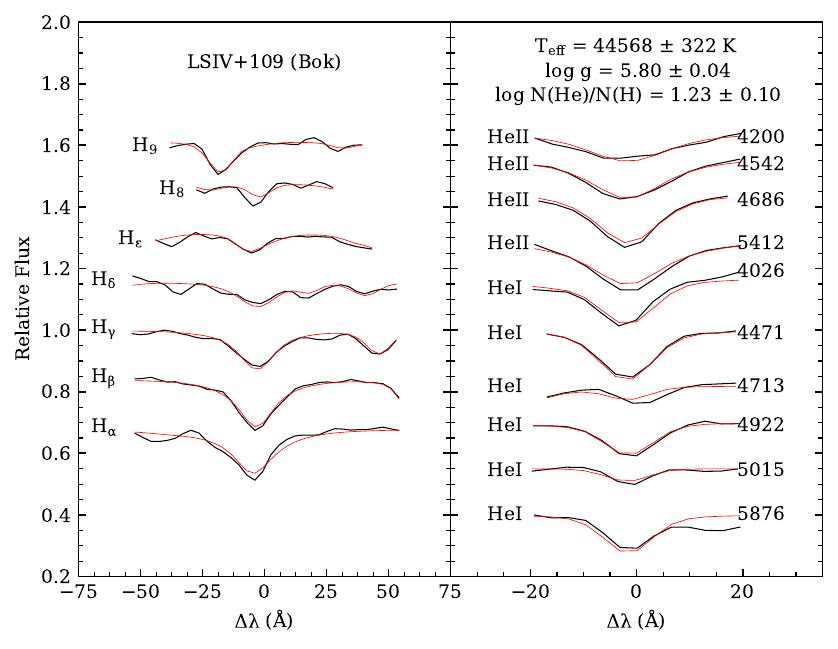}}\vspace{1pt}  
     \caption{Bok spectrum (black) and best fit solution (red) for four stars in our sample. PB7352 is a typical hydrogen-rich sdB and PG1708+602 is a hot sdO. PG1247+554 is an iHe-sdB located below the EHB in the Kiel diagram. LSIV+10$^{\circ}$4 is a He-sdO where the \ion{He}{ii} Pickering series blends with the Balmer lines.
     }
     \label{fig:bok_fits}
\end{figure*}

\subsubsection{Atmospheric parameters}\label{sec:res:bok:atmo}

All of the Bok spectra were fit following the method described in Sect.~\ref{sec:method:spectro_fit}. From the inspection of the spectral fits, five peculiar objects stood out and were removed from the sample because their spectra could not be modeled adequately. They are listed in Appendix~\ref{app:removed} and their spectra are shown in Fig.~\ref{fig:app:peculiar}. 
In Figs.~\ref{fig:Kiel} and \ref{fig:Teff_He} we show the distribution of the hot subdwarfs from the Bok sample in the \teff$-$log~$g$ plane (hereafter Kiel diagram) and in the \teff$-$He plane. Given the low resolution of the Bok spectra, the helium abundance in the most He-poor stars (\heh\ $\sim -4$) should be considered as an upper limit. Figure~\ref{fig:bok_fits} shows the Bok spectrum and best-fitting model for four stars of various \teff\ and helium abundances. The best fit for all stars in the Bok sample are available as supplementary figures online (see Sect.~\ref{sec:data_end}).
The atmospheric parameters obtained for all stars in the Bok sample are listed in Table~\ref{table_res_all}. This table also includes the stellar parameters obtained from the parallax and SED fits, as well as additional information on the pulsation and binary properties of each star. An extended version of Table~\ref{table_res_all}, with additional columns, is only available online at the CDS (see Sect.~\ref{sec:data_end}). In Table~\ref{table_res_short} we present an excerpt of Table~\ref{table_res_all} for 15 stars.

\begin{table*}[h]
    \centering
    \scriptsize
    \caption{\label{table_res_short} Excerpt from the table of results for 15 stars}
    \begin{tabular}{l c c c c c c c c c c c}        
    \toprule\toprule
    \noalign{\vskip4bp}
Star & Type & \teff\ & log~$g$ & \heh\  & $R$ & $L$ & $M$ & IR-excess & $Gaia$ \texttt{ruwe} & Pulsation & Binarity \\ 
 & & (K) & (cm s$^{-2}$) &  & ($R_\odot$) & ($L_\odot$) & ($M_\odot$) & &  &  &  \\
\hline
PG0001+275 & sdB & $27019${\raisebox{0.5ex}{\tiny$^{+42}_{-44}$}} & $5.58${\raisebox{0.5ex}{\tiny$^{+0.01}_{-0.01}$}} &$-2.85${\raisebox{0.5ex}{\tiny$^{+0.03}_{-0.03}$}} & $0.187${\raisebox{0.5ex}{\tiny$^{+0.005}_{-0.004}$}} &$16.9${\raisebox{0.5ex}{\tiny$^{+1.1}_{-1.0}$}} & $0.48${\raisebox{0.5ex}{\tiny$^{+0.08}_{-0.07}$}} & no & 1.01  & g & binary \\ 
PG0004+133 & sdB & $28650${\raisebox{0.5ex}{\tiny$^{+71}_{-149}$}} & $5.46${\raisebox{0.5ex}{\tiny$^{+0.01}_{-0.01}$}} &$-1.88${\raisebox{0.5ex}{\tiny$^{+0.02}_{-0.02}$}} & $0.204${\raisebox{0.5ex}{\tiny$^{+0.004}_{-0.004}$}} &$25.2${\raisebox{0.5ex}{\tiny$^{+1.5}_{-1.5}$}} & $0.43${\raisebox{0.5ex}{\tiny$^{+0.07}_{-0.06}$}} & no & 0.93  & no & binary \\ 
PG0009+036 & MS-B & $18698${\raisebox{0.5ex}{\tiny$^{+188}_{-221}$}} & $4.55${\raisebox{0.5ex}{\tiny$^{+0.03}_{-0.03}$}} &$-1.90${\raisebox{0.5ex}{\tiny$^{+0.04}_{-0.03}$}} & $1.884${\raisebox{0.5ex}{\tiny$^{+0.392}_{-0.283}$}} &$390.3${\raisebox{0.5ex}{\tiny$^{+182.4}_{-110.5}$}} & $4.66${\raisebox{0.5ex}{\tiny$^{+2.37}_{-1.43}$}} & no & 1.01  & ... & ... \\ 
PG0011+221 & sdO & $42039${\raisebox{0.5ex}{\tiny$^{+179}_{-176}$}} & $5.48${\raisebox{0.5ex}{\tiny$^{+0.02}_{-0.02}$}} &$-3.19${\raisebox{0.5ex}{\tiny$^{+0.05}_{-0.05}$}} & $0.191${\raisebox{0.5ex}{\tiny$^{+0.007}_{-0.007}$}} &$103.1${\raisebox{0.5ex}{\tiny$^{+12.1}_{-10.8}$}} & $0.41${\raisebox{0.5ex}{\tiny$^{+0.08}_{-0.07}$}} & no & 0.97  & no & binary \\ 
PG0011+283 & sdB & $24742${\raisebox{0.5ex}{\tiny$^{+152}_{-176}$}} & $5.47${\raisebox{0.5ex}{\tiny$^{+0.02}_{-0.01}$}} &$-3.61${\raisebox{0.5ex}{\tiny$^{+0.08}_{-0.09}$}} & $0.176${\raisebox{0.5ex}{\tiny$^{+0.004}_{-0.004}$}} &$10.5${\raisebox{0.5ex}{\tiny$^{+0.8}_{-0.7}$}} & $0.33${\raisebox{0.5ex}{\tiny$^{+0.06}_{-0.05}$}} & no & 0.68  & g & single \\ 
PG0014+068 & sdB & $35638${\raisebox{0.5ex}{\tiny$^{+102}_{-178}$}} & $5.93${\raisebox{0.5ex}{\tiny$^{+0.01}_{-0.02}$}} &$-1.62${\raisebox{0.5ex}{\tiny$^{+0.02}_{-0.02}$}} & $0.116${\raisebox{0.5ex}{\tiny$^{+0.018}_{-0.014}$}} &$19.5${\raisebox{0.5ex}{\tiny$^{+6.6}_{-4.5}$}} & $0.42${\raisebox{0.5ex}{\tiny$^{+0.16}_{-0.11}$}} & yes & 1.15  & p & ... \\ 
PG0032+247 & sdO & $38221${\raisebox{0.5ex}{\tiny$^{+147}_{-159}$}} & $5.63${\raisebox{0.5ex}{\tiny$^{+0.02}_{-0.02}$}} &$-3.20${\raisebox{0.5ex}{\tiny$^{+0.07}_{-0.07}$}} & $0.172${\raisebox{0.5ex}{\tiny$^{+0.008}_{-0.008}$}} &$56.7${\raisebox{0.5ex}{\tiny$^{+6.9}_{-6.0}$}} & $0.46${\raisebox{0.5ex}{\tiny$^{+0.10}_{-0.08}$}} & no & 0.99  & ... & single \\ 
PG0033+266 & sdB & $26706${\raisebox{0.5ex}{\tiny$^{+158}_{-108}$}} & $5.58${\raisebox{0.5ex}{\tiny$^{+0.02}_{-0.02}$}} &$-2.45${\raisebox{0.5ex}{\tiny$^{+0.03}_{-0.03}$}} & $0.182${\raisebox{0.5ex}{\tiny$^{+0.007}_{-0.007}$}} &$15.2${\raisebox{0.5ex}{\tiny$^{+1.4}_{-1.3}$}} & $0.46${\raisebox{0.5ex}{\tiny$^{+0.08}_{-0.07}$}} & no & 1.11  & no & single \\ 
PG0039+135 & He-sdO & $47860${\raisebox{0.5ex}{\tiny$^{+145}_{-156}$}} & $5.97${\raisebox{0.5ex}{\tiny$^{+0.03}_{-0.03}$}} &$1.03${\raisebox{0.5ex}{\tiny$^{+0.08}_{-0.06}$}} & $0.195${\raisebox{0.5ex}{\tiny$^{+0.008}_{-0.007}$}} &$180.6${\raisebox{0.5ex}{\tiny$^{+30.2}_{-26.3}$}} & $1.31${\raisebox{0.5ex}{\tiny$^{+0.29}_{-0.24}$}} & no & 0.95  & no & ... \\ 
PG0057+155 & sdB & $34736${\raisebox{0.5ex}{\tiny$^{+65}_{-91}$}} & $5.75${\raisebox{0.5ex}{\tiny$^{+0.01}_{-0.01}$}} &$-1.68${\raisebox{0.5ex}{\tiny$^{+0.01}_{-0.01}$}} & $0.143${\raisebox{0.5ex}{\tiny$^{+0.003}_{-0.003}$}} &$26.9${\raisebox{0.5ex}{\tiny$^{+1.9}_{-1.7}$}} & $0.42${\raisebox{0.5ex}{\tiny$^{+0.07}_{-0.06}$}} & no & 1.03  & no & single \\ 
PG0101+039 & sdB & $27108${\raisebox{0.5ex}{\tiny$^{+51}_{-72}$}} & $5.53${\raisebox{0.5ex}{\tiny$^{+0.01}_{-0.01}$}} &$-2.75${\raisebox{0.5ex}{\tiny$^{+0.03}_{-0.03}$}} & $0.187${\raisebox{0.5ex}{\tiny$^{+0.004}_{-0.004}$}} &$17.0${\raisebox{0.5ex}{\tiny$^{+1.0}_{-1.0}$}} & $0.43${\raisebox{0.5ex}{\tiny$^{+0.07}_{-0.06}$}} & no & 0.85  & g & sd+WD \\ 
PG0105+276 & He-sdO & $64031${\raisebox{0.5ex}{\tiny$^{+800}_{-710}$}} & $5.47${\raisebox{0.5ex}{\tiny$^{+0.04}_{-0.04}$}} &$0.42${\raisebox{0.5ex}{\tiny$^{+0.03}_{-0.00}$}} & $0.275${\raisebox{0.5ex}{\tiny$^{+0.046}_{-0.035}$}} &$1150.5${\raisebox{0.5ex}{\tiny$^{+546.4}_{-357.0}$}} & $0.82${\raisebox{0.5ex}{\tiny$^{+0.36}_{-0.23}$}} & no & 0.94  & ... & ... \\ 
PG0108+195 & sdO & $45672${\raisebox{0.5ex}{\tiny$^{+284}_{-326}$}} & $6.11${\raisebox{0.5ex}{\tiny$^{+0.03}_{-0.03}$}} &$-2.13${\raisebox{0.5ex}{\tiny$^{+0.05}_{-0.05}$}} & $0.122${\raisebox{0.5ex}{\tiny$^{+0.006}_{-0.005}$}} &$58.0${\raisebox{0.5ex}{\tiny$^{+9.4}_{-8.1}$}} & $0.69${\raisebox{0.5ex}{\tiny$^{+0.16}_{-0.13}$}} & no & 0.93  & no & single \\ 
PG0123+159 & sdB & $29215${\raisebox{0.5ex}{\tiny$^{+104}_{-133}$}} & $5.62${\raisebox{0.5ex}{\tiny$^{+0.02}_{-0.02}$}} &$-2.41${\raisebox{0.5ex}{\tiny$^{+0.03}_{-0.04}$}} & $0.185${\raisebox{0.5ex}{\tiny$^{+0.008}_{-0.007}$}} &$22.4${\raisebox{0.5ex}{\tiny$^{+2.2}_{-1.9}$}} & $0.52${\raisebox{0.5ex}{\tiny$^{+0.10}_{-0.08}$}} & no & 1.09  & no & single \\ 
PG0133+114 & sdB & $29112${\raisebox{0.5ex}{\tiny$^{+104}_{-145}$}} & $5.68${\raisebox{0.5ex}{\tiny$^{+0.01}_{-0.02}$}} &$-2.43${\raisebox{0.5ex}{\tiny$^{+0.02}_{-0.02}$}} & $0.155${\raisebox{0.5ex}{\tiny$^{+0.005}_{-0.005}$}} &$15.5${\raisebox{0.5ex}{\tiny$^{+1.3}_{-1.2}$}} & $0.42${\raisebox{0.5ex}{\tiny$^{+0.07}_{-0.06}$}} & no & 0.83  & no & sd+WD \\ 
\bottomrule
\end{tabular}
\tablefoot{
Uncertainties on the atmospheric parameters (\teff, \logg, and, \heh) are only statisticals.
}
\end{table*}

\begin{figure*}
\sidecaption
%\resizebox{\hsize}{!}{
   \includegraphics[width=12cm]{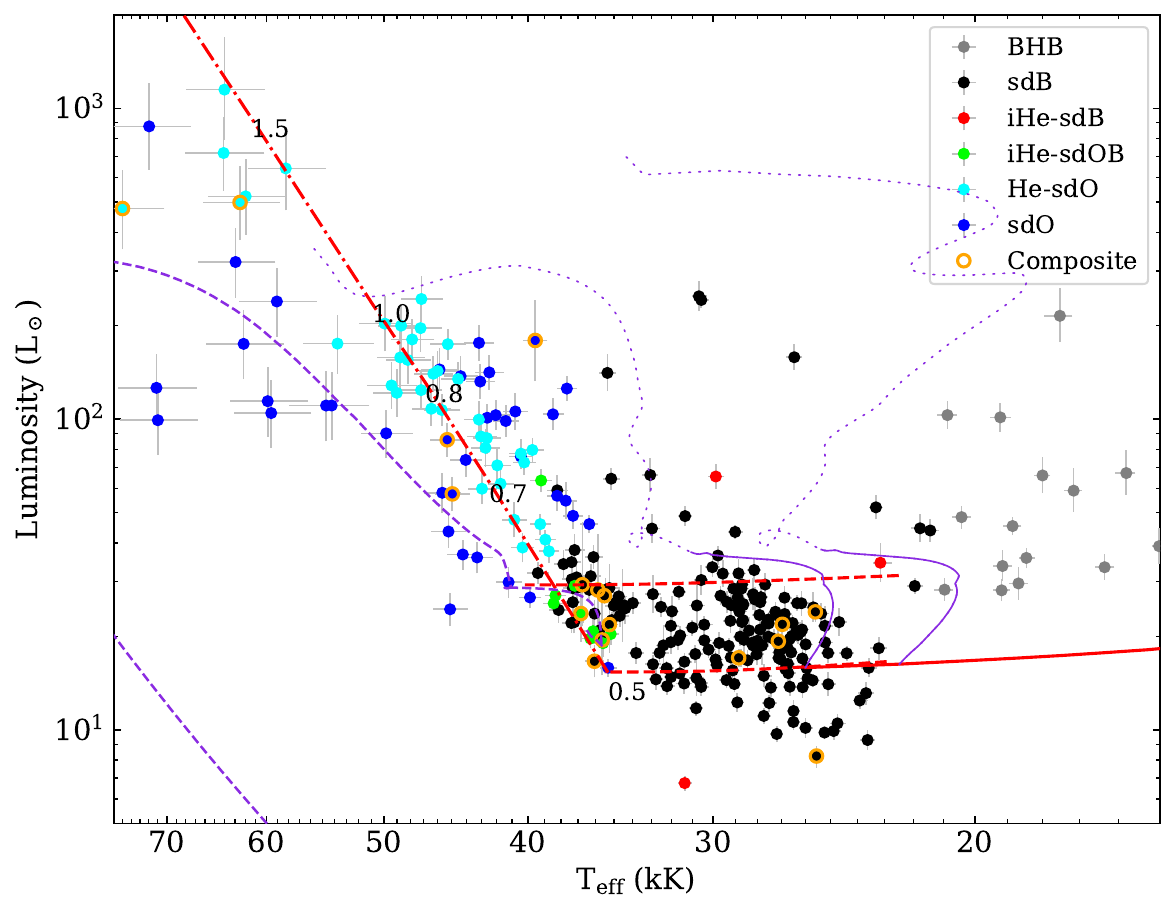}
     \caption{Luminosity of our stars versus their \teff\ (i.e., HRD). The different spectral types are color-coded as in the previous figures. We indicate the composite objects with an additional orange circle around the symbols.  
     The evolutionary tracks are the same as in the Kiel diagram (Fig.~\ref{fig:Kiel}).
     The four luminous post-AGB stars mentioned in Sect.~\ref{sec:res:bok:atmo} are outside the luminosity range shown here.
     }
     \label{fig:HRD}
\end{figure*}

\paragraph{Evolutionary models.} 
On the Kiel diagram (Fig.~\ref{fig:Kiel}) and some other figures in this work, we include evolutionary tracks that we describe here. 
The zero-age and terminal-age extreme horizontal branches (respectively ZAEHB and TAEHB) shown in Figs.~\ref{fig:Kiel},~\ref{fig:HRD}, and \ref{fig:Teff_mass} are those predicted from static stellar models computed with STELUM for a fixed core mass of 0.47~\msun\ and various hydrogen envelope thicknesses (see \citealt{2024A&A...686A..65B} and Sect. 5.2 for a detailed description of the models). 
We extended the ZAEHB to the cooler HB region with a model taken from the BaSTI database \citep{2018ApJ...856..125H}\footnote{\url{http://basti-iac.oa-teramo.inaf.it/}} for solar metallicity. Two EHB evolutionary tracks from BaSTI and one late flasher model track \citep{2008A&A...491..253M} are also included. The two BaSTI tracks show that the cooler sdBs, having a thicker hydrogen envelope, decrease in \logg\ during their post-EHB evolution. In contrast, sdBs at the hot end of the EHB, modeled as the products of a delayed helium flash, have very little hydrogen envelope left and evolve towards hotter temperatures at almost constant \logg\ (see also \citealt{2017A&A...599A..54X}).
Finally, we also indicate the zero age helium main sequence (HeZAMS, \citealt{1971AcA....21....1P}) that is located at the hot end of the EHB. 

\paragraph{Spectral classification.}
We divided the stars into different groups based on their helium abundance and effective temperature, following the classification presented in Fig. 4.1.4 of \citet{2024PhDT........36D}.
Besides bona fide hot subdwarf stars, our sample also includes some cooler objects that are mostly BHB stars, with ten potential main sequence (MS) B stars\footnote{These are not included in the figures.} (see Sect.~\ref{sec:res:bok:stellar}). Our classification uses the following categories (see also Fig.~\ref{fig:Teff_He}): BHB for the cooler HB objects (\teff\ < 21 kK), sdB and sdO for the objects that have an hydrogen-rich atmosphere, iHe-sdOB and iHe-sdB for the few intermediate-helium objects with \heh\ between about $-$1.0 and 0.6, and, finally, He-sdOs for the hot stars (\teff\ $\gtrsim$ 38 kK) with a helium dominated atmosphere. 

The solar helium abundance has been traditionally used as a divide between helium-poor and helium-rich sdBs, however from our distribution we see that the helium abundances in none of our sdBs reach the solar value. Instead, the abundances clearly plateau at about \heh=$-$1.4. This plateau of the sdB helium abundances is also conspicuous among the 500 pc sample (Dawson et al., in prep.) and consistent with the early results of \citet{2003A&A...400..939E}. Thus, we believe that the separation between the H-rich sdBs and the intermediate-helium subdwarfs should be made at a value of \heh=$-$1.2, rather than $-$1.0.
 
We note that our classification is intended to examine and compare mass distributions by class.
Thus the separation between the sdBs and sdOs mainly aims at separating the typical hydrogen-rich subdwarfs on the EHB (sdBs) from their hotter evolved post-EHB counterparts (sdOs) and is not strictly based on effective temperature or strength of the \ion{He}{ii} lines. The transition from sdB to sdO is gradual, and both classes are connected from an evolutionary perspective \citep{2009ARA&A..47..211H}. 
The separation between the cool sdBs and the BHB is also ambiguous.
The position of the Momany jump separating the BHB stars from the EHB is known to be around 18$-$20 kK \citep{1976ApJ...204..804N,momany02,brown16}. When looking at our stars in Figs.~\ref{fig:Kiel} and \ref{fig:Teff_He} we see instead a discontinuity around 23~kK, with the cooler stars being found at lower log~$g$, close to the TAEHB in Fig.~\ref{fig:Kiel}, and clustering at a helium abundance close to \heh\ = $-$2 (Fig.~\ref{fig:Teff_He}). Whether the few stars between 20 and 23 kK are sdBs or BHBs remains uncertain.

\paragraph{Peculiar objects.}

Our sample contains four luminous hot stars (50$-$60~kK) with \logg\ $\leq$ 4.5 (indicated in Fig.~\ref{fig:Kiel}). These are the two very similar extreme-He stars \object{BD+37$\degr$1977} and \object{BD+37$\degr$442}, which likely have surface gravities slightly below the limit of our model grid at 50 kK (i.e., log~$g <$ 4.25, see \citealt{2020MNRAS.496..718J} and references therein), the He-sdO \object{KPD2322+4933} (\heh=$0.9$), and the sdO \object{BD$-3\degr$2179} (\heh=$-$0.6). These luminous objects are likely associated with the post-AGB phase. Another interesting object is the He-sdO \object{GSC03214-02615} (\teff=43\,kK, \logg=5.7, \heh=0.44) that is a candidate magnetic sdO. Its spectrum, displayed in Fig.~\ref{fig:app:peculiar}, shows the strong absorption feature around 4620~\AA, which is still of unidentified origin. This feature is present in other stars with similar atmospheric parameters, whether they show the Zeeman splitting indicative of magnetic field or not \citep{2024A&A...691A.165D}.

\subsubsection{Stellar parameters}\label{sec:res:bok:stellar}

\begin{figure}
\resizebox{\hsize}{!}{
   \includegraphics{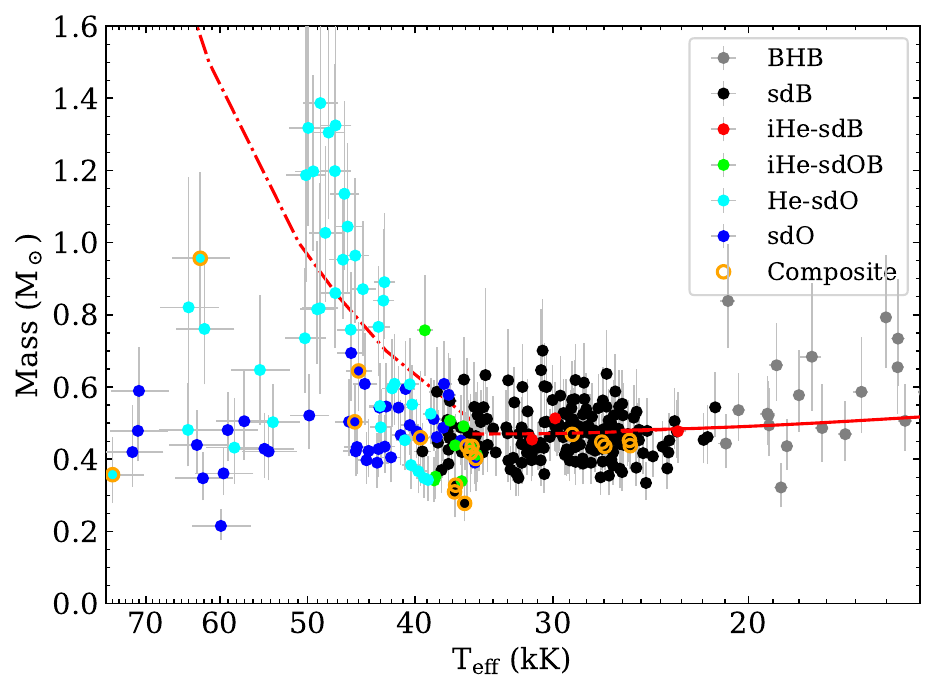}
}
     \caption{Mass as a function of \teff\ for the stars in the Bok sample. The spectral types are indicated following the same color scheme as in Fig.~\ref{fig:HRD}. The masses expected from the STELUM and BaSTI models for the ZAEHB are indicated by dashed and solid red lines respectively. Masses expected from the ZAHeMS are indicated with the dashed-dotted line.
     }
     \label{fig:Teff_mass}
\end{figure}

Having determined the atmospheric parameters of our stars, we proceed with the fit of the SEDs as described in Sect.~\ref{sec:method:SED}. 
The main goal of the SED fits is to derive reliable stellar parameters for the stars in our sample.
This requires the $Gaia$ parallax to be well measured. Thus, in the figures and the analyses involving $R$, $L$, and $M$, we always exclude stars with a parallax error larger than 20\% and those with \texttt{ruwe}\footnote{the re-normalized unit-weight error is a measure of the astrometric goodness of fit.} $>$ 1.4 \citep{El-Badry2021}.
We also inspected the SED fit result for each star and removed stars that could not be properly reproduced. For example, the SED of \object{PB5450} (also known as EQ Psc) is strongly affected by the irradiation of its companion (see \citealt{2019MNRAS.489.1556B}). 
We show in Fig.~\ref{fig:HRD} the distribution of the 300 remaining stars in terms of luminosity versus \teff\ (the HRD). 
From the masses derived with the parallaxes and SED fits, we identified ten stars among the BHB sample that have masses above 3~\msun, indicating that they are more likely to be MS objects although the less massive ones might be intermediate-mass stripped stars \citep{2022NatAs...6.1414I,2023MNRAS.525.5121V}. 
From the SED fits, we identified 29 stars with an IR excess consistent with MS companions of \teff\ between 3000 and 5900 K (see Fig.~\ref{fig:app:HR_comp}), which are marked with an additional orange circle\footnote{We note that many of the stars with an IR-excess have \texttt{ruwe} $>$ 1.4 and thus do not appear in the Figs.~\ref{fig:HRD} and \ref{fig:Teff_mass}. However they are all listed in Table~\ref{table_composites}.} in Figs.~\ref{fig:Kiel} to \ref{fig:Teff_mass}.
The properties of these systems are listed in Table~\ref{table_composites} and several are discussed in Appendix~\ref{App:composites}. 
The composite systems include a few known short and long period binaries (e.g. \citealt{2000MNRAS.311..877M,1998ApJ...502..394S,2019MNRAS.482.4592V}). 
We note that the spectra of these composite systems were fitted in the same way as all stars, only with a model atmosphere for the hot subdwarf component. For most of them, we do not see contamination from the companion in the Bok spectrum. This is the case for PG0014+068, among others (see Appendix~\ref{sec:appD}). Only in stars with more luminous G-type companions, such as PG1701+359 we do see spectral features from the companion, like the Mg triplet lines ($\sim$5170~\AA). The wavelength regions contaminated by features from the companion are ignored during the spectral fit. Nevertheless, the atmospheric parameters obtained for the stars with the strongest IR-excess might be less accurate than for the other stars.

On the HRD, most of the sdBs are found on the theoretical HB, but a noticeable fraction are located below the ZAEHB. These ``underluminous'' stars are further discussed in Sect.~\ref{sec:diss:mass:b_ehb}. The luminous sdBs and the sdOs are consistent with a post-EHB phase as suggested by the evolutionary tracks. Interestingly, the hydrogen-rich sdOs are mainly divided into two groups, found on each side of the ZAHeMS. The group on the hot side of the ZAHeMS (the hottest sdOs) is consistent with the evolutionary track starting at the hot end of the EHB, meaning they probably evolved from the hottest sdBs (those with the smallest hydrogen envelope, often called sdOBs), while the cooler post-EHB stars may be the progeny of cooler sdBs. This division of the sdOs into two regions is also visible in the Kiel diagram (Fig.~\ref{fig:Kiel}). 
Another noteworthy feature visible in both the HR and Kiel diagrams is a gap in the \teff\ distribution of sdB stars around 33$-$34\,kK, corresponding to a luminosity of about 20 $L_{\odot}$.
Such a drop in the density of sdBs along the EHB was also reported in \citet{2022A&A...661A.113G} who noticed a void of stars around 33~kK and log~$g$ = 5.7 in the Kiel diagram of their hot subdwarfs sample. 
This separation of the sdBs into two regions, with one populating the very hot end of the EHB is reminiscent of the dichotomy found by \citet{2017A&A...599A..54X} in their evolutionary models. In their models, the hotter sdBs are those with almost no H-rich envelope left because they experienced flash-mixing in their atmosphere due to a delayed He-flash. This is essentially the late-flasher scenario, first presented in \citet{1996ApJ...466..359D} to explain the population of blue hook stars in some massive globular clusters. As for the cooler sdBs, they follow the canonical formation scenario where the He-flash happens at the tip of the RGB. We note that the presence of a gap between the canonical EHB stars and the hotter He-enriched blue-hook stars is also observed in $\omega$~Cen and NGC\,2808 \citep{brown16} and happens around 32$-$33~kK \citep{shotglas1}. 
Finally, the position of the majority of He-sdOs is remarkably consistent with that of the helium main sequence of \citet{1971AcA....21....1P}. Following the theoretical masses along this sequence, our He-sdOs would have masses that increase with \teff, mainly from 0.6 to 1.0 \msun. 

Figure~\ref{fig:Teff_mass} shows that our masses do not have a trend with \teff\footnote{This could not be achieved for the MMT sample, see Sect.~\ref{sec:res:mmt:params}.}, with the  
exception of the He-sdOs. 
This is the expected behavior because the effective temperatures of EHB stars are determined by the extent of their hydrogen envelopes, which are always negligible in mass ($\lesssim$0.02~\msun) compared to the core. Even for the BHB stars, the increase in mass is very small as shown by the BaSTI track. 

 \begin{figure}
\resizebox{\hsize}{!}{
   \includegraphics{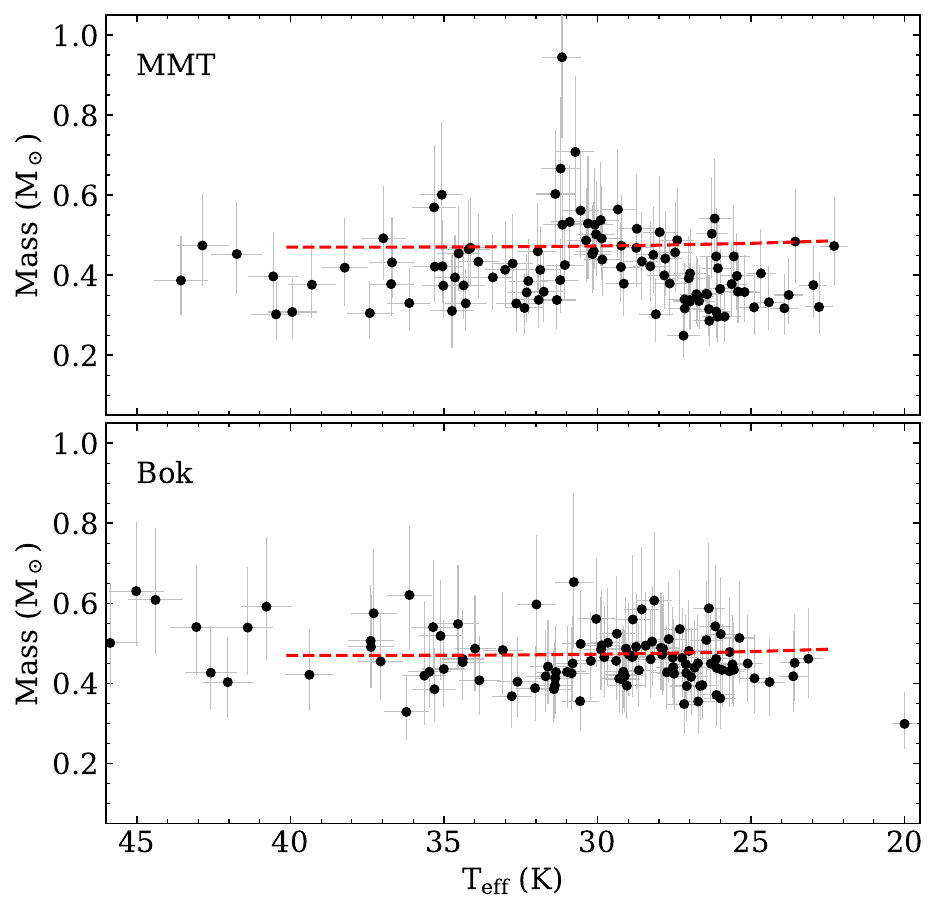}}
     \caption{Mass versus \teff\ for the stars in the MMT sample. The masses obtained when using the atmospheric parameters from the MMT spectra (top) and the Bok spectra (bottom). The dashed line shows the canonical, 0.47~\msun, expected for EHB stars.
     }
     \label{fig:MMT_mass}
\end{figure}

\begin{figure}
\resizebox{\hsize}{!}{
   \includegraphics{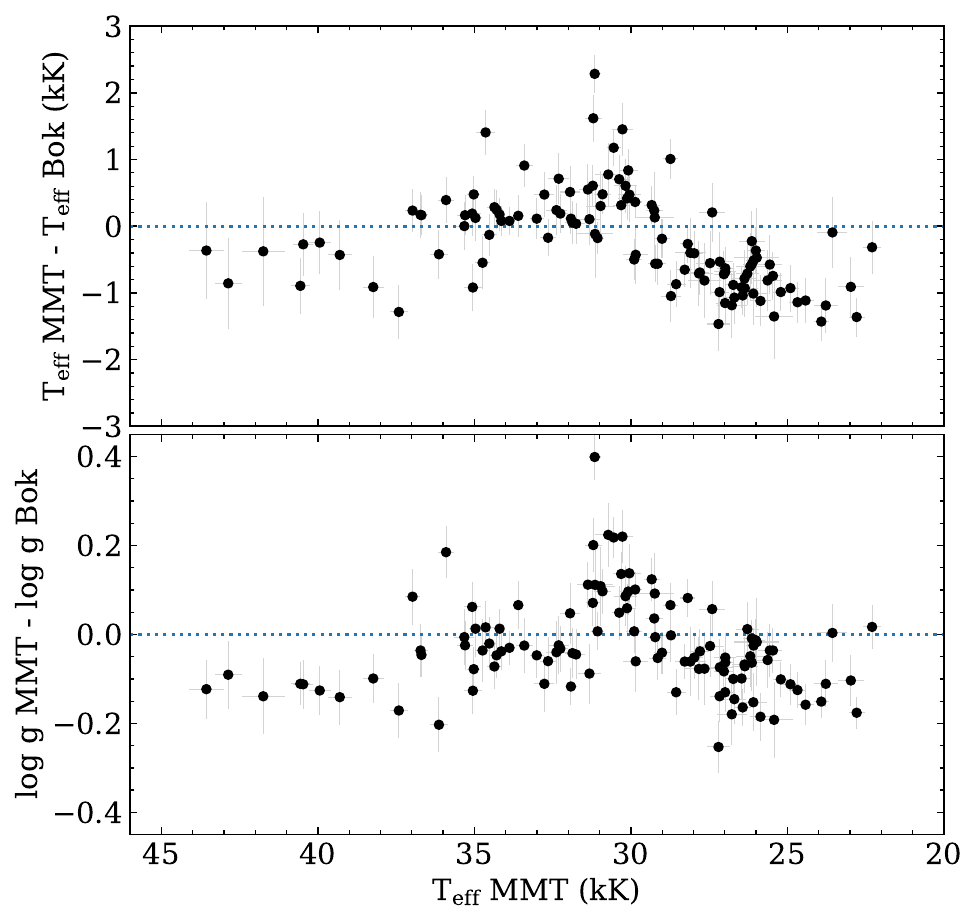}}
     \caption{Comparison between the atmospheric parameters, \teff\ and \logg, obtained from the MMT and Bok spectra. }
     \label{fig:MMT_comp}
\end{figure}

For the He-sdOs, we observe the expected mass increase for stars between 40 and 50~kK. However, compared to the masses inferred from their positions near the ZAHeMS on the HRD, the masses obtained from the SEDs appear underestimated at the cool end and overestimated at the hot end. This discrepancy likely arises from a \teff-dependent bias in our \logg\ measurements. Indeed, in the Kiel diagram, the cool He-sdOs are found above the ZAHeMS while the hot He-sdOs are slightly below it (Fig.~\ref{fig:Kiel}), which is somewhat inconsistent with their more precise location on the HRD.
The spectra of the He-sdOs remain challenging to model, and it is possible that their atmospheric parameters, especially log~$g$, still suffer from some systematics.

 \begin{figure*}
\resizebox{\hsize}{!}{
   \includegraphics{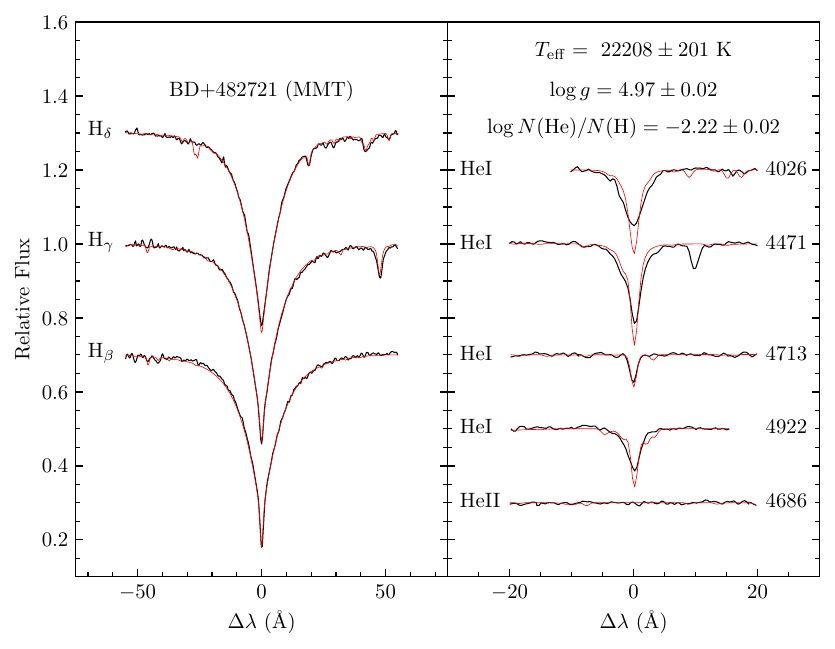}
   \includegraphics{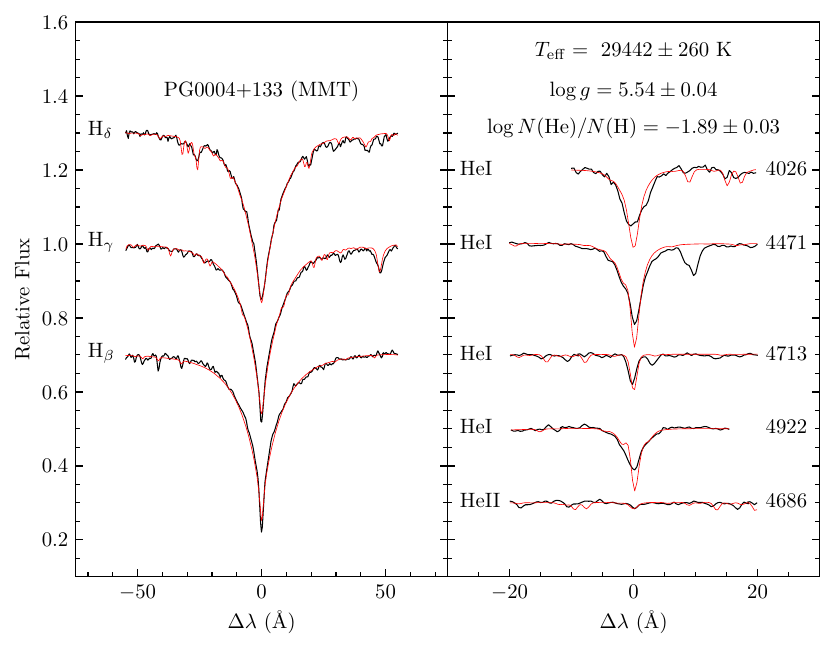}}\vspace{1pt}
\resizebox{\hsize}{!}{   
    \includegraphics{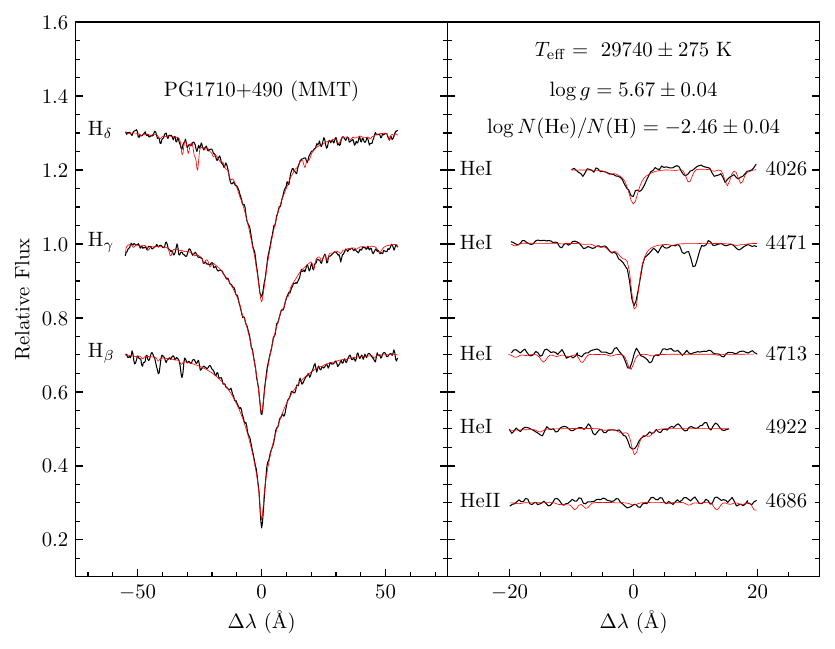}
   \includegraphics{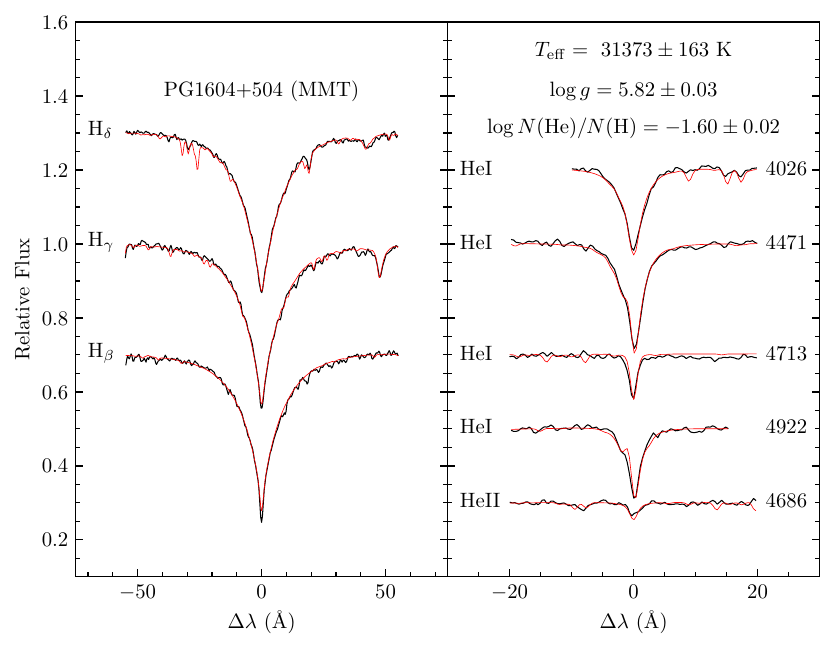}}\vspace{1pt}  
     \caption{Upper panel: Best fit of the Balmer and helium lines in selected stars from the MMT sample. Top: BD+48\degr2721, a known \het\ star, and PG0004+133. The mismatch between the predicted and observed He lines is due to the stratification of helium in the atmosphere. Bottom: PG1710+490, a known \het\ star with no indication of stratification (also according to \citealt{2018A&A...618A..86S}) and PG1604+504, that shows no indication of stratification.
     }
     \label{fig:He_strat}
\end{figure*}

\subsection{The MMT Sample}\label{sec:res:mmt}
 \subsubsection{Atmospheric and stellar parameters}\label{sec:res:mmt:params}

 The MMT spectra were fit in the same way as described earlier, and from their atmospheric parameters we also derived the stellar parameters via the SED fitting method. Although these spectra have a higher resolution than the Bok spectra, they cover a smaller wavelength range (4000$-$4950 \AA) that includes only three Balmer lines (H$_\beta$, H$_\gamma$, and H$_\delta$). All stars in the MMT sample are shown in the MMT spectral atlas included in Fig.~D.1. When inspecting the resulting masses, we noticed an obvious mass trend as seen in the top panel of Fig.~\ref{fig:MMT_mass}. There is a significant increase in masses for the stars around 30~kK. The MMT sample consists of hydrogen-rich sdBs and sdOs only, thus no visible mass dependency over \teff\ is expected, as seen for the masses derived with the atmospheric parameters obtained from the Bok spectra of these same stars (Fig.~\ref{fig:MMT_mass}, bottom panel).

We then inspected the differences in atmospheric parameters obtained from the MMT and the Bok spectra for the 116 stars present in both samples. From the results shown in Fig.~\ref{fig:MMT_comp}, we observe a trend in \teff\ and log~$g$ for stars between 25 and 30 kK. Generally speaking, the MMT spectra of the ``cool'' sdBs are fit with lower \teff\ and lower \logg\ compared to the Bok spectra. The opposite happens for the sdBs hotter than $\sim$35~kK; their MMT spectra result in higher \teff\ and higher \logg\ than their Bok spectra. 
We thus conclude that the restricted wavelength range of the MMT spectra does not allow us to derive atmospheric parameters that are accurate enough to obtain reliable masses from the SED fits, even if the spectra are of excellent quality. This highlights the importance of the high Balmer lines in the determination of the atmospheric parameters, especially that of the surface gravity \citep{1994A&A...282..151H}. Nevertheless, there are things to be learned from the MMT spectra. We discuss the presence of helium stratification in the following subsection and the particular case of PG0215+183 is presented in App.~\ref{app:pg0215}.

\begin{figure}
\resizebox{\hsize}{!}{
   \includegraphics{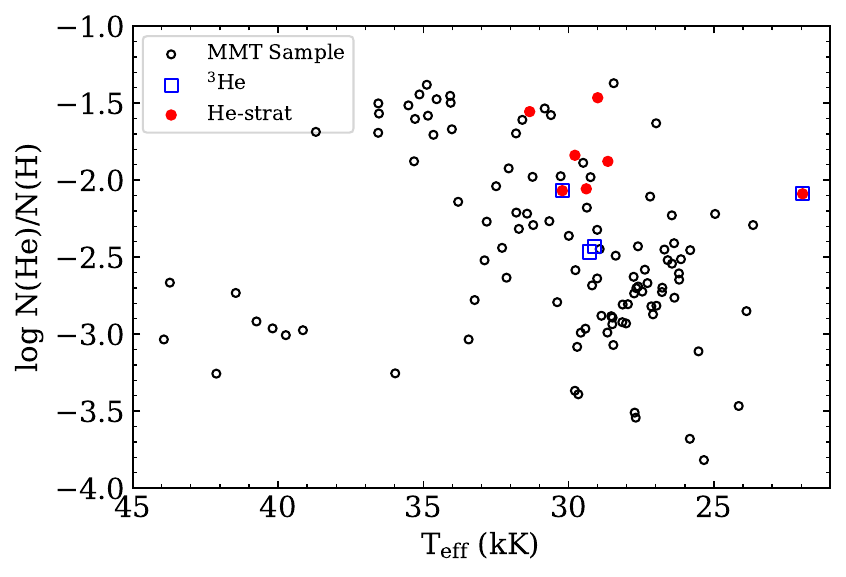}}
     \caption{Position of the stars from the MMT sample in the \teff$-$He plane. Stars showing indication of He-stratification in their MMT spectrum are indicated with filled red circles. Stars having \het\ in their atmosphere are indicated with blue squares (from \citealt{2018A&A...618A..86S} and \citealt{2013A&A...557A.122G}).
     }
     \label{fig:He_teff_strat}
\end{figure}

\subsubsection{Helium stratification}\label{sec:res:mmt:strat}

While examining the best-fit solutions of the MMT spectra, we noticed that the helium lines in a few stars were poorly reproduced (see Fig.~\ref{fig:He_strat} for two examples). One of these stars is \object{BD+48\degr2721}, which is a cool sdB, or possibly a BHB, known to show the \het\ anomaly \citep{2001AN....322..401E}.
The \het\ anomaly implies that neutral He absorption lines feature contributions from both the \het\ and $^4$He isotopes. 
In addition, \citet{2018A&A...618A..86S} found that the strongest \ion{He}{i} lines in BD+48\degr2721 could not be properly reproduced by the model spectra: the line cores are shallower in the observations. This phenomenon is attributed to a vertical stratification of He in the stellar atmosphere. In fact, the vertical stratification and the presence of \het\ are both a manifestation of atomic diffusion taking place in the atmosphere.
Besides this object, \citet{2018A&A...618A..86S} also detected He stratification in three \het\ sdB stars with \teff\ between 26.5 and 29~kK. 

The mismatch between the observed and modeled He lines in the spectrum of BD+48\degr2721 is caused by He stratification. Similar mismatches are seen in \object{BD+42\degr3250}, \object{PG0004+133} (see Fig.~\ref{fig:He_strat}), \object{PG1738+505}, \object{PG0242+132}, and possibly \object{PG1512+244} and \object{PG1519+640}, indicating that they also have He stratification. Figure~\ref{fig:He_teff_strat} shows the position of these stars in the \teff-helium plane.
The \teff\ of these six objects are between 28 and 31 kK, which is consistent with the \teff\ range of the known \het\ and stratified stars \citep{2018A&A...618A..86S,2013A&A...557A.122G}.

It is likely that the six new He-stratified objects are also enriched in \het, but the isotopic shift of the He lines in the MMT spectral range is too small (0.2-0.3 \AA) to be detectable at a resolution of 1~\AA. Up until now, He stratification has only been reported in \het\ stars. However, not all \het\ stars show evidence of stratification. Our sample also includes two known \het\ stars that do not show evidence of He stratification: \object{PG0133+114}, and \object{PG1710+490} (see Fig.~\ref{fig:He_strat}). The latter was included in the analysis of \citet{2018A&A...618A..86S} while the former was only reported by \citet{2001AN....322..401E}. 
In light of these results, it appears that the stratification of helium, and possibly of other elements \citep{2013A&A...549A.110G}, and also the presence of \het, is more common than previously thought in sdBs with \teff\ around 28$-$31~kK.

\section{Discussion}\label{sec:diss}

\subsection{Mass distributions}\label{sec:diss:mass}

\begin{figure}
\resizebox{\hsize}{!}{
   \includegraphics{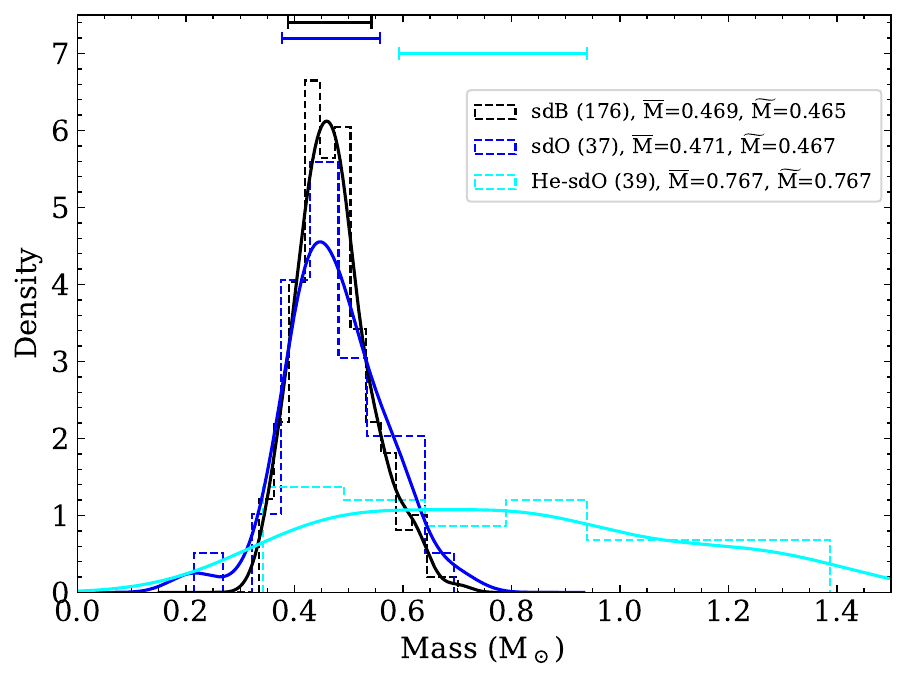}}
     \caption{Mass distribution obtained from parallaxes and SED fits for the sdBs, sdOs, and He-sdOs in our sample. The dashed lines show the normalized histograms for each spectral group and the solid curves are the associated kernel density function. The stars with a composite SED were excluded. In the legend, we indicated the number of stars in parenthesis as well as the average ($\overline{M}$) and median ($\widetilde{M}$) mass for each spectral type. The error bars on top are indicative of the mass uncertainty for each spectral type (see $M_{\rm err}$ in Table~\ref{table_masses}). 
     }
     \label{fig:Mass_dist}
\end{figure}

Using the masses presented in Sect.~\ref{sec:res:bok:stellar}, we now examine the mass distribution of our hot subdwarfs. 
For this discussion,
we always exclude the stars that were fitted with an IR excess, unless stated otherwise.
We first concentrate on the three most populated spectral types: sdB, sdO and He-sdO.
We show in Fig.~\ref{fig:Mass_dist} the mass distributions for these three spectral types.
Along with the mass distributions, we also calculated the average mass ($\overline{M}$) as a weighted mean for each spectral type. Because the uncertainties ($\epsilon$) are proportional to the mass itself, we used the relative uncertainty as weight, leading to:
\begingroup
\setlength{\abovedisplayskip}{4pt}
\setlength{\belowdisplayskip}{4pt}
\[\overline{M} = {\frac{\sum w_i M_i}{\sum w_i}}\textnormal{, where }  w_i =\frac{M_i^2}{\epsilon_i^2}.\]
\endgroup
We also computed the median mass ($\widetilde{M}$) of the distributions. Both of these values are reported in Fig.~\ref{fig:Mass_dist} and in Table~\ref{table_masses}. In the table, we include three additional diagnostics: the median value of the uncertainties ($M_{\rm err}$), the standard deviation ($\sigma$) of the mass distribution, and its 68\% interquartile range ($Q_{16}$-$Q_{84}$). Both $M_{\rm err}$ and $\sigma$ are expressed as a percentage to remove the correlation between the masses and their absolute uncertainties. 

As shown in Fig.~\ref{fig:Mass_dist}, the sdB and sdO distributions are very similar and peak around 0.47~\msun. 
This agrees with theoretical expectations, since the two types are thought to be evolutionarily linked and have comparable average masses close to this (canonical) value. 
Our mean and median values for the sdBs are also consistent with the 0.47~\msun\ obtained by \citet{2012A&A...539A..12F} from 22 sdBs with masses derived from asteroseismology and the light curve analysis of binary systems.
Our distribution is wider than that of \citet{2012A&A...539A..12F}, but masses obtained from asteroseismology and light curve modeling typically have smaller uncertainties than those derived from SED fits and parallaxes. 
The typical uncertainty on the individual masses ($M_{\rm err}$) and the standard deviation ($\sigma$) of the sdO distribution are similar, while for the sdBs we found $\sigma$ to be smaller than the typical error. This means that the broadening of our sdB and sdO distributions can be fully explained by the uncertainties.
The consistency between our sdB and sdO mass distribution is a significant improvement compared to the results of \citet{Lei2023_mass}, who obtained a relatively flat mass distribution for their sdO stars. Their unrealistic sdO masses are most likely caused by the lack of metallic opacities in the model atmospheres they used to derive the atmospheric parameters and compute the SEDs of their hot sdOs. It is especially important to take into account both NLTE effects and line-blanketing when modeling the spectra of hot sdO stars \citep{Werner1996,2013ApJ...773...84L,2015A&A...579A..39L}.

Intriguingly, the mass distribution of the He-sdOs is very different from that of the H-rich hot subdwarfs: it is much broader, with 68\% of the stellar masses falling between 0.48 and 1.18~\msun, and it does not show a distinct peak. The dispersion is significantly larger than the typical uncertainties, suggesting the presence of an intrinsic dispersion.
The weighted mean and median mass of the He-sdOs are around 0.77~\msun, which is significantly larger than the canonical core-helium burning mass. However, caution is advised when assessing the masses derived for the He-sdOs. As mentioned in Sect.~\ref{sec:res:bok:stellar}, the masses obtained from SED fits and parallaxes of the He-sdOs are somewhat different from those expected from the position of the stars in the HRD. 
We also calculated masses for the He-sdOs from their position on the HRD, meaning that we projected each star onto the nearest point along the ZAHeMS. Uncertainties in \teff\ and $L$ were taken into account to estimate the corresponding uncertainties on the masses, which are notably lower than those obtained from the SED fits. The HRD masses are included in Table 1 and we also include their statistics in Table~\ref{table_masses}. Masses from the HRD were not computed for the 3 He-sdOs that are post-AGB objects (see Sect.~\ref{sec:res:bok:atmo}). The He-sdO masses obtained from the HRD are mostly distributed between 0.6 and 1~\msun\ without any distinct peak. The four He-sdOs hotter than 58~kK have masses around 1.4~\msun.
Analyses of He-sdO samples including mass estimates are still scarce in the literature, but 
the mass distribution of He-sdOs shown in \citet[Fig. 4.1.7]{2024PhDT........36D} also has an extended high-mass tail with values above 1~\msun. On the other hand, the position of He-sdOs from the SALT spectroscopic survey \citep{2021MNRAS.501..623J} in the HRD shown in \citet{2024A&A...691A.165D} suggests that they have masses between 0.6 and 1.0 \msun\ according to the helium main sequence of \citet{1971AcA....21....1P}. This is very similar to what is seen in our sample, although this might not be surprising given that the model atmospheres and analysis method used in \citet{2024A&A...691A.165D} are the same as in this work. Given the systematic uncertainties on the log~$g$ values of the He-sdOs, it is likely that the masses derived from the HRD positions are more accurate.
In any case, both the masses obtained from the SED fits and the position of He-sdO stars in the HRD indicate that they are, on average, more massive and span a wider mass range than the hydrogen-rich sdBs and sdOs. The higher masses of the He-sdOs and their broader distribution are fully consistent with a stellar merger origin (see, e.g. \citealt{2003MNRAS.341..669H,2011MNRAS.410..984J,2000MNRAS.313..671S,2012MNRAS.419..452Z}). This hypothesis, although long suspected, was recently further supported by the discovery of several magnetic He-sdOs \citep{2024A&A...691A.165D,2022A&A...658L...9D,2022MNRAS.515.2496P} and the low fraction of short-period binaries among He-sdOs \citep{2022A&A...661A.113G,2025MNRAS.537.2079S}.

As for the BHB stars, they are expected to be slightly more massive than canonical H-rich subdwarfs, with masses between 0.50 and 0.55 \msun, depending on the metallicity. From our observations, we derive a median mass of about 0.54~\msun, which is in good agreement with the expectation.
Finally considering the iHe-sdOBs and iHe-sdBs, our samples are too small, with nine and three stars respectively, to derive a statistically significant mass distribution.  Saying that, the average mass of these stars appears to be consistent with the canonical hot subdwarf mass of around 0.5~\msun. While the three iHe-sdBs are rather scattered on the Kiel and HR diagrams, their masses are all fairly close to the canonical values (see Fig.~\ref{fig:Teff_mass}). The iHe-sdB that is significantly below the ZAEHB in both the Kiel and HR diagrams is \object{PG1247+554} (also known as GD\,319), whose best-fit is shown in Fig.~\ref{fig:bok_fits}. This is a particular system comprising an iHe-sdB in a 0.6 day period binary and cool MS star located about 1$\arcsec$ away \citep{2000MNRAS.311..877M,USNO_2001}, however the MS star is most likely a background star (S. Geier, priv. comm. 2025). The hot subdwarf and the MS star are resolved by $Gaia$ and our SED fit is solely based on the $Gaia$ XP spectra of the hot subdwarf.
Although this star was a promising low-mass hot subdwarf candidate from its location on the HRD, the normal mass that we obtained from the SED fit does not match this interpretation.

\setcounter{table}{1}
\begin{table}%[h]
\footnotesize
\caption{Mass properties by spectral group and binary type in our sample.}
\label{table_masses}      
\centering                    
\begin{tabular}{l c c c c c c}        
\toprule\toprule
\noalign{\vskip4bp}
Type & $N$star & $\overline{M}$ & $\widetilde{M}$ &  $M_{\rm err}$ & $\sigma$ & $Q_{16}$-$Q_{84}$\\ 
 &  & [\msun] & [\msun] & [\%] & [\%] & [\msun] \\
 &  & (1) & (2) & (3) & (4) & (5)\\ 
\noalign{\vskip2bp}
 \midrule
 & \multicolumn{6}{c}{Spectral groups} \\
  \midrule
% \noalign{\vskip3bp}
sdB & 176 & 0.469 & 0.465 & 16.2 & 14.4 & 0.408-0.542 \\
sdO & 37 & 0.471 & 0.467 & 19.3 & 18.8 & 0.403-0.554 \\
He-sdO (SED) & 39 & 0.767 & 0.767 & 20.4 & 39.4 & 0.482-1.183 \\
He-sdO (HRD) & 36 & 0.753 & 0.845 & 4.2 & 24.9 & 0.696-0.964 \\
iHe-sdOB & 9 & 0.449 & 0.432 & 18.2 & 28.4 & 0.344-0.502 \\
BHB & 19 & 0.565 & 0.536 & 19.0 & 23.7 & 0.466-0.690 \\
 
 \midrule
 & \multicolumn{6}{c}{Binary types} \\
  \midrule
% \noalign{\vskip3bp}
sd+WD & 38 & 0.452 & 0.442 & 16.2 & 12.8 & 0.399-0.522 \\
sd+dM & 15 & 0.499 & 0.504 & 15.9 & 14.9 & 0.411-0.573 \\
single & 133 & 0.468 & 0.470 & 17.1 & 15.8 & 0.413-0.524 \\
binary & 92 & 0.469 & 0.460 & 16.2 & 15.7 & 0.400-0.551 \\
\bottomrule
\end{tabular}
\tablefoot{
 (1) Weighted average. (2) Median. (3) Median of the individual mass uncertainties. (4) Standard deviation. (5) Interquartile range, 16$^{th}$-84$^{th}$ percentile
}

\end{table}

\begin{figure}
\resizebox{\hsize}{!}{
   \includegraphics{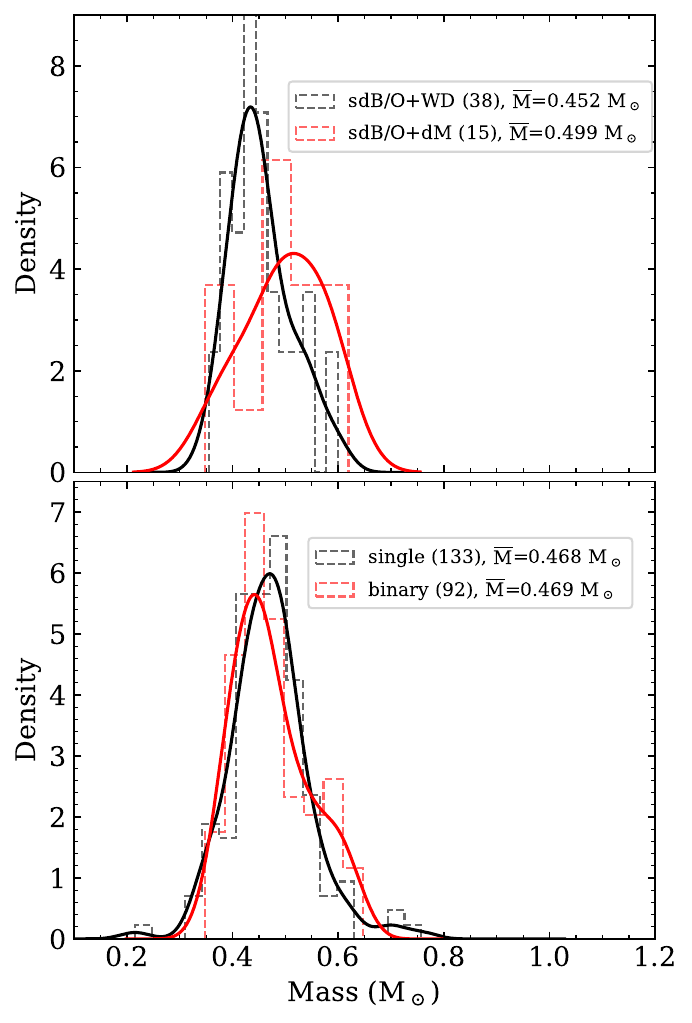}}
     \caption{Top: mass distributions of the hot subdwarf components in close binary systems with WD and dM companions. For most of the systems, the type of binarity was classified by \citet{2022A&A...666A.182S}. The weighted average mass of both categories is also indicated. Bottom: mass distribution for hot subdwarfs that are candidate single stars compared to those found in close binary systems.} 
     \label{fig:mass_dist_dM_WD}
\end{figure}

\subsection{Mass distribution and binarity}\label{sec:diss:mass:binary}

To investigate a possible dependence of mass on binarity in the stars of our Bok sample, we divided them into binary groups based on information available from the literature. Notably, \citet{2022A&A...666A.182S} 
classified many hot subdwarf binaries, mostly sdBs but also a few sdOs with \teff $\sim$40 kK, into two different categories based on their TESS light-curve properties: sdB/O+dM (HW Vir systems and reflection effects) and sdB/O+WD (ellipsoidal and beaming effects). For some of these systems, the authors estimated masses from parallaxes and SED fits and found that the mass distribution of hot subdwarfs in sdB/O+WD systems peaks at a lower mass than those having a M-dwarf (dM) companion. 

To investigate this topic further, we cross-matched the stars in our sample with the stars listed in Table A.3 and A.4 of \citet{2022A&A...666A.182S} and extracted binary information for a few additional targets of ours from other sources: PG0101+039 \citep{2002MNRAS.333..231M,2008A&A...477L..13G}, PG2345+318 \citep{2004Ap&SS.291..267G}, KIC007668647 \citep{2014A&A...570A.129T}, and KIC011558725 \citep{2012A&A...544A...1T} as sdB+WD systems and PG1438-029 \citep{2004Ap&SS.291..267G}, HS2231+2441 \citep{2008ASPC..392..221O}, FBS1531+381 \citep{2010ApJ...708..253F}, and UVO1758+36 (Schaffenroth priv. comm. 2025) as sdB+dM systems. This yielded a subsample of
38 sdB/O+WD and 15 sdB/O+dM systems\footnote{Those are 15 systems that do not show IR-excess in their SED.}. 
We show the resulting mass distributions in the top panel of Fig.~\ref{fig:mass_dist_dM_WD}. 
Interestingly, there is a distinct difference between the mean mass and the mass distribution of the two types of binaries (see also Table~\ref{table_masses}). The sdB/O+WD systems have an average mass close to 0.45~\msun, while the sdB/O+dM have a slightly larger average mass of 0.5~\msun. This is in line with the finding of \citet{2022A&A...666A.182S} that the hot subdwarfs with dM companions appear to be, on average, slightly more massive than those with WD companions. To verify whether the difference is significant, we ran 1000 iterations of the Kolmogorov-Smirnoff test (KS-test) using masses drawn from a normal distribution based on the individual errors and we obtained an average $p$-value ($p_{\rm KS}$) of 0.30. This means that if the two underlying distributions were the same, we would get the observed difference at least 30\% of the time. Thus we cannot conclude that the difference between the two mass distributions are significant. In this case, we are likely limited by the small number of sdB/O+dM, only 15, identified in our sample. It is also worth noting that in sdB/O+dM systems, the atmospheric parameters of the hot subdwarf vary with the phase when the reflection effect is sufficiently strong, adding additional uncertainties on the atmospheric parameters derived from a single spectrum (see, e.g. \citealt{2004A&A...420..251H,2010ApJ...708..253F}).

We also looked at the mass distribution of non-composite binaries, essentially meaning close binaries with companions that are detected only from radial velocity or light-curve variations (without producing IR-excess), to compare it with that of the apparently single stars. For this comparison, we only considered the He-poor and intermediate-He hot subdwarfs, thus excluding the He-sdOs and BHBs.
To build our list of close binary stars, we started with the 53 systems identified previously (the sdB/O+WD and sdB/O+dM), to which we added known binaries from radial velocity surveys in the literature (e.g. \citealt{1998ApJ...502..394S,2001MNRAS.326.1391M,2011MNRAS.415.1381C,2003MNRAS.338..752M,2022A&A...661A.113G,2025A&A...693A.121H}) and suspected binaries from the RV measurements of the MMT spectra and from on-going RV monitoring (F. Mattig, priv. comm. 2025). 
This way, we increased our number of binary stars to 92 objects.
The 133 remaining stars, for which we have no evidence of binarity, or no information about their RVs, are considered as single star candidates. 
The resulting mass distributions for binaries and single hot subdwarfs are shown in the bottom panel of Fig.~\ref{fig:mass_dist_dM_WD} and their characteristics are listed in Table~\ref{table_masses}. 
Both distributions appear very similar and our series of KS-tests did not find a statistically significant difference between the two distributions ($p_{\rm KS}=0.57$); however, this does not necessarily imply that the underlying distributions are identical. 

The formation of single hot subdwarfs remains challenging to explain from a theoretical point of view. 
One potential formation channel is via the merger of two low-mass stars (e.g. two He-core WDs, CO-core and He-core WDs, or low-mass MS + He-core WD; \citealt{2003MNRAS.341..669H,2008ApJ...687L..99P,2016MNRAS.463.2756H}). According to \citet[see their Fig.~12]{2003MNRAS.341..669H}, the predicted mass distribution for sdBs originating from the merger channel is broader and shifted to higher masses (0.5$-$0.55~\msun) compared to those formed via common-envelope ejection. If a majority of our candidate single He-poor sdB/Os were merger products, we would expect their mass distribution to be different to that of He-poor sdB/Os in close binary systems, formed via the common envelope channel.
However, given the similarity of the two mass distributions, we find no evidence that a substantial fraction of the He-poor single star candidates are formed via the merger channel.

The close binary fraction of our hot subdwarf sample is 41\%\footnote{This is when excluding the composite systems, some of which also are close-binary systems, such as PG0940+068 and PG1101+249 (see App.~\ref{App:composites}). When including all 27 composite systems (the two He-sdOs composites being excluded) it reaches 48\%. Our binary fraction is not corrected for inclination effets} (92/225).
Close binary fractions from the literature ranges from 30\% \citep{2022A&A...661A.113G,2025A&A...693A.121H} to 48\% \citep{2001MNRAS.326.1391M,2011MNRAS.415.1381C}, with the SPY sample giving an intermediate value of 39\% for sdBs \citep{2004Ap&SS.291..321N}. The number of close binaries identified in our sample thus seems reasonable and although there is
certainly some binaries remaining in our single star sample, for example due to low inclination, or because the companion could be too faint or distant to be detectable. However, they are unlikely to be numerous enough to change our conclusion concerning their mass distributions. It is also worth mentioning that there are hints that the iHe-sdBs and iHe-sdOBs have different binary properties than the rest of the hydrogen-rich hot subdwarfs (see the discussions in \citealt{2022A&A...661A.113G} and \citealt{2024PhDT........36D}), but we decided to keep them in our hydrogen-rich sample as they are only 12 stars and they represent less than 5\% of our sample.

\subsection{Low-mass hot subdwarfs}\label{sec:diss:mass:b_ehb}

\begin{figure}
\resizebox{\hsize}{!}{
   \includegraphics{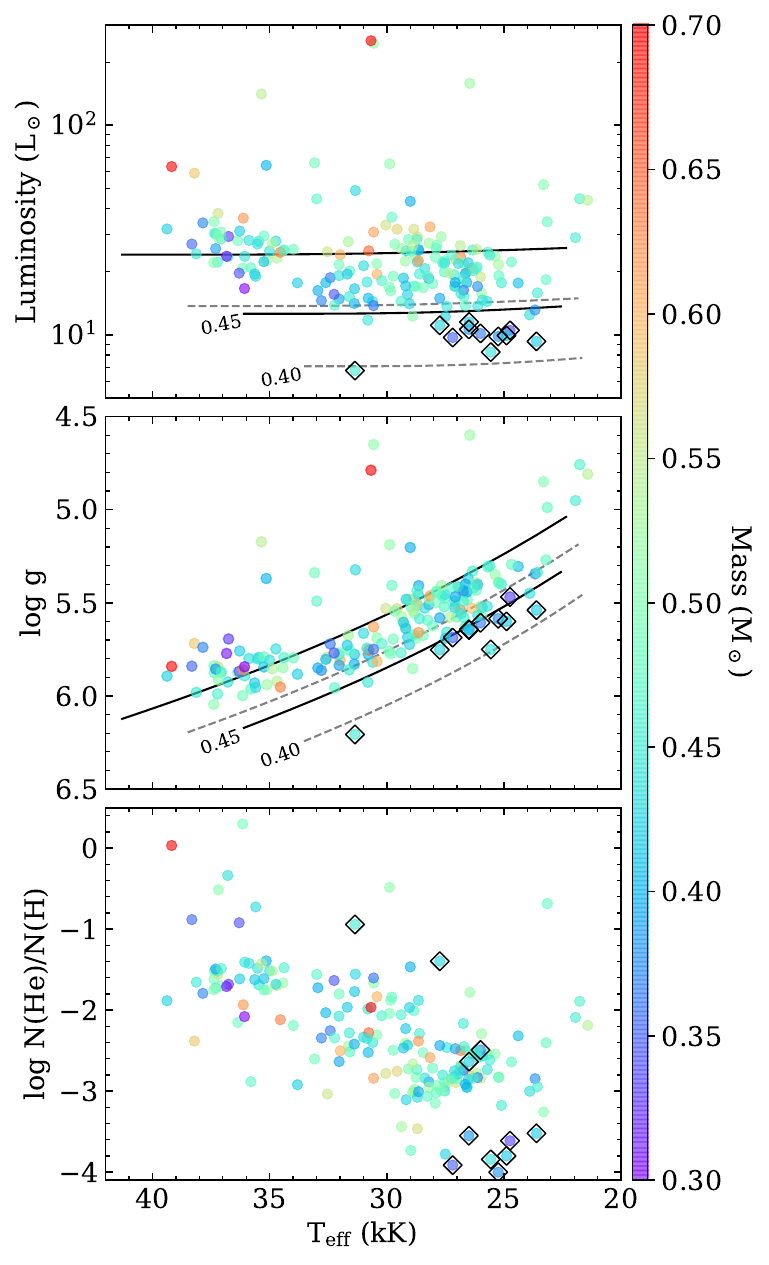}}
     \caption{Position of the underluminous sdBs (black diamonds) in the HRD (top), Kiel diagram (middle), and \teff $-$\heh\ plane. The underluminous sdBs were selected from their position below the ZAEHB for a core mass of 0.45 \msun. The ZAEHB and TAEHB computed with STELUM for core masses of 0.45~\msun\ (solid lines) and 0.40~\msun\ (dashed lines) are indicated. All stars are color-coded with the mass obtained from the SED fits. We only show the stars with spectral types that can be associated with the EHB region: sdB, iHe-sdB, and iHe-sdOB.}
     \label{fig:below_ehb}
\end{figure} 

Most hot subdwarfs are the progeny of low-mass MS stars ($\sim$0.8$-$2~\msun) that experienced a helium flash, meaning that helium ignition happened under degenerate conditions. This occurs when the helium core reaches a mass in the range of 0.46$-$0.51~\msun, almost regardless of the progenitor mass for MS stars below $\sim$ 1.6 \msun\ (see Fig.~1 of \citealt{2013EPJWC..4303002M} and chapter~6 of \citealt{10.1088/2514-3433/adcf15}). Instead, the exact mass of the He-core at the onset of the He-flash depends on the metallicity (and helium content) of the progenitor, higher metallicity stars experiencing the He-flash at a lower core mass 
\citep{1967ApJ...147..624I,1993ApJ...419..596D,SalarisCassisi2005}. The He-flash may occur at a slightly lower (by about 0.01~\msun) core mass if it is delayed until after the progenitor has left the RGB \citep{2001ApJ...562..368B,2008A&A...491..253M}. Hot subwdarfs less massive than 0.45~\msun\ are unlikely to have undergone a He-flash. For an sdB with $M$ $\lesssim$ 0.45~\msun\ to be burning helium in its core, it must have evolved from a higher mass progenitor ($\sim$2.3$-$3.5 \msun) that ignited helium quiescently under non-degenerate or semi-degenerate conditions. In such cases, helium ignition can happen at core masses that are as low as $\sim$0.33~\msun\ \citep{SalarisCassisi2005,2002MNRAS.336..449H,2008A&A...490..243H,2024MNRAS.52711184A,2025MNRAS.539.3273R}.

Any such low-mass sdBs should be mostly found below the ZAEHB in the Kiel and HR diagrams. Because the luminosity is less affected by systematic uncertainties than the surface gravity, as explained in Sect.~\ref{sec:method:SED}, we selected our candidate low-mass sdBs from the HRD. We found 11 stars with a luminosity lower than the STELUM ZAEHB track for a 0.45~\msun\ core (see Fig.~\ref{fig:below_ehb}, top panel). We refer to these as "underluminous" sdBs. 
Their positions in the HRD are consistent with masses between 0.40 and 0.45 \msun. The luminosity of the theoretical EHB band increases with increasing core mass and this effect is visible when color-coding the stars according to their mass (Fig.~\ref{fig:below_ehb}): we see a general increase of the masses with luminosity.
On the Kiel diagram (Fig.~\ref{fig:below_ehb}, middle panel), the 11 underluminous sdBs are located below or very close to the ZAEHB. 

To verify whether the masses of the underluminous sdBs are indeed lower than the canonical mass, we computed the weighted average and median mass for the 10 underluminous sdBs that have no IR excess\footnote{PG0250+186 is the only underluminous sdB with IR excess.}. We obtained a weighted average and median mass of 0.40~\msun.
As a comparison sample, we used the remaining 178 non-composite stars classified as sdBs, iHe-sdBs, and iHe-sdOBs (shown as coloured circles in Fig.~\ref{fig:below_ehb}). We restrict our comparison sample to these spectral types, because the sdOs and He-sdOs are not positioned on the EHB, thus they cannot be found in the region below the ZAEHB. For this comparison sample, we obtained a weighted average and median mass of 0.47~\msun. 
This confirms that the underluminous sdBs are, on average, less massive than stars on the canonical EHB. With the exception of PG1247+554 (see Sect.~\ref{sec:diss:mass}), these underluminous sdBs are good candidates for originating from massive progenitors ($\sim$ 2$-$3.5 \msun).  
Our fraction of underluminous sdBs is 5.7\% (11/190). This fraction is lower than that obtained by Dawson et al. (in prep.) in their 500 pc sample, which is around 11\%\footnote{Note that if we use the same criteria as Dawson et al. to select our underluminous sdBs (i.e. log $L/L_{\odot} \leq 1.05 $), we are left with six stars only, leading to an even lower fraction of 3\%.}.

An interesting feature of the underluminous sdBs is that more than half of them are found to be at the ``low-helium'' end of the \teff $-$He sequence (see Fig.~\ref{fig:below_ehb}, bottom panel). A similar behavior is also seen in the 500 pc sample of Dawson et al. (in prep.).
Another remarkable peculiarity of the most helium-poor sdBs is the weakness of their metal lines. This is most noticeable when looking at the MMT spectral atlas in Fig~\ref{fig:spectra_all}: the spectra with weak or absent \ion{He}{i} lines also have very weak metallic lines compared to other spectra of stars at similar \teff. Three underluminous stars with especially low helium are seen on the first panel of Fig.~\ref{fig:spectra_all}: PG1111$-$077, PG0856+121, and, PG0250+189. 
The low-helium sdBs are mainly found below the ZAEHB, consequently they have higher surface gravity than other sdBs at the same temperature. This likely affects the diffusion processes in the atmosphere, because a higher surface gravity means a stronger gravitational pull on the individual atoms. This could explain why low helium and low metal abundances appear to be correlated.

\subsection{Pulsators}\label{sec:diss:puls}

\begin{figure}
\resizebox{\hsize}{!}{
   \includegraphics{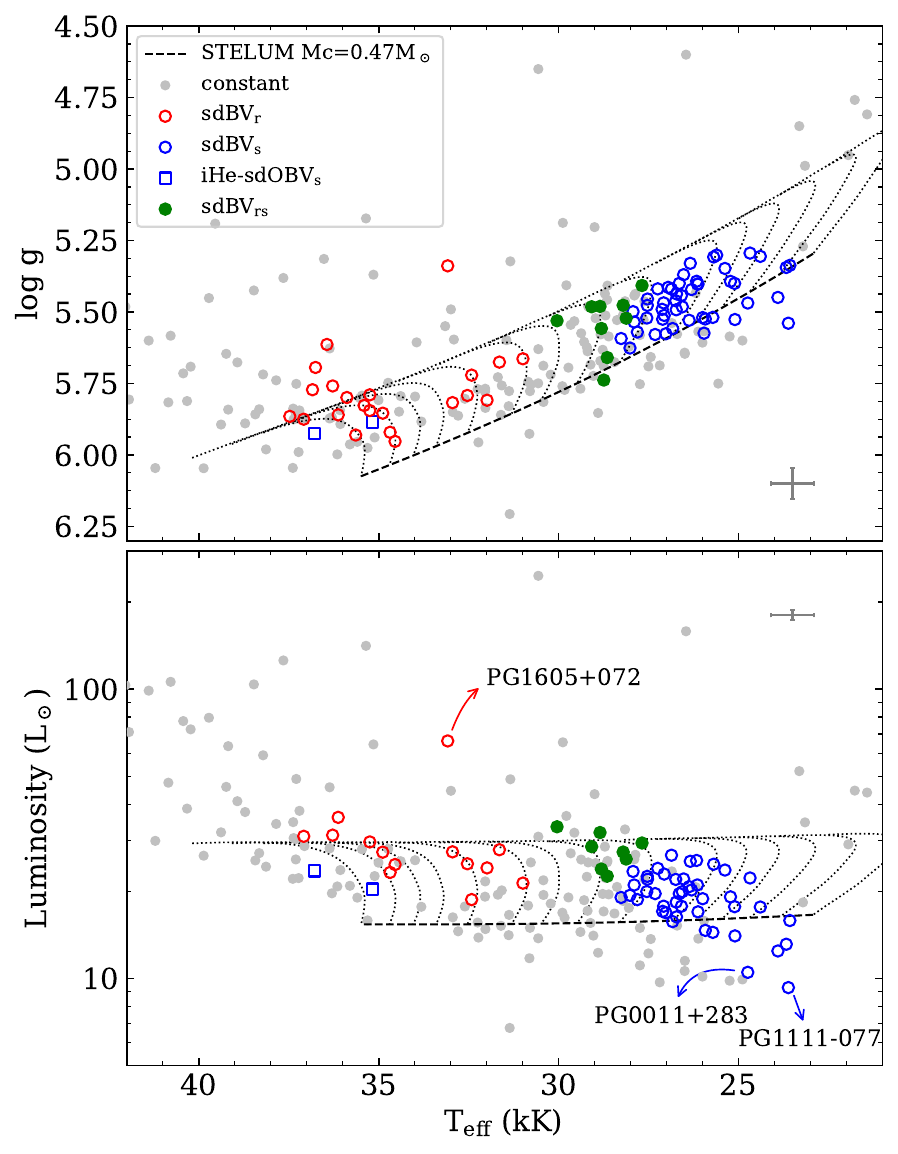}}
     \caption{ Position of the pulsating stars in the Kiel (top) and HRD (bottom) diagram. The different types of pulsators are indicated with various symbols: rapid pulsators sdBV$_{\rm r}$ (open red circles), slow pulsators sdBV$_{\rm s}$ (open blue circles), and hybrid pulsators sdBV$_{\rm rs}$ (green filled circles). Stars found to be constant are indicated with grey filled circles. Feige~46 and LS\,IV\,-14$^{\circ}$116 (blue open squares) are peculiar slow pulsators belonging to the small class of iHe-sdOBV (V366 Aqr). The ZAEHB (dashed line) is indicated for a core mass of 0.47~\msun\ and corresponding evolutionary sequences of the core-helium burning phase are shown for different H-envelope masses (dotted lines). Stars with a poor parallax measurement are excluded in the bottom panel. Representative errors are shown in each panel.
     }
     \label{fig:Kiel_puls}
\end{figure} 

Figure~\ref{fig:Kiel_puls} shows the hot subdwarfs in the Bok sample for which we have information on the pulsation status, obtained according to the methodology described in Sect.~\ref{id_puls}. We plot them in the Kiel diagram (top panel) and in the HRD (bottom panel). 
The different types of pulsators are indicated with various symbols, with the sdBV$_{\rm r}$ being the rapid $p$-mode pulsators, sdBV$_{\rm s}$ the slow $g$-mode pulsators, and sdBV$_{\rm rs}$ the hybrid pulsators showing both $p$- and $g$-modes \citep{kilkenny10}. As in the Kiel and HR diagrams (Figs.~\ref{fig:Kiel} and ~\ref{fig:HRD}), we observe a lack of sdBV$_{\rm r}$ pulsators around 33 kK
(see Sect.~\ref{sec:res:bok:stellar} for more details about this feature). Feige 46 and LS\,IV\,-14$^{\circ}$116 are two known slow pulsators belonging to the very small group of iHe-sdOBV (or V366 Aqr). This class of pulsator only includes three members so far, which all
share very similar pulsation and atmospheric properties \citep{2011ApJ...734...59G,2019A&A...623L..12L,2020MNRAS.499.3738O}, including the extreme heavy-metal abundances characteristic of iHe-sdOBs. 

The observational instability regions for the different types of pulsations are well defined, and separated, in both diagrams shown in Fig.~\ref{fig:Kiel_puls}. The gravity modes are excited from $\sim$23.5~kK to about $\sim$30~kK in the hottest hybrid pulsators, while pressure modes are found in the 27.5$-$38~kK range. Our distribution of rapid pulsators is similar to that presented in \citet{2024A&A...686A..65B}, with the difference that we do not see any pure $p$-mode pulsators below 31~kK. However, the number of rapid pulsators in the sample of \citet{2024A&A...686A..65B} is about two times larger than ours, so this could be a selection effect. The constant and pulsating stars are similarly distributed in the Kiel diagram, but in the HRD we noticed that at a given effective temperature, the pulsators are preferentially located at higher luminosities compared to the non-pulsating stars. This trend may be related to the time it takes to build-up the iron/nickel reservoir in the Z-bump region (responsible for the driving of the modes; \citealt{charpinet97,fontaine03,jeffery06,jeffery07}) through competing radiative levitation, gravitational settling and other processes \citep{fontaine97,fontaine06,theado09,2011MNRAS.418..195H}. Stars too close to the ZAEHB may not yet have had time to accumulate enough metals and trigger pulsations.

We also find that within their instability strip, the slow pulsators are very common, and only a few sdB stars colder than 28~kK are found to be non-pulsating from our ground-based and TESS light curves. This is in line with the early estimate of 75\% for the fraction of $g$-mode pulsators \citep{2011ApJ...734...59G}, which was further supported by results from the Kepler survey of compact pulsators \citep{2011MNRAS.414.2860O}, although based on a smaller number of stars (12 pulsators out of 16 sdBs). 
In comparison, the rapid pulsators are much less common among the hotter sdB stars. This is especially remarkable taking into account the generally much lower amplitude of g-modes ($\lesssim$ 0.1\%) compared to p-modes ($\lesssim$ 1\%) and the fact that the latter should be easier to find. The fraction of $p$-mode pulsators is estimated to be around 10\% \citep{2010A&A...513A...6O}.

The distribution of pulsators on Fig.~\ref{fig:Kiel_puls} is compatible with the canonical 0.47~\msun\ EHB. However, three pulsators stand out as being farther away from the canonical mass EHB. Two of them are found below the ZAEHB and are among the low-mass sdB candidates discussed in Sect.~\ref{sec:diss:mass:b_ehb}: PG0011+283 and PG1111$-$077. Both of these stars also have a very low helium abundance (\heh\ $\lesssim$ $-$3.5) and their SED masses are smaller than 0.45~\msun. We believe they are good candidates for having evolved from a higher-mass progenitor and, as such, are interesting objects for future asteroseismic analyses. The third outlier is PG1605+075, which is located well above the TAEHB \citep{1999A&A...348L..25H}, indicating that the star is in the post-EHB evolutionary phase, which is supported by the canonical mass derived from the SED fit (0.49 $\pm$ 0.1~\msun). It has unusually high-amplitudes pulsations with five modes above 1\%, and a very rich pulsation spectrum \citep{1999MNRAS.303..525K}. Several hypotheses have been proposed to explain its unusual pulsation properties, without a convincing picture emerging (richness of the pulsation spectrum explained by fast rotation or by linear combinations between a few high-amplitude pulsation modes; \citealt{2008PhDT........66V,2010Ap&SS.329..217V}). 
Nine sdB pulsators in our Bok sample have masses and radii derived by asteroseismic modeling (PG0014+068, PG1047+003, PG1219+534, Feige 48, PG1325+102, PG0911+456, Balloon090100001, PG1336-018, and KPD0629-0016;
\citealt{2012A&A...539A..12F}). We compared the masses and radii from our SED fits to those seismically derived. All measurements agree quite well, within or close to 1-sigma of the SED errors. The only exception is PG0911+456 \citep{2007A&A...476.1317R}, for which radius and mass estimates are a bit different from asteroseismology, but still within 2-sigma of the SED errors.

In Fig.~\ref{fig:Kiel_puls} we noticed that the sdBV$_{\rm r}$ stars are mainly located above the ZAEHB, and are instead found closer to the TAEHB for a core-mass of 0.47~\msun. This is especially notable for those hotter than 33~kK and was also reported in \citet{2024A&A...686A..65B} from their sample of TESS $p$-mode pulsators. However, it is not a property unique to the pulsators, but common to all stars located at the very hot end of the EHB: they sit at higher luminosities than their cooler counterparts. Such a shift could be explained if these stars have slightly higher than canonical masses ($\sim$0.50~\msun), but this requires further investigation.

\section{Summary and conclusions}\label{sec:end}

We analyzed the spectra of more than 326 relatively bright hot subluminous stars observed with the 2.3m Bok telescope at the Kitt Peak observatory.
The combination of size, quality, and homogeneity of this spectroscopic sample is unprecedented. We performed the analysis of the stars in two steps. First, we fit the observed spectra with state-of-the-art model atmospheres, synthetic spectra, and fitting techniques, to derive the atmospheric parameters of the stars, i.e. their \teff, \logg, and helium abundance. Secondly, we performed SED fits using parallaxes from \textit{Gaia} DR3 and magnitudes retrieved from various photometric catalogs to derive the stellar parameters radius $R$, luminosity $L$, and mass $M$. 
The sample includes a wide variety of hot subdwarfs that we separate into different categories based on their atmospheric parameters: sdBs, sdOs, iHe-sdBs, iHe-sdOBs, He-sdOs, and BHBs. From our SED fits we also identified 29 composite systems with an IR-excess indicative of a MS companion.
For a subset of 116 stars from the Bok sample, we analyzed additional spectra taken with the MMT telescope. Compared to the Bok spectra, the MMT spectra have a higher resolution but a shorter wavelength coverage.

We summarize below the conclusions drawn from the various aspects of our analysis:

\begin{itemize}

\item Spectra must cover the high Balmer lines to constrain the atmospheric parameters with enough accuracy to derive meaningful spectroscopic masses from parallaxes and SED fits.
This is the case for our Bok spectra but not for the MMT spectra, which only cover H$_\beta$, H$_\gamma$, and H$_\delta$.
As a result, the atmospheric parameters derived from the MMT spectra lead to mass estimates that show unrealistic trends with \teff.
In contrast, the masses obtained from the atmospheric parameters of the Bok spectra yield nearly constant masses across the \teff\ range of the sdBs and sdOs (Sect.~\ref{sec:res:mmt:params}), which constitutes an important reliability check.

\item The distribution of sdBs along the EHB is not continuous. There is a paucity of stars around 33~kK that is clearly visible in both the Kiel and the HR diagrams (Figs.~\ref{fig:Kiel},~\ref{fig:HRD}). This is also observed among pulsators, with a void of sdBV$_{\rm r}$ pulsators around 33 kK (Fig.~\ref{fig:Kiel_puls}). Such a discontinuity was also noticed in the sdB sample of \citet{2022A&A...661A.113G} and is recovered from the evolutionary sequences computed by \citet{2017A&A...599A..54X}. According to these sequences, the stars on opposite sides of the gap experienced the He-flash at different times during their evolution: the hotter ones when they were already contracting on the WD cooling track (the hot- or late-flasher scenario), the cooler ones already at the tip of the RGB.
A late-flasher origin for the sdB hotter than $\sim$33~kK is consistent with the fact that we derive, on average, canonical masses for these sdBs.

\item The helium abundances in our sdBs follow the well-documented trend of helium increasing with \teff, first noticed by \citet{2003A&A...400..939E}. In the hottest sdBs, the helium abundances plateau around \heh\ =~$-$1.4 (Fig.~\ref{fig:Teff_He}), not quite reaching the solar value of \heh\ =~$-$1.0, which is commonly used to distinguish between H-rich and He-rich (or intermediate-He) subdwarfs. Based on this observation, we believe that a helium abundance of \heh\ =~$-$1.2 constitutes a more appropriate separator between the H-rich and intermediate-He classes (Sect.~\ref{sec:res:bok:atmo}). We find that the low \teff\ and low helium (\heh\ $\lesssim -3$) tail of the He$-$\teff\ relation is sparsely populated and the most He-poor sdBs show very weak metal lines in their MMT spectra (Fig.~\ref{fig:spectra_all}), if any. In addition, many of the most He-poor sdBs are also located below the ZAEHB for a core-mass of 0.45~\msun.

\item Our sample includes more than 80 pulsating hot subdwarfs and we found the different types of pulsators ($g$-mode, $p$-mode, and hybrid) to be well separated in both the Kiel and the HR diagrams. We noted that at a given \teff\, the pulsating sdBs are preferentially found at higher luminosity than their constant counterparts. This trend is possibly related to the time it takes for the iron/nickel reservoir to build-up in the Z-bump region at a sufficient level, through the action of radiative levitation (in competition with other processes), to start driving effectively pulsations.

\item Among the stars with MMT spectra, we identified seven objects that show hints of helium stratification from the inspection of the fit of their helium lines, six of which were not known to exhibit such features. They have \teff\ in a narrow range between 28 and 31 kK, in line with the few stars already known to display helium stratification  \citep{2018A&A...618A..86S,2013A&A...557A.122G}. Unfortunately, the resolution of our MMT spectra is too low to detect the presence of \het. We believe that stratification, and the presence of \het, is more common than suspected in this temperature range, but detecting it requires at least medium spectral resolution, sufficiently high $S/N$, and a detailed inspection of the spectral fits (Sect.~\ref{sec:res:mmt:strat}).

\item The H-rich sdBs and sdOs have a similar mass distribution, with a median value around 0.47~\msun\ (Table~\ref{table_masses}), in good agreement with the theoretical expectation for the canonical mass at the He-core flash. The dispersions of the sdB and sdO mass distributions are consistent with the uncertainties of the individual mass measurements, which are around 18\% ($\sim$0.08~\msun), suggesting no significant intrinsic scatter in the masses of these stars. We see no significant low-mass or high-mass tail in their mass distributions (Sect.~\ref{sec:diss:mass}, Fig.~\ref{fig:Mass_dist}). Combined with the fact that there is no significant difference between the mass distribution of the close binaries (identified from RV variations) and the single hot subdwarfs, we believe that the merger channels do not significantly contribute to the formation of hydrogen-rich hot subdwarfs (Sect.~\ref{sec:diss:mass:binary}).

\item Candidates low-mass sdBs ($M< 0.45$ \msun) selected from their position in the HRD only represent a small fraction ($< 6$\%) of our EHB population (Sect~\ref{sec:diss:mass:b_ehb}). Their masses from parallax and SED fits are, on average, lower than canonical and they are likely the progeny of intermediate-mass MS stars (2$-$3.5 \msun) that ignited helium in a non-degenerate core. 

\item The positions of the He-sdOs in the HRD follow the theoretical helium main sequence remarkably well (Fig.~\ref{fig:HRD}). According to their position along this sequence, the majority of them have masses between 0.6 and 1.0~\msun, indicating a wider mass range than derived for their H-rich counterparts. The masses obtained from the parallax and SED fits also span a wide range, with 68\% of them ($Q_{16}$-$Q_{84}$) falling within 0.48$-$1.14~\msun\ and a median mass of 0.78~\msun. Their mass dispersion is larger than expected from the individual uncertainties alone (Table~\ref{table_masses}). A dominant contribution by merger channels is the most likely explanation for this wide mass distribution and relatively large average mass (Sect.~\ref{sec:diss:mass}).

\end{itemize}

\section{Data availability}\label{sec:data_end}

An extended version of Table~\ref{table_res_all} including additional information (i.e. columns) for each star is only available online at the CDS via anonymous ftp to \url{cdsarc.u-strasbg.fr (XXXX)} or via \url{http://cdsarc.u-strasbg.fr/viz-bin/cat/J/A+A/XXX/zzz}.
All of the Bok spectra analyzed here are available as ascii files at the CDS.
Additional figures in Appendix~\ref{sec:appD} can be found online (on Zenodo, include link). This includes the MMT spectral atlas, and figures showing the best-fit solutions for the Bok spectra and SED fits of the stars in our sample.

\begin{acknowledgements}
We are forever grateful to Gilles Fontaine, who started this project with Betsy Green two decades ago. Thanks to the data and records that he kept, we were able to bring his work to completion and achieve much more than would have been possible in 2005. G.F. would have thanked Pierre Bergeron for his help in the very early phase of this project. We are thankful for those who shared with us their unpublished data to help us track down pulsators and RV variable stars: W. Zong, S. Geier, F. Mattig, H. Dawson, and V. Schaffenroth.
M.L. acknowledges funding from the Deutsche Forschungsgemeinschaft (grant LA 4383/4-1). 
M.D. was supported by the Deutsches Zentrum für Luft- und Raumfahrt (DLR) through grant 50-OR-2304. V.V.G. is a F.R.S.-FNRS Research Associate.
S.C. acknowledge support from the Centre National d’Études Spatiales (CNES, France), focused on the mission TESS. 
This work has made use of data from the European Space Agency (ESA) mission {\it Gaia} (\url{https://www.cosmos.esa.int/gaia}), processed by the {\it Gaia} Data Processing and Analysis Consortium (DPAC, \url{https://www.cosmos.esa.int/web/gaia/dpac/consortium}). Funding for the DPAC has been provided by national institutions, in particular the institutions participating in the {\it Gaia} Multilateral Agreement.
This work has made use of IRAF, which was distributed by the National Optical Astronomy Observatory, USA, which is operated by the Association of Universities for Research in Astronomy, Inc., under a cooperative agreement with the National Science Foundation.
%This publication makes use of data products from the Two Micron All Sky Survey, which is a joint project of the University of Massachusetts and the Infrared Processing and Analysis Center/California Institute of Technology, funded by the National Aeronautics and Space Administration and the National Science Foundation.
 This research has made use of NASA’s Astrophysics Data System Bibliographic Services, of the SIMBAD database \citep{Simbad}, operated at CDS, Strasbourg, France, and of the VizieR catalogue access tool \citep{vizier}, CDS, Strasbourg Astronomical Observatory, France (DOI : 10.26093/cds/vizier). This research has made use of TOPCAT \citep{topcat} and the \textsc{python} packages pandas \citep{reback2020pandas} and \textsc{matplotlib} \citep{Hunter:2007}. 
 
\end{acknowledgements}

   \bibliographystyle{aa} % style aa.bst
%   \bibliography{ref} % your references Yourfile.bib

\begin{thebibliography}{179}
\expandafter\ifx\csname natexlab\endcsname\relax\def\natexlab#1{#1}\fi

\bibitem[{{Ahmad} {et~al.}(2004){Ahmad}, {Jeffery}, \& {Fullerton}}]{Ahmad2004}
{Ahmad}, A., {Jeffery}, C.~S., \& {Fullerton}, A.~W. 2004, \aap, 418, 275

\bibitem[{{Alam} {et~al.}(2015){Alam}, {Albareti}, {Allende Prieto}, {Anders},
  {Anderson}, {Anderton}, {Andrews}, {Armengaud}, {Aubourg}, {Bailey}, {Basu},
  {Bautista}, {Beaton}, {Beers}, {Bender}, {Berlind}, {Beutler}, {Bhardwaj},
  {Bird}, {Bizyaev}, {Blake}, {Blanton}, {Blomqvist}, {Bochanski}, {Bolton},
  {Bovy}, {Shelden Bradley}, {Brandt}, {Brauer}, {Brinkmann}, {Brown},
  {Brownstein}, {Burden}, {Burtin}, {Busca}, {Cai}, {Capozzi}, {Carnero
  Rosell}, {Carr}, {Carrera}, {Chambers}, {Chaplin}, {Chen}, {Chiappini},
  {Chojnowski}, {Chuang}, {Clerc}, {Comparat}, {Covey}, {Croft}, {Cuesta},
  {Cunha}, {da Costa}, {Da Rio}, {Davenport}, {Dawson}, {De Lee}, {Delubac},
  {Deshpande}, {Dhital}, {Dutra-Ferreira}, {Dwelly}, {Ealet}, {Ebelke},
  {Edmondson}, {Eisenstein}, {Ellsworth}, {Elsworth}, {Epstein}, {Eracleous},
  {Escoffier}, {Esposito}, {Evans}, {Fan}, {Fern{\'a}ndez-Alvar}, {Feuillet},
  {Filiz Ak}, {Finley}, {Finoguenov}, {Flaherty}, {Fleming}, {Font-Ribera},
  {Foster}, {Frinchaboy}, {Galbraith-Frew}, {Garc{\'\i}a},
  {Garc{\'\i}a-Hern{\'a}ndez}, {Garc{\'\i}a P{\'e}rez}, {Gaulme}, {Ge},
  {G{\'e}nova-Santos}, {Georgakakis}, {Ghezzi}, {Gillespie}, {Girardi},
  {Goddard}, {Gontcho}, {Gonz{\'a}lez Hern{\'a}ndez}, {Grebel}, {Green},
  {Grieb}, {Grieves}, {Gunn}, {Guo}, {Harding}, {Hasselquist}, {Hawley},
  {Hayden}, {Hearty}, {Hekker}, {Ho}, {Hogg}, {Holley-Bockelmann}, {Holtzman},
  {Honscheid}, {Huber}, {Huehnerhoff}, {Ivans}, {Jiang}, {Johnson},
  {Kinemuchi}, {Kirkby}, {Kitaura}, {Klaene}, {Knapp}, {Kneib}, {Koenig},
  {Lam}, {Lan}, {Lang}, {Laurent}, {Le Goff}, {Leauthaud}, {Lee}, {Lee},
  {Licquia}, {Liu}, {Long}, {L{\'o}pez-Corredoira}, {Lorenzo-Oliveira},
  {Lucatello}, {Lundgren}, {Lupton}, {Mack}, {Mahadevan}, {Maia}, {Majewski},
  {Malanushenko}, {Malanushenko}, {Manchado}, {Manera}, {Mao}, {Maraston},
  {Marchwinski}, {Margala}, {Martell}, {Martig}, {Masters}, {Mathur},
  {McBride}, {McGehee}, {McGreer}, {McMahon}, {M{\'e}nard}, {Menzel},
  {Merloni}, {M{\'e}sz{\'a}ros}, {Miller}, {Miralda-Escud{\'e}}, {Miyatake},
  {Montero-Dorta}, {More}, {Morganson}, {Morice-Atkinson}, {Morrison},
  {Mosser}, {Muna}, {Myers}, {Nandra}, {Newman}, {Neyrinck}, {Nguyen},
  {Nichol}, {Nidever}, {Noterdaeme}, {Nuza}, {O'Connell}, {O'Connell},
  {O'Connell}, {Ogando}, {Olmstead}, {Oravetz}, {Oravetz}, {Osumi}, {Owen},
  {Padgett}, {Padmanabhan}, {Paegert}, {Palanque-Delabrouille}, \&
  {Pan}}]{sdss3}
{Alam}, S., {Albareti}, F.~D., {Allende Prieto}, C., {et~al.} 2015, \apjs, 219,
  12

\bibitem[{{Arancibia-Rojas} {et~al.}(2024){Arancibia-Rojas}, {Zorotovic},
  {Vu{\v{c}}kovi{\'c}}, {Bobrick}, {Vos}, \&
  {Piraino-Cerda}}]{2024MNRAS.52711184A}
{Arancibia-Rojas}, E., {Zorotovic}, M., {Vu{\v{c}}kovi{\'c}}, M., {et~al.}
  2024, \mnras, 527, 11184

\bibitem[{{Asplund} {et~al.}(2009){Asplund}, {Grevesse}, {Sauval}, \&
  {Scott}}]{asplund09}
{Asplund}, M., {Grevesse}, N., {Sauval}, A.~J., \& {Scott}, P. 2009, \araa, 47,
  481

\bibitem[{{Baran} {et~al.}(2024){Baran}, {Charpinet}, {{\O}stensen}, {Reed},
  {Van Grootel}, {Lyu}, {Telting}, \& {N{\'e}meth}}]{2024A&A...686A..65B}
{Baran}, A.~S., {Charpinet}, S., {{\O}stensen}, R.~H., {et~al.} 2024, \aap,
  686, A65

\bibitem[{{Baran} {et~al.}(2019){Baran}, {Telting}, {Jeffery}, {{\O}stensen},
  {Vos}, {Reed}, {V{\r{A}}­ckovi{\'c}}, \& {}}]{2019MNRAS.489.1556B}
{Baran}, A.~S., {Telting}, J.~H., {Jeffery}, C.~S., {et~al.} 2019, \mnras, 489,
  1556

\bibitem[{{Barlow} {et~al.}(2022){Barlow}, {Corcoran}, {Parker}, {Kupfer},
  {N{\'e}meth}, {Hermes}, {Lopez}, {Frondorf}, {Vestal}, \&
  {Holden}}]{2022ApJ...928...20B}
{Barlow}, B.~N., {Corcoran}, K.~A., {Parker}, I.~M., {et~al.} 2022, \apj, 928,
  20

\bibitem[{{Beauchamp} {et~al.}(1997){Beauchamp}, {Wesemael}, \&
  {Bergeron}}]{1997ApJS..108..559B}
{Beauchamp}, A., {Wesemael}, F., \& {Bergeron}, P. 1997, \apjs, 108, 559

\bibitem[{{B{\'e}dard} {et~al.}(2020){B{\'e}dard}, {Bergeron}, {Brassard}, \&
  {Fontaine}}]{2020ApJ...901...93B}
{B{\'e}dard}, A., {Bergeron}, P., {Brassard}, P., \& {Fontaine}, G. 2020, \apj,
  901, 93

\bibitem[{{Bianchi} {et~al.}(2017){Bianchi}, {Shiao}, \& {Thilker}}]{galex}
{Bianchi}, L., {Shiao}, B., \& {Thilker}, D. 2017, \apjs, 230, 24

\bibitem[{{Blanchette} {et~al.}(2008){Blanchette}, {Chayer}, {Wesemael},
  {Fontaine}, {Fontaine}, {Dupuis}, {Kruk}, \& {Green}}]{2008ApJ...678.1329B}
{Blanchette}, J.~P., {Chayer}, P., {Wesemael}, F., {et~al.} 2008, \apj, 678,
  1329

\bibitem[{{Brassard} {et~al.}(2001){Brassard}, {Fontaine}, {Bill{\`e}res},
  {Charpinet}, {Liebert}, \& {Saffer}}]{2001ApJ...563.1013B}
{Brassard}, P., {Fontaine}, G., {Bill{\`e}res}, M., {et~al.} 2001, \apj, 563,
  1013

\bibitem[{{Brown} {et~al.}(2016){Brown}, {Cassisi}, {D'Antona}, {Salaris},
  {Milone}, {Dalessandro}, {Piotto}, {Renzini}, {Sweigart}, {Bellini},
  {Ortolani}, {Sarajedini}, {Aparicio}, {Bedin}, {Anderson}, {Pietrinferni}, \&
  {Nardiello}}]{brown16}
{Brown}, T.~M., {Cassisi}, S., {D'Antona}, F., {et~al.} 2016, \apj, 822, 44

\bibitem[{{Brown} {et~al.}(2001){Brown}, {Sweigart}, {Lanz}, {Landsman}, \&
  {Hubeny}}]{2001ApJ...562..368B}
{Brown}, T.~M., {Sweigart}, A.~V., {Lanz}, T., {Landsman}, W.~B., \& {Hubeny},
  I. 2001, \apj, 562, 368

\bibitem[{Butler \& Giddings(1985)}]{butler85}
Butler, K. \& Giddings, J. 1985, College London

\bibitem[{{Charpinet} {et~al.}(2005{\natexlab{a}}){Charpinet}, {Fontaine},
  {Brassard}, {Bill{\`e}res}, {Green}, \& {Chayer}}]{2005ASPC..334..619C}
{Charpinet}, S., {Fontaine}, G., {Brassard}, P., {et~al.} 2005{\natexlab{a}},
  in Astronomical Society of the Pacific Conference Series, Vol. 334, 14th
  European Workshop on White Dwarfs, ed. D.~{Koester} \& S.~{Moehler}, 619

\bibitem[{{Charpinet} {et~al.}(1997){Charpinet}, {Fontaine}, {Brassard},
  {Chayer}, {Rogers}, {Iglesias}, \& {Dorman}}]{charpinet97}
{Charpinet}, S., {Fontaine}, G., {Brassard}, P., {et~al.} 1997, \apjl, 483,
  L123

\bibitem[{{Charpinet} {et~al.}(2002){Charpinet}, {Fontaine}, {Brassard}, \&
  {Dorman}}]{2002ApJS..139..487C}
{Charpinet}, S., {Fontaine}, G., {Brassard}, P., \& {Dorman}, B. 2002, \apjs,
  139, 487

\bibitem[{{Charpinet} {et~al.}(2005{\natexlab{b}}){Charpinet}, {Fontaine},
  {Brassard}, {Green}, \& {Chayer}}]{2005A&A...437..575C}
{Charpinet}, S., {Fontaine}, G., {Brassard}, P., {Green}, E.~M., \& {Chayer},
  P. 2005{\natexlab{b}}, \aap, 437, 575

\bibitem[{{Charpinet} {et~al.}(2010){Charpinet}, {Green}, {Baglin}, {Van
  Grootel}, {Fontaine}, {Vauclair}, {Chaintreuil}, {Weiss}, {Michel},
  {Auvergne}, {Catala}, {Samadi}, \& {Baudin}}]{2010A&A...516L...6C}
{Charpinet}, S., {Green}, E.~M., {Baglin}, A., {et~al.} 2010, \aap, 516, L6

\bibitem[{{Copperwheat} {et~al.}(2011){Copperwheat}, {Morales-Rueda}, {Marsh},
  {Maxted}, \& {Heber}}]{2011MNRAS.415.1381C}
{Copperwheat}, C.~M., {Morales-Rueda}, L., {Marsh}, T.~R., {Maxted}, P.~F.~L.,
  \& {Heber}, U. 2011, \mnras, 415, 1381

\bibitem[{{{\c{S}}ener} \& {Jeffery}(2014)}]{Sener2014}
{{\c{S}}ener}, H.~T. \& {Jeffery}, C.~S. 2014, \mnras, 440, 2676

\bibitem[{{Culpan} {et~al.}(2024){Culpan}, {Dorsch}, {Geier}, {Pelisoli},
  {Heber}, {Kub{\'a}tov{\'a}}, \& {Cabezas}}]{2024A&A...685A.134C}
{Culpan}, R., {Dorsch}, M., {Geier}, S., {et~al.} 2024, \aap, 685, A134

\bibitem[{{D'Cruz} {et~al.}(1996){D'Cruz}, {Dorman}, {Rood}, \&
  {O'Connell}}]{1996ApJ...466..359D}
{D'Cruz}, N.~L., {Dorman}, B., {Rood}, R.~T., \& {O'Connell}, R.~W. 1996, \apj,
  466, 359

\bibitem[{{De Angeli} {et~al.}(2023){De Angeli}, {Weiler}, {Montegriffo},
  {Evans}, {Riello}, {Andrae}, {Carrasco}, {Busso}, {Burgess}, {Cacciari},
  {Davidson}, {Harrison}, {Hodgkin}, {Jordi}, {Osborne}, {Pancino},
  {Altavilla}, {Barstow}, {Bailer-Jones}, {Bellazzini}, {Brown}, {Castellani},
  {Cowell}, {Delchambre}, {De Luise}, {Diener}, {Fabricius}, {Fouesneau},
  {Fr{\'e}mat}, {Gilmore}, {Giuffrida}, {Hambly}, {Hidalgo}, {Holland},
  {Kostrzewa-Rutkowska}, {van Leeuwen}, {Lobel}, {Marinoni}, {Miller},
  {Pagani}, {Palaversa}, {Piersimoni}, {Pulone}, {Ragaini}, {Rainer},
  {Richards}, {Rixon}, {Ruz-Mieres}, {Sanna}, {Sarro}, {Rowell}, {Sordo},
  {Walton}, \& {Yoldas}}]{gaia_spectra}
{De Angeli}, F., {Weiler}, M., {Montegriffo}, P., {et~al.} 2023, \aap, 674, A2

\bibitem[{{Dorman} {et~al.}(1993){Dorman}, {Rood}, \&
  {O'Connell}}]{1993ApJ...419..596D}
{Dorman}, B., {Rood}, R.~T., \& {O'Connell}, R.~W. 1993, \apj, 419, 596

\bibitem[{{Dorsch}(2024)}]{2024PhDT........36D}
{Dorsch}, M. 2024, PhD thesis, Friedrich Alexander University of
  Erlangen-Nuremberg, Germany

\bibitem[{{Dorsch} {et~al.}(2024){Dorsch}, {Jeffery}, {Philip Monai}, {Tout},
  {Snowdon}, {Monageng}, {Scott}, {Miszalski}, \&
  {Woolf}}]{2024A&A...691A.165D}
{Dorsch}, M., {Jeffery}, C.~S., {Philip Monai}, A., {et~al.} 2024, \aap, 691,
  A165

\bibitem[{{Dorsch} {et~al.}(2022){Dorsch}, {Reindl}, {Pelisoli}, {Heber},
  {Geier}, {Istrate}, \& {Justham}}]{2022A&A...658L...9D}
{Dorsch}, M., {Reindl}, N., {Pelisoli}, I., {et~al.} 2022, \aap, 658, L9

\bibitem[{{Downes}(1986)}]{1986ApJS...61..569D}
{Downes}, R.~A. 1986, \apjs, 61, 569

\bibitem[{{Driebe} {et~al.}(1998){Driebe}, {Schoenberner}, {Bloecker}, \&
  {Herwig}}]{1998A&A...339..123D}
{Driebe}, T., {Schoenberner}, D., {Bloecker}, T., \& {Herwig}, F. 1998, \aap,
  339, 123

\bibitem[{{Edelmann}(2003)}]{2003PhDT........48E}
{Edelmann}, H. 2003, PhD thesis, Friedrich Alexander University of
  Erlangen-Nuremberg, Germany

\bibitem[{{Edelmann} {et~al.}(2003){Edelmann}, {Heber}, {Hagen}, {Lemke},
  {Dreizler}, {Napiwotzki}, \& {Engels}}]{2003A&A...400..939E}
{Edelmann}, H., {Heber}, U., {Hagen}, H.~J., {et~al.} 2003, \aap, 400, 939

\bibitem[{{Edelmann} {et~al.}(2004){Edelmann}, {Heber}, {Lisker}, \&
  {Green}}]{2004Ap&SS.291..315E}
{Edelmann}, H., {Heber}, U., {Lisker}, T., \& {Green}, E.~M. 2004, \apss, 291,
  315

\bibitem[{{Edelmann} {et~al.}(2001){Edelmann}, {Heber}, \&
  {Napiwotzki}}]{2001AN....322..401E}
{Edelmann}, H., {Heber}, U., \& {Napiwotzki}, R. 2001, Astronomische
  Nachrichten, 322, 401

\bibitem[{{El-Badry} {et~al.}(2021){El-Badry}, {Rix}, \&
  {Heintz}}]{El-Badry2021}
{El-Badry}, K., {Rix}, H.-W., \& {Heintz}, T.~M. 2021, \mnras, 506, 2269

\bibitem[{{Feige}(1958)}]{1958ApJ...128..267F}
{Feige}, J. 1958, \apj, 128, 267

\bibitem[{{Filiz} {et~al.}(2024){Filiz}, {Werner}, {Rauch}, \&
  {Reindl}}]{Filiz2024}
{Filiz}, S., {Werner}, K., {Rauch}, T., \& {Reindl}, N. 2024, \aap, 691, A290

\bibitem[{{Fitzpatrick} {et~al.}(2019){Fitzpatrick}, {Massa}, {Gordon},
  {Bohlin}, \& {Clayton}}]{2019ApJ...886..108F}
{Fitzpatrick}, E.~L., {Massa}, D., {Gordon}, K.~D., {Bohlin}, R., \& {Clayton},
  G.~C. 2019, \apj, 886, 108

\bibitem[{{Fitzpatrick} {et~al.}(2014){Fitzpatrick}, {Olsen}, {Economou},
  {Stobie}, {Beers}, {Dickinson}, {Norris}, {Saha}, {Seaman}, {Silva},
  {Swaters}, {Thomas}, \& {Valdes}}]{Fitzpatrick2014}
{Fitzpatrick}, M.~J., {Olsen}, K., {Economou}, F., {et~al.} 2014, in Society of
  Photo-Optical Instrumentation Engineers (SPIE) Conference Series, Vol. 9149,
  Observatory Operations: Strategies, Processes, and Systems V, ed. A.~B.
  {Peck}, C.~R. {Benn}, \& R.~L. {Seaman}, 91491T

\bibitem[{{Flewelling}(2018)}]{pan-starrsdr2}
{Flewelling}, H. 2018, in American Astronomical Society Meeting Abstracts, Vol.
  231, American Astronomical Society Meeting Abstracts \#231, 436.01

\bibitem[{{Fontaine} {et~al.}(2019){Fontaine}, {Bergeron}, {Brassard},
  {Charpinet}, {Randall}, {Van Grootel}, {Latour}, \&
  {Green}}]{2019ApJ...880...79F}
{Fontaine}, G., {Bergeron}, P., {Brassard}, P., {et~al.} 2019, \apj, 880, 79

\bibitem[{{Fontaine} {et~al.}(2006){Fontaine}, {Brassard}, {Charpinet}, \&
  {Chayer}}]{fontaine06}
{Fontaine}, G., {Brassard}, P., {Charpinet}, S., \& {Chayer}, P. 2006, \memsai,
  77, 49

\bibitem[{{Fontaine} {et~al.}(2003){Fontaine}, {Brassard}, {Charpinet},
  {Green}, {Chayer}, {Bill{\`e}res}, \& {Randall}}]{fontaine03}
{Fontaine}, G., {Brassard}, P., {Charpinet}, S., {et~al.} 2003, \apj, 597, 518

\bibitem[{{Fontaine} {et~al.}(2012){Fontaine}, {Brassard}, {Charpinet},
  {Green}, {Randall}, \& {Van Grootel}}]{2012A&A...539A..12F}
{Fontaine}, G., {Brassard}, P., {Charpinet}, S., {et~al.} 2012, \aap, 539, A12

\bibitem[{{Fontaine} \& {Chayer}(1997)}]{fontaine97}
{Fontaine}, G. \& {Chayer}, P. 1997, in The Third Conference on Faint Blue
  Stars, ed. A.~G.~D. {Philip}, J.~{Liebert}, R.~{Saffer}, \& D.~S. {Hayes},
  169

\bibitem[{{Fontaine} {et~al.}(2014){Fontaine}, {Green}, {Brassard}, {Latour},
  \& {Chayer}}]{2014ASPC..481...83F}
{Fontaine}, G., {Green}, E., {Brassard}, P., {Latour}, M., \& {Chayer}, P.
  2014, in Astronomical Society of the Pacific Conference Series, Vol. 481, 6th
  Meeting on Hot Subdwarf Stars and Related Objects, ed. V.~{van Grootel},
  E.~{Green}, G.~{Fontaine}, \& S.~{Charpinet}, 83

\bibitem[{{For} {et~al.}(2010){For}, {Green}, {Fontaine}, {Drechsel}, {Shaw},
  {Dittmann}, {Fay}, {Francoeur}, {Laird}, {Moriyama}, {Morris},
  {Rodr{\'\i}guez-L{\'o}pez}, {Sierchio}, {Story}, {Strom}, {Wang}, {Adams},
  {Bolin}, {Eskew}, \& {Chayer}}]{2010ApJ...708..253F}
{For}, B.~Q., {Green}, E.~M., {Fontaine}, G., {et~al.} 2010, \apj, 708, 253

\bibitem[{{Geier}(2013)}]{2013A&A...549A.110G}
{Geier}, S. 2013, \aap, 549, A110

\bibitem[{{Geier} {et~al.}(2022){Geier}, {Dorsch}, {Pelisoli}, {Reindl},
  {Heber}, \& {Irrgang}}]{2022A&A...661A.113G}
{Geier}, S., {Dorsch}, M., {Pelisoli}, I., {et~al.} 2022, \aap, 661, A113

\bibitem[{{Geier} {et~al.}(2013){Geier}, {Heber}, {Edelmann}, {Morales-Rueda},
  {Kilkenny}, {O'Donoghue}, {Marsh}, \& {Copperwheat}}]{2013A&A...557A.122G}
{Geier}, S., {Heber}, U., {Edelmann}, H., {et~al.} 2013, \aap, 557, A122

\bibitem[{{Geier} {et~al.}(2024){Geier}, {Heber}, {Irrgang}, {Dorsch},
  {Bastian}, {Neunteufel}, {Kupfer}, {Bloemen}, {Kreuzer}, {M{\"o}ller},
  {Schindewolf}, {Schneider}, {Ziegerer}, {Pelisoli}, {Schaffenroth}, {Barlow},
  {Raddi}, {Geier}, {Reindl}, {Rauch}, {Nemeth}, \&
  {G{\"a}nsicke}}]{2024A&A...690A.368G}
{Geier}, S., {Heber}, U., {Irrgang}, A., {et~al.} 2024, \aap, 690, A368

\bibitem[{{Geier} {et~al.}(2008){Geier}, {Nesslinger}, {Heber}, {Randall},
  {Edelmann}, \& {Green}}]{2008A&A...477L..13G}
{Geier}, S., {Nesslinger}, S., {Heber}, U., {et~al.} 2008, \aap, 477, L13

\bibitem[{{Giddings}(1981)}]{giddings81}
{Giddings}, J.~R. 1981, PhD thesis, -

\bibitem[{{Gigosos} \& {Gonz{\'a}lez}(2009)}]{stark4471}
{Gigosos}, M.~A. \& {Gonz{\'a}lez}, M.~{\'A}. 2009, \aap, 503, 293

\bibitem[{{Green} {et~al.}(2008){Green}, {Fontaine}, {Hyde}, {For}, \&
  {Chayer}}]{2008ASPC..392...75G}
{Green}, E.~M., {Fontaine}, G., {Hyde}, E.~A., {For}, B.~Q., \& {Chayer}, P.
  2008, in Astronomical Society of the Pacific Conference Series, Vol. 392, Hot
  Subdwarf Stars and Related Objects, ed. U.~{Heber}, C.~S. {Jeffery}, \&
  R.~{Napiwotzki}, 75

\bibitem[{{Green} {et~al.}(2003){Green}, {Fontaine}, {Reed}, {Callerame},
  {Seitenzahl}, {White}, {Hyde}, {{\O}stensen}, {Cordes}, {Brassard}, {Falter},
  {Jeffery}, {Dreizler}, {Schuh}, {Giovanni}, {Edelmann}, {Rigby}, \&
  {Bronowska}}]{2003ApJ...583L..31G}
{Green}, E.~M., {Fontaine}, G., {Reed}, M.~D., {et~al.} 2003, \apjl, 583, L31

\bibitem[{{Green} {et~al.}(2004){Green}, {For}, {Hyde}, {Seitenzahl},
  {Callerame}, {White}, {Young}, {Huff}, {Mills}, \&
  {Steinfadt}}]{2004Ap&SS.291..267G}
{Green}, E.~M., {For}, B., {Hyde}, E.~A., {et~al.} 2004, \apss, 291, 267

\bibitem[{{Green} {et~al.}(2005){Green}, {For}, \&
  {Hyde}}]{2005ASPC..334..363G}
{Green}, E.~M., {For}, B.~Q., \& {Hyde}, E.~A. 2005, in Astronomical Society of
  the Pacific Conference Series, Vol. 334, 14th European Workshop on White
  Dwarfs, ed. D.~{Koester} \& S.~{Moehler}, 363

\bibitem[{{Green} {et~al.}(2011){Green}, {Guvenen}, {O'Malley}, {O'Connell},
  {Baringer}, {Villareal}, {Carleton}, {Fontaine}, {Brassard}, \&
  {Charpinet}}]{2011ApJ...734...59G}
{Green}, E.~M., {Guvenen}, B., {O'Malley}, C.~J., {et~al.} 2011, \apj, 734, 59

\bibitem[{{Green} {et~al.}(1986){Green}, {Schmidt}, \&
  {Liebert}}]{1986ApJS...61..305G}
{Green}, R.~F., {Schmidt}, M., \& {Liebert}, J. 1986, \apjs, 61, 305

\bibitem[{{Hall} \& {Jeffery}(2016)}]{2016MNRAS.463.2756H}
{Hall}, P.~D. \& {Jeffery}, C.~S. 2016, \mnras, 463, 2756

\bibitem[{{Han} {et~al.}(2003){Han}, {Podsiadlowski}, {Maxted}, \&
  {Marsh}}]{2003MNRAS.341..669H}
{Han}, Z., {Podsiadlowski}, P., {Maxted}, P.~F.~L., \& {Marsh}, T.~R. 2003,
  \mnras, 341, 669

\bibitem[{{Han} {et~al.}(2002){Han}, {Podsiadlowski}, {Maxted}, {Marsh}, \&
  {Ivanova}}]{2002MNRAS.336..449H}
{Han}, Z., {Podsiadlowski}, P., {Maxted}, P.~F.~L., {Marsh}, T.~R., \&
  {Ivanova}, N. 2002, \mnras, 336, 449

\bibitem[{{He} {et~al.}(2025){He}, {Meng}, {Lei}, {Yan}, \&
  {Lan}}]{2025A&A...693A.121H}
{He}, R., {Meng}, X., {Lei}, Z., {Yan}, H., \& {Lan}, S. 2025, \aap, 693, A121

\bibitem[{{Heber}(1991)}]{1991IAUS..145..363H}
{Heber}, U. 1991, in IAU Symposium, Vol. 145, Evolution of Stars: the
  Photospheric Abundance Connection, ed. G.~{Michaud} \& A.~V. {Tutukov}, 363

\bibitem[{{Heber}(2009)}]{2009ARA&A..47..211H}
{Heber}, U. 2009, \araa, 47, 211

\bibitem[{{Heber}(2016)}]{2016PASP..128h2001H}
{Heber}, U. 2016, \pasp, 128, 082001

\bibitem[{{Heber}(2024)}]{2024arXiv241011663H}
{Heber}, U. 2024, arXiv e-prints, arXiv:2410.11663

\bibitem[{{Heber} {et~al.}(2004){Heber}, {Drechsel}, {{\O}stensen}, {Karl},
  {Napiwotzki}, {Altmann}, {Cordes}, {Solheim}, {Voss}, {Koester}, \&
  {Folkes}}]{2004A&A...420..251H}
{Heber}, U., {Drechsel}, H., {{\O}stensen}, R., {et~al.} 2004, \aap, 420, 251

\bibitem[{{Heber} {et~al.}(2018){Heber}, {Irrgang}, \&
  {Schaffenroth}}]{2018OAst...27...35H}
{Heber}, U., {Irrgang}, A., \& {Schaffenroth}, J. 2018, Open Astronomy, 27, 35

\bibitem[{{Heber} {et~al.}(1999){Heber}, {Reid}, \&
  {Werner}}]{1999A&A...348L..25H}
{Heber}, U., {Reid}, I.~N., \& {Werner}, K. 1999, \aap, 348, L25

\bibitem[{{Hidalgo} {et~al.}(2018){Hidalgo}, {Pietrinferni}, {Cassisi},
  {Salaris}, {Mucciarelli}, {Savino}, {Aparicio}, {Silva Aguirre}, \&
  {Verma}}]{2018ApJ...856..125H}
{Hidalgo}, S.~L., {Pietrinferni}, A., {Cassisi}, S., {et~al.} 2018, \apj, 856,
  125

\bibitem[{{Houck} \& {Denicola}(2000)}]{houck00}
{Houck}, J.~C. \& {Denicola}, L.~A. 2000, in Astronomical Society of the
  Pacific Conference Series, Vol. 216, Astronomical Data Analysis Software and
  Systems IX, ed. N.~{Manset}, C.~{Veillet}, \& D.~{Crabtree}, 591

\bibitem[{{Hu} {et~al.}(2008){Hu}, {Dupret}, {Aerts}, {Nelemans}, {Kawaler},
  {Miglio}, {Montalban}, \& {Scuflaire}}]{2008A&A...490..243H}
{Hu}, H., {Dupret}, M.~A., {Aerts}, C., {et~al.} 2008, \aap, 490, 243

\bibitem[{{Hu} {et~al.}(2011){Hu}, {Tout}, {Glebbeek}, \&
  {Dupret}}]{2011MNRAS.418..195H}
{Hu}, H., {Tout}, C.~A., {Glebbeek}, E., \& {Dupret}, M.~A. 2011, \mnras, 418,
  195

\bibitem[{{Hubeny} {et~al.}(1994){Hubeny}, {Hummer}, \&
  {Lanz}}]{1994A&A...282..151H}
{Hubeny}, I., {Hummer}, D.~G., \& {Lanz}, T. 1994, \aap, 282, 151

\bibitem[{{Hubeny} \& {Lanz}(2011{\natexlab{a}})}]{2011ascl.soft09022H}
{Hubeny}, I. \& {Lanz}, T. 2011{\natexlab{a}}, {Synspec: General Spectrum
  Synthesis Program}, Astrophysics Source Code Library, record ascl:1109.022

\bibitem[{{Hubeny} \& {Lanz}(2011{\natexlab{b}})}]{2011ascl.soft09021H}
{Hubeny}, I. \& {Lanz}, T. 2011{\natexlab{b}}, {TLUSTY: Stellar Atmospheres,
  Accretion Disks, and Spectroscopic Diagnostics}, Astrophysics Source Code
  Library, record ascl:1109.021

\bibitem[{{Hubeny} \& {Lanz}(2017{\natexlab{a}})}]{2017arXiv170601859H}
{Hubeny}, I. \& {Lanz}, T. 2017{\natexlab{a}}, arXiv e-prints, arXiv:1706.01859

\bibitem[{{Hubeny} \& {Lanz}(2017{\natexlab{b}})}]{2017arXiv170601935H}
{Hubeny}, I. \& {Lanz}, T. 2017{\natexlab{b}}, arXiv e-prints, arXiv:1706.01935

\bibitem[{{Hubeny} \& {Lanz}(2017{\natexlab{c}})}]{2017arXiv170601937H}
{Hubeny}, I. \& {Lanz}, T. 2017{\natexlab{c}}, arXiv e-prints, arXiv:1706.01937

\bibitem[{Hunter(2007)}]{Hunter:2007}
Hunter, J.~D. 2007, Computing in Science \& Engineering, 9, 90

\bibitem[{{Husser} {et~al.}(2013){Husser}, {Wende-von Berg}, {Dreizler},
  {Homeier}, {Reiners}, {Barman}, \& {Hauschildt}}]{husser2013}
{Husser}, T.-O., {Wende-von Berg}, S., {Dreizler}, S., {et~al.} 2013, \aap,
  553, A6

\bibitem[{{Iben}(1967)}]{1967ApJ...147..624I}
{Iben}, Jr., I. 1967, \apj, 147, 624

\bibitem[{{Irrgang} {et~al.}(2021){Irrgang}, {Geier}, {Heber}, {Kupfer},
  {El-Badry}, \& {Bloemen}}]{2021A&A...650A.102I}
{Irrgang}, A., {Geier}, S., {Heber}, U., {et~al.} 2021, \aap, 650, A102

\bibitem[{{Irrgang} {et~al.}(2018){Irrgang}, {Kreuzer}, {Heber}, \&
  {Brown}}]{irrgang2018}
{Irrgang}, A., {Kreuzer}, S., {Heber}, U., \& {Brown}, W. 2018, \aap, 615, L5

\bibitem[{{Irrgang} {et~al.}(2014){Irrgang}, {Przybilla}, {Heber}, {B{\"o}ck},
  {Hanke}, {Nieva}, \& {Butler}}]{irrgang2014}
{Irrgang}, A., {Przybilla}, N., {Heber}, U., {et~al.} 2014, \aap, 565, A63

\bibitem[{{Irrgang} {et~al.}(2022){Irrgang}, {Przybilla}, \&
  {Meynet}}]{2022NatAs...6.1414I}
{Irrgang}, A., {Przybilla}, N., \& {Meynet}, G. 2022, Nature Astronomy, 6, 1414

\bibitem[{{Istrate} {et~al.}(2016){Istrate}, {Marchant}, {Tauris}, {Langer},
  {Stancliffe}, \& {Grassitelli}}]{2016A&A...595A..35I}
{Istrate}, A.~G., {Marchant}, P., {Tauris}, T.~M., {et~al.} 2016, \aap, 595,
  A35

\bibitem[{{Jeffery}(2020)}]{2020MNRAS.496..718J}
{Jeffery}, C.~S. 2020, \mnras, 496, 718

\bibitem[{{Jeffery} {et~al.}(2021){Jeffery}, {Miszalski}, \&
  {Snowdon}}]{2021MNRAS.501..623J}
{Jeffery}, C.~S., {Miszalski}, B., \& {Snowdon}, E. 2021, \mnras, 501, 623

\bibitem[{{Jeffery} \& {Saio}(2006)}]{jeffery06}
{Jeffery}, C.~S. \& {Saio}, H. 2006, \mnras, 371, 659

\bibitem[{{Jeffery} \& {Saio}(2007)}]{jeffery07}
{Jeffery}, C.~S. \& {Saio}, H. 2007, \mnras, 378, 379

\bibitem[{{Justham} {et~al.}(2011){Justham}, {Podsiadlowski}, \&
  {Han}}]{2011MNRAS.410..984J}
{Justham}, S., {Podsiadlowski}, P., \& {Han}, Z. 2011, \mnras, 410, 984

\bibitem[{{Kilkenny} {et~al.}(2010){Kilkenny}, {Fontaine}, {Green}, \&
  {Schuh}}]{kilkenny10}
{Kilkenny}, D., {Fontaine}, G., {Green}, E.~M., \& {Schuh}, S. 2010,
  Information Bulletin on Variable Stars, 5927, 1

\bibitem[{{Kilkenny} {et~al.}(1999){Kilkenny}, {Koen}, {O'Donoghue}, {van Wyk},
  {Larson}, {Shobbrook}, {Sullivan}, {Burleigh}, {Dobbie}, \&
  {Kawaler}}]{1999MNRAS.303..525K}
{Kilkenny}, D., {Koen}, C., {O'Donoghue}, D., {et~al.} 1999, \mnras, 303, 525

\bibitem[{{Koen}(2011)}]{2011MNRAS.415.3042K}
{Koen}, C. 2011, \mnras, 415, 3042

\bibitem[{{Kurucz}(1996)}]{Kurucz96}
{Kurucz}, R.~L. 1996, in Astronomical Society of the Pacific Conference Series,
  Vol. 108, M.A.S.S., Model Atmospheres and Spectrum Synthesis, ed. S.~J.
  {Adelman}, F.~{Kupka}, \& W.~W. {Weiss}, 2

\bibitem[{{Lara} {et~al.}(2012){Lara}, {Gonz{\'a}lez}, \&
  {Gigosos}}]{stark4922}
{Lara}, N., {Gonz{\'a}lez}, M.~{\'A}., \& {Gigosos}, M.~A. 2012, \aap, 542, A75

\bibitem[{{Latour} {et~al.}(2013){Latour}, {Fontaine}, {Chayer}, \&
  {Brassard}}]{2013ApJ...773...84L}
{Latour}, M., {Fontaine}, G., {Chayer}, P., \& {Brassard}, P. 2013, \apj, 773,
  84

\bibitem[{{Latour} {et~al.}(2015){Latour}, {Fontaine}, {Green}, \&
  {Brassard}}]{2015A&A...579A..39L}
{Latour}, M., {Fontaine}, G., {Green}, E.~M., \& {Brassard}, P. 2015, \aap,
  579, A39

\bibitem[{{Latour} {et~al.}(2019){Latour}, {Green}, \&
  {Fontaine}}]{2019A&A...623L..12L}
{Latour}, M., {Green}, E.~M., \& {Fontaine}, G. 2019, \aap, 623, L12

\bibitem[{{Latour} {et~al.}(2018){Latour}, {Randall}, {Calamida}, {Geier}, \&
  {Moehler}}]{shotglas1}
{Latour}, M., {Randall}, S.~K., {Calamida}, A., {Geier}, S., \& {Moehler}, S.
  2018, \aap, 618, A15

\bibitem[{{Lei} {et~al.}(2023){Lei}, {He}, {N{\'e}meth}, {Zou}, {Xiao}, {Yang},
  \& {Zhao}}]{Lei2023_mass}
{Lei}, Z., {He}, R., {N{\'e}meth}, P., {et~al.} 2023, \apj, 953, 122

\bibitem[{{Lindegren} {et~al.}(2021){Lindegren}, {Bastian}, {Biermann},
  {Bombrun}, {de Torres}, {Gerlach}, {Geyer}, {Hern{\'a}ndez}, {Hilger},
  {Hobbs}, {Klioner}, {Lammers}, {McMillan}, {Ramos-Lerate},
  {Steidelm{\"u}ller}, {Stephenson}, \& {van Leeuwen}}]{Lindegren2021}
{Lindegren}, L., {Bastian}, U., {Biermann}, M., {et~al.} 2021, \aap, 649, A4

\bibitem[{{Mason} {et~al.}(2001){Mason}, {Wycoff}, {Hartkopf}, {Douglass}, \&
  {Worley}}]{USNO_2001}
{Mason}, B.~D., {Wycoff}, G.~L., {Hartkopf}, W.~I., {Douglass}, G.~G., \&
  {Worley}, C.~E. 2001, \aj, 122, 3466

\bibitem[{{Maxted} {et~al.}(2001){Maxted}, {Heber}, {Marsh}, \&
  {North}}]{2001MNRAS.326.1391M}
{Maxted}, P.~F.~L., {Heber}, U., {Marsh}, T.~R., \& {North}, R.~C. 2001,
  \mnras, 326, 1391

\bibitem[{{Maxted} {et~al.}(2002){Maxted}, {Marsh}, {Heber}, {Morales-Rueda},
  {North}, \& {Lawson}}]{2002MNRAS.333..231M}
{Maxted}, P.~F.~L., {Marsh}, T.~R., {Heber}, U., {et~al.} 2002, \mnras, 333,
  231

\bibitem[{{Maxted} {et~al.}(2000){Maxted}, {Moran}, {Marsh}, \&
  {Gatti}}]{2000MNRAS.311..877M}
{Maxted}, P.~F.~L., {Moran}, C.~K.~J., {Marsh}, T.~R., \& {Gatti}, A.~A. 2000,
  \mnras, 311, 877

\bibitem[{{Miller Bertolami} {et~al.}(2008){Miller Bertolami}, {Althaus},
  {Unglaub}, \& {Weiss}}]{2008A&A...491..253M}
{Miller Bertolami}, M.~M., {Althaus}, L.~G., {Unglaub}, K., \& {Weiss}, A.
  2008, \aap, 491, 253

\bibitem[{{Moehler} \& {Heber}(1998)}]{1998A&A...335..985M}
{Moehler}, S. \& {Heber}, U. 1998, \aap, 335, 985

\bibitem[{{Momany} {et~al.}(2002){Momany}, {Piotto}, {Recio-Blanco}, {Bedin},
  {Cassisi}, \& {Bono}}]{momany02}
{Momany}, Y., {Piotto}, G., {Recio-Blanco}, A., {et~al.} 2002, \apjl, 576, L65

\bibitem[{{Moni Bidin} {et~al.}(2011){Moni Bidin}, {Villanova}, {Piotto},
  {Moehler}, \& {D'Antona}}]{monibidin2011}
{Moni Bidin}, C., {Villanova}, S., {Piotto}, G., {Moehler}, S., \& {D'Antona},
  F. 2011, \apjl, 738, L10

\bibitem[{{Montalb{\'a}n} \& {Noels}(2013)}]{2013EPJWC..4303002M}
{Montalb{\'a}n}, J. \& {Noels}, A. 2013, in European Physical Journal Web of
  Conferences, Vol.~43, European Physical Journal Web of Conferences, 03002

\bibitem[{{Morales-Rueda} {et~al.}(2003){Morales-Rueda}, {Maxted}, {Marsh},
  {North}, \& {Heber}}]{2003MNRAS.338..752M}
{Morales-Rueda}, L., {Maxted}, P.~F.~L., {Marsh}, T.~R., {North}, R.~C., \&
  {Heber}, U. 2003, \mnras, 338, 752

\bibitem[{{Napiwotzki} {et~al.}(2004){Napiwotzki}, {Karl}, {Lisker}, {Heber},
  {Christlieb}, {Reimers}, {Nelemans}, \& {Homeier}}]{2004Ap&SS.291..321N}
{Napiwotzki}, R., {Karl}, C.~A., {Lisker}, T., {et~al.} 2004, \apss, 291, 321

\bibitem[{{Naslim} {et~al.}(2013){Naslim}, {Jeffery}, {Hibbert}, \&
  {Behara}}]{2013MNRAS.434.1920N}
{Naslim}, N., {Jeffery}, C.~S., {Hibbert}, A., \& {Behara}, N.~T. 2013, \mnras,
  434, 1920

\bibitem[{{Newell} \& {Graham}(1976)}]{1976ApJ...204..804N}
{Newell}, B. \& {Graham}, J.~A. 1976, \apj, 204, 804

\bibitem[{{Nieva} \& {Przybilla}(2007)}]{nieva2007}
{Nieva}, M.~F. \& {Przybilla}, N. 2007, \aap, 467, 295

\bibitem[{Noels-Grotsch \& Miglio(2025)}]{10.1088/2514-3433/adcf15}
Noels-Grotsch, A. \& Miglio, A. 2025, The Golden Gift of Red Giants, 2514-3433
  (IOP Publishing)

\bibitem[{{Norris} {et~al.}(2011){Norris}, {Wright}, {Wade}, {Mahadevan}, \&
  {Gettel}}]{2011ApJ...743...88N}
{Norris}, J.~M., {Wright}, J.~T., {Wade}, R.~A., {Mahadevan}, S., \& {Gettel},
  S. 2011, \apj, 743, 88

\bibitem[{{Ochsenbein} {et~al.}(2000){Ochsenbein}, {Bauer}, \&
  {Marcout}}]{vizier}
{Ochsenbein}, F., {Bauer}, P., \& {Marcout}, J. 2000, \aaps, 143, 23

\bibitem[{{Ostensen} {et~al.}(2005){Ostensen}, {Heber}, \&
  {Maxted}}]{2005ASPC..334..435O}
{Ostensen}, R., {Heber}, U., \& {Maxted}, P. 2005, in Astronomical Society of
  the Pacific Conference Series, Vol. 334, 14th European Workshop on White
  Dwarfs, ed. D.~{Koester} \& S.~{Moehler}, 435

\bibitem[{{{\O}stensen} {et~al.}(2001){{\O}stensen}, {Heber}, {Silvotti},
  {Solheim}, {Dreizler}, \& {Edelmann}}]{2001A&A...378..466O}
{{\O}stensen}, R., {Heber}, U., {Silvotti}, R., {et~al.} 2001, \aap, 378, 466

\bibitem[{{{\O}stensen} {et~al.}(2020){{\O}stensen}, {Jeffery}, {Saio},
  {Hermes}, {Telting}, {Vu{\v{c}}kovi{\'c}}, {Vos}, {Baran}, \&
  {Reed}}]{2020MNRAS.499.3738O}
{{\O}stensen}, R.~H., {Jeffery}, C.~S., {Saio}, H., {et~al.} 2020, \mnras, 499,
  3738

\bibitem[{{{\O}stensen} {et~al.}(2008){{\O}stensen}, {Oreiro}, {Hu},
  {Drechsel}, \& {Heber}}]{2008ASPC..392..221O}
{{\O}stensen}, R.~H., {Oreiro}, R., {Hu}, H., {Drechsel}, H., \& {Heber}, U.
  2008, in Astronomical Society of the Pacific Conference Series, Vol. 392, Hot
  Subdwarf Stars and Related Objects, ed. U.~{Heber}, C.~S. {Jeffery}, \&
  R.~{Napiwotzki}, 221

\bibitem[{{{\O}stensen} {et~al.}(2010){{\O}stensen}, {Oreiro}, {Solheim},
  {Heber}, {Silvotti}, {Gonz{\'a}lez-P{\'e}rez}, {Ulla}, {P{\'e}rez
  Hern{\'a}ndez}, {Rodr{\'\i}guez-L{\'o}pez}, \&
  {Telting}}]{2010A&A...513A...6O}
{{\O}stensen}, R.~H., {Oreiro}, R., {Solheim}, J.~E., {et~al.} 2010, \aap, 513,
  A6

\bibitem[{{{\O}stensen} {et~al.}(2011){{\O}stensen}, {Silvotti}, {Charpinet},
  {Oreiro}, {Bloemen}, {Baran}, {Reed}, {Kawaler}, {Telting}, {Green},
  {O'Toole}, {Aerts}, {G{\"a}nsicke}, {Marsh}, {Breedt}, {Heber}, {Koester},
  {Quint}, {Kurtz}, {Rodr{\'\i}guez-L{\'o}pez}, {Vu{\v{c}}kovi{\'c}},
  {Ottosen}, {Frimann}, {Somero}, {Wilson}, {Thygesen}, {Lindberg}, {Kjeldsen},
  {Christensen-Dalsgaard}, {Allen}, {McCauliff}, \&
  {Middour}}]{2011MNRAS.414.2860O}
{{\O}stensen}, R.~H., {Silvotti}, R., {Charpinet}, S., {et~al.} 2011, \mnras,
  414, 2860

\bibitem[{{Paczy{\'n}ski}(1971)}]{1971AcA....21....1P}
{Paczy{\'n}ski}, B. 1971, \actaa, 21, 1

\bibitem[{pandas~development team(2020)}]{reback2020pandas}
pandas~development team, T. 2020, pandas-dev/pandas: Pandas

\bibitem[{{Pelisoli} {et~al.}(2022){Pelisoli}, {Dorsch}, {Heber},
  {G{\"a}nsicke}, {Geier}, {Kupfer}, {N{\'e}meth}, {Scaringi}, \&
  {Schaffenroth}}]{2022MNRAS.515.2496P}
{Pelisoli}, I., {Dorsch}, M., {Heber}, U., {et~al.} 2022, \mnras, 515, 2496

\bibitem[{{Pereira}(2011)}]{2011PhDT.......261P}
{Pereira}, C. 2011, PhD thesis, Queens University Belfast, Ireland

\bibitem[{{Politano} {et~al.}(2008){Politano}, {Taam}, {van der Sluys}, \&
  {Willems}}]{2008ApJ...687L..99P}
{Politano}, M., {Taam}, R.~E., {van der Sluys}, M., \& {Willems}, B. 2008,
  \apjl, 687, L99

\bibitem[{{Przybilla}(2005)}]{przybilla05}
{Przybilla}, N. 2005, \aap, 443, 293

\bibitem[{{Przybilla} \& {Butler}(2004)}]{przybilla04}
{Przybilla}, N. \& {Butler}, K. 2004, \apj, 609, 1181

\bibitem[{{Przybilla} {et~al.}(2006){Przybilla}, {Butler}, {Becker}, \&
  {Kudritzki}}]{przybilla2006}
{Przybilla}, N., {Butler}, K., {Becker}, S.~R., \& {Kudritzki}, R.~P. 2006,
  \aap, 445, 1099

\bibitem[{{Przybilla} {et~al.}(2011){Przybilla}, {Nieva}, \&
  {Butler}}]{przybilla2011}
{Przybilla}, N., {Nieva}, M.-F., \& {Butler}, K. 2011, in Journal of Physics
  Conference Series, Vol. 328, Journal of Physics Conference Series, 012015

\bibitem[{{Randall} {et~al.}(2007){Randall}, {Green}, {Van Grootel},
  {Fontaine}, {Charpinet}, {Lesser}, {Brassard}, {Sugimoto}, {Chayer}, {Fay},
  {Wroblewski}, {Daniel}, {Story}, \& {Fitzgerald}}]{2007A&A...476.1317R}
{Randall}, S.~K., {Green}, E.~M., {Van Grootel}, V., {et~al.} 2007, \aap, 476,
  1317

\bibitem[{{Ricker} {et~al.}(2014){Ricker}, {Winn}, {Vanderspek}, {Latham},
  {Bakos}, {Bean}, {Berta-Thompson}, {Brown}, {Buchhave}, {Butler}, {Butler},
  {Chaplin}, {Charbonneau}, {Christensen-Dalsgaard}, {Clampin}, {Deming},
  {Doty}, {De Lee}, {Dressing}, {Dunham}, {Endl}, {Fressin}, {Ge}, {Henning},
  {Holman}, {Howard}, {Ida}, {Jenkins}, {Jernigan}, {Johnson}, {Kaltenegger},
  {Kawai}, {Kjeldsen}, {Laughlin}, {Levine}, {Lin}, {Lissauer}, {MacQueen},
  {Marcy}, {McCullough}, {Morton}, {Narita}, {Paegert}, {Palle}, {Pepe},
  {Pepper}, {Quirrenbach}, {Rinehart}, {Sasselov}, {Sato}, {Seager},
  {Sozzetti}, {Stassun}, {Sullivan}, {Szentgyorgyi}, {Torres}, {Udry}, \&
  {Villasenor}}]{Tess2014}
{Ricker}, G.~R., {Winn}, J.~N., {Vanderspek}, R., {et~al.} 2014, in Society of
  Photo-Optical Instrumentation Engineers (SPIE) Conference Series, Vol. 9143,
  Space Telescopes and Instrumentation 2014: Optical, Infrared, and Millimeter
  Wave, ed. J.~M. {Oschmann}, Jr., M.~{Clampin}, G.~G. {Fazio}, \& H.~A.
  {MacEwen}, 914320

\bibitem[{{Riello} {et~al.}(2021){Riello}, {De Angeli}, {Evans}, {Montegriffo},
  {Carrasco}, {Busso}, {Palaversa}, {Burgess}, {Diener}, {Davidson}, {Rowell},
  {Fabricius}, {Jordi}, {Bellazzini}, {Pancino}, {Harrison}, {Cacciari}, {van
  Leeuwen}, {Hambly}, {Hodgkin}, {Osborne}, {Altavilla}, {Barstow}, {Brown},
  {Castellani}, {Cowell}, {De Luise}, {Gilmore}, {Giuffrida}, {Hidalgo},
  {Holland}, {Marinoni}, {Pagani}, {Piersimoni}, {Pulone}, {Ragaini}, {Rainer},
  {Richards}, {Sanna}, {Walton}, {Weiler}, \& {Yoldas}}]{gaia_edr3}
{Riello}, M., {De Angeli}, F., {Evans}, D.~W., {et~al.} 2021, \aap, 649, A3

\bibitem[{{Rodr{\'\i}guez-Segovia} \& {Ruiter}(2025)}]{2025MNRAS.539.3273R}
{Rodr{\'\i}guez-Segovia}, N. \& {Ruiter}, A.~J. 2025, \mnras, 539, 3273

\bibitem[{{Saffer} {et~al.}(1994){Saffer}, {Bergeron}, {Koester}, \&
  {Liebert}}]{1994ApJ...432..351S}
{Saffer}, R.~A., {Bergeron}, P., {Koester}, D., \& {Liebert}, J. 1994, \apj,
  432, 351

\bibitem[{{Saffer} {et~al.}(1998){Saffer}, {Livio}, \&
  {Yungelson}}]{1998ApJ...502..394S}
{Saffer}, R.~A., {Livio}, M., \& {Yungelson}, L.~R. 1998, \apj, 502, 394

\bibitem[{{Saio} \& {Jeffery}(2000)}]{2000MNRAS.313..671S}
{Saio}, H. \& {Jeffery}, C.~S. 2000, \mnras, 313, 671

\bibitem[{{Saio} \& {Jeffery}(2002)}]{2002MNRAS.333..121S}
{Saio}, H. \& {Jeffery}, C.~S. 2002, \mnras, 333, 121

\bibitem[{Salaris \& Cassisi(2005)}]{SalarisCassisi2005}
Salaris, M. \& Cassisi, S. 2005, Evolution of Stars and Stellar Populations
  (Chichester, UK: John Wiley \& Sons)

\bibitem[{{Sandage}(1962)}]{1962ApJ...135..333S}
{Sandage}, A. 1962, \apj, 135, 333

\bibitem[{{Schaffenroth} {et~al.}(2022){Schaffenroth}, {Pelisoli}, {Barlow},
  {Geier}, \& {Kupfer}}]{2022A&A...666A.182S}
{Schaffenroth}, V., {Pelisoli}, I., {Barlow}, B.~N., {Geier}, S., \& {Kupfer},
  T. 2022, \aap, 666, A182

\bibitem[{{Schneider}(2022)}]{schneider_thesis}
{Schneider}, D. 2022, PhD thesis, Friedrich-Alexander-Universität
  Erlangen-Nürnberg

\bibitem[{{Schneider} {et~al.}(2018){Schneider}, {Irrgang}, {Heber}, {Nieva},
  \& {Przybilla}}]{2018A&A...618A..86S}
{Schneider}, D., {Irrgang}, A., {Heber}, U., {Nieva}, M.~F., \& {Przybilla}, N.
  2018, \aap, 618, A86

\bibitem[{{Silvotti} {et~al.}(2002){Silvotti}, {{\O}stensen}, {Heber},
  {Solheim}, {Dreizler}, \& {Altmann}}]{2002A&A...383..239S}
{Silvotti}, R., {{\O}stensen}, R., {Heber}, U., {et~al.} 2002, \aap, 383, 239

\bibitem[{{Silvotti} {et~al.}(2020){Silvotti}, {Ostensen}, \&
  {Telting}}]{2020arXiv200204545S}
{Silvotti}, R., {Ostensen}, R.~H., \& {Telting}, J.~H. 2020, arXiv e-prints,
  arXiv:2002.04545

\bibitem[{{Skrutskie} {et~al.}(2006){Skrutskie}, {Cutri}, {Stiening},
  {Weinberg}, {Schneider}, {Carpenter}, {Beichman}, {Capps}, {Chester},
  {Elias}, {Huchra}, {Liebert}, {Lonsdale}, {Monet}, {Price}, {Seitzer},
  {Jarrett}, {Kirkpatrick}, {Gizis}, {Howard}, {Evans}, {Fowler}, {Fullmer},
  {Hurt}, {Light}, {Kopan}, {Marsh}, {McCallon}, {Tam}, {Van Dyk}, \&
  {Wheelock}}]{2mass}
{Skrutskie}, M.~F., {Cutri}, R.~M., {Stiening}, R., {et~al.} 2006, \aj, 131,
  1163

\bibitem[{{Snowdon} {et~al.}(2025){Snowdon}, {Jeffery}, {Schlagenhauf}, \&
  {Dorsch}}]{2025MNRAS.537.2079S}
{Snowdon}, E.~J., {Jeffery}, C.~S., {Schlagenhauf}, S., \& {Dorsch}, M. 2025,
  \mnras, 537, 2079

\bibitem[{{Stark} \& {Wade}(2003)}]{2003AJ....126.1455S}
{Stark}, M.~A. \& {Wade}, R.~A. 2003, \aj, 126, 1455

\bibitem[{{Sweigart}(1987)}]{1987ApJS...65...95S}
{Sweigart}, A.~V. 1987, \apjs, 65, 95

\bibitem[{{Taylor}(2005)}]{topcat}
{Taylor}, M.~B. 2005, in Astronomical Society of the Pacific Conference Series,
  Vol. 347, Astronomical Data Analysis Software and Systems XIV, ed.
  P.~{Shopbell}, M.~{Britton}, \& R.~{Ebert}, 29

\bibitem[{{Telting} {et~al.}(2014){Telting}, {Baran}, {Nemeth}, {{\O}stensen},
  {Kupfer}, {Macfarlane}, {Heber}, {Aerts}, \& {Geier}}]{2014A&A...570A.129T}
{Telting}, J.~H., {Baran}, A.~S., {Nemeth}, P., {et~al.} 2014, \aap, 570, A129

\bibitem[{{Telting} {et~al.}(2012){Telting}, {{\O}stensen}, {Baran}, {Bloemen},
  {Reed}, {Oreiro}, {Farris}, {Ottosen}, {Aerts}, {Kawaler}, {Heber}, {Prins},
  {Green}, {Kalomeni}, {O'Toole}, {Mullally}, {Sanderfer}, {Smith}, \&
  {Kjeldsen}}]{2012A&A...544A...1T}
{Telting}, J.~H., {{\O}stensen}, R.~H., {Baran}, A.~S., {et~al.} 2012, \aap,
  544, A1

\bibitem[{{Th{\'e}ado} {et~al.}(2009){Th{\'e}ado}, {Vauclair}, {Alecian}, \&
  {LeBlanc}}]{theado09}
{Th{\'e}ado}, S., {Vauclair}, S., {Alecian}, G., \& {LeBlanc}, F. 2009, \apj,
  704, 1262

\bibitem[{{Tody}(1986)}]{1986SPIE..627..733T}
{Tody}, D. 1986, in Society of Photo-Optical Instrumentation Engineers (SPIE)
  Conference Series, Vol. 627, Instrumentation in astronomy VI, ed. D.~L.
  {Crawford}, 733

\bibitem[{{Tody}(1993)}]{1993ASPC...52..173T}
{Tody}, D. 1993, in Astronomical Society of the Pacific Conference Series,
  Vol.~52, Astronomical Data Analysis Software and Systems II, ed. R.~J.
  {Hanisch}, R.~J.~V. {Brissenden}, \& J.~{Barnes}, 173

\bibitem[{{Tremblay} \& {Bergeron}(2009)}]{2009ApJ...696.1755T}
{Tremblay}, P.~E. \& {Bergeron}, P. 2009, \apj, 696, 1755

\bibitem[{{Uzundag} {et~al.}(2024){Uzundag}, {Krzesinski}, {Pelisoli},
  {Nemeth}, {Silvotti}, {Vuckovic}, {Dawson}, \& {Geier}}]{2024yCat..36840118U}
{Uzundag}, M., {Krzesinski}, J., {Pelisoli}, I., {et~al.} 2024, {VizieR Online
  Data Catalog: Pulsating hot subdwarf B stars (Uzundag+, 2024)}, VizieR
  On-line Data Catalog: J/A+A/684/A118. Originally published in:
  2024A\&A...684A.118U

\bibitem[{{Van Grootel}(2008)}]{2008PhDT........66V}
{Van Grootel}, V. 2008, PhD thesis, University of Montreal, Canada

\bibitem[{{Van Grootel} {et~al.}(2010){Van Grootel}, {Charpinet}, {Fontaine},
  \& {Brassard}}]{2010Ap&SS.329..217V}
{Van Grootel}, V., {Charpinet}, S., {Fontaine}, G., \& {Brassard}, P. 2010,
  \apss, 329, 217

\bibitem[{{Villase{\~n}or} {et~al.}(2023){Villase{\~n}or}, {Lennon}, {Picco},
  {Shenar}, {Marchant}, {Langer}, {Dufton}, {Nardini}, {Evans}, {Bodensteiner},
  {de Mink}, {G{\"o}tberg}, {Soszy{\'n}ski}, {Taylor}, \&
  {Sana}}]{2023MNRAS.525.5121V}
{Villase{\~n}or}, J.~I., {Lennon}, D.~J., {Picco}, A., {et~al.} 2023, \mnras,
  525, 5121

\bibitem[{{Vos} {et~al.}(2018){Vos}, {N{\'e}meth}, {Vu{\v{c}}kovi{\'c}},
  {{\O}stensen}, \& {Parsons}}]{Vos2018}
{Vos}, J., {N{\'e}meth}, P., {Vu{\v{c}}kovi{\'c}}, M., {{\O}stensen}, R., \&
  {Parsons}, S. 2018, \mnras, 473, 693

\bibitem[{{Vos} {et~al.}(2019){Vos}, {Vu{\v{c}}kovi{\'c}}, {Chen}, {Han},
  {Boudreaux}, {Barlow}, {{\O}stensen}, \& {N{\'e}meth}}]{2019MNRAS.482.4592V}
{Vos}, J., {Vu{\v{c}}kovi{\'c}}, M., {Chen}, X., {et~al.} 2019, \mnras, 482,
  4592

\bibitem[{{Vu{\v{c}}kovi{\'c}} {et~al.}(2007){Vu{\v{c}}kovi{\'c}}, {Aerts},
  {{\"O}stensen}, {Nelemans}, {Hu}, {Jeffery}, {Dhillon}, \&
  {Marsh}}]{2007A&A...471..605V}
{Vu{\v{c}}kovi{\'c}}, M., {Aerts}, C., {{\"O}stensen}, R., {et~al.} 2007, \aap,
  471, 605

\bibitem[{{Webbink}(1984)}]{1984ApJ...277..355W}
{Webbink}, R.~F. 1984, \apj, 277, 355

\bibitem[{{Wenger} {et~al.}(2000){Wenger}, {Ochsenbein}, {Egret}, {Dubois},
  {Bonnarel}, {Borde}, {Genova}, {Jasniewicz}, {Lalo{\"e}}, {Lesteven}, \&
  {Monier}}]{Simbad}
{Wenger}, M., {Ochsenbein}, F., {Egret}, D., {et~al.} 2000, \aaps, 143, 9

\bibitem[{{Werner}(1996)}]{Werner1996}
{Werner}, K. 1996, \apjl, 457, L39

\bibitem[{{Werner} {et~al.}(2022){Werner}, {Reindl}, {Geier}, \&
  {Pritzkuleit}}]{werner2022}
{Werner}, K., {Reindl}, N., {Geier}, S., \& {Pritzkuleit}, M. 2022, \mnras,
  511, L66

\bibitem[{{Williams} {et~al.}(2001){Williams}, {McGraw}, \&
  {Grashuis}}]{2001PASP..113..490W}
{Williams}, T., {McGraw}, J.~T., \& {Grashuis}, R. 2001, \pasp, 113, 490

\bibitem[{{Xiong} {et~al.}(2017){Xiong}, {Chen}, {Podsiadlowski}, {Li}, \&
  {Han}}]{2017A&A...599A..54X}
{Xiong}, H., {Chen}, X., {Podsiadlowski}, P., {Li}, Y., \& {Han}, Z. 2017,
  \aap, 599, A54

\bibitem[{{Zhang} \& {Jeffery}(2012)}]{2012MNRAS.419..452Z}
{Zhang}, X. \& {Jeffery}, C.~S. 2012, \mnras, 419, 452

\bibitem[{{Zong} {et~al.}(2016){Zong}, {Charpinet}, {Vauclair}, {Giammichele},
  \& {Van Grootel}}]{2016A&A...585A..22Z}
{Zong}, W., {Charpinet}, S., {Vauclair}, G., {Giammichele}, N., \& {Van
  Grootel}, V. 2016, \aap, 585, A22

\end{thebibliography}


% \clearpage
% \newpage
\begin{appendix} 

\section{Stars with composite SED}\label{App:composites}

\begin{figure}
\resizebox{\hsize}{!}{
   \includegraphics{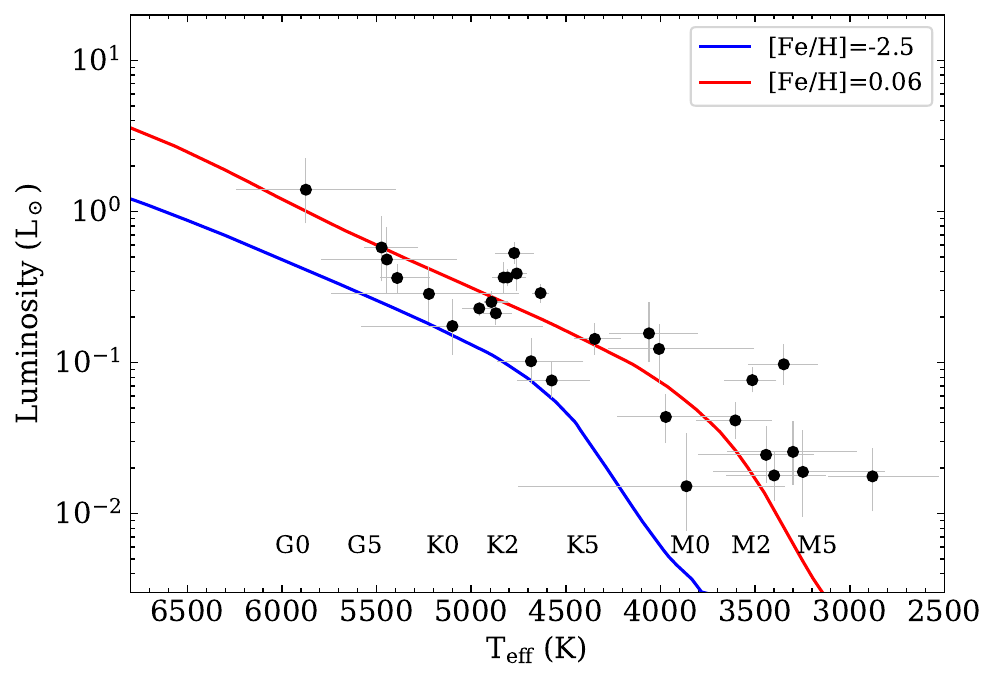}}
     \caption{HR diagram for the stellar companions of the hot subdwarfs with composite spectra. The parameters are derived from the fits of SED. The solid lines show the expected parameters for MS stars at [Fe/H] = 0.06 (red) and $-$2.5 (blue) as obtained from BaSTI isochrones.}
     \label{fig:app:HR_comp}
\end{figure}

Twenty nine stars in our sample were found to have IR excess that could be reproduced by a MS companion. The stars are listed in Table~\ref{table_composites}, where the atmospheric parameters of the hot subdwarfs are indicated, as well as the \teff, radius $R$ and luminosity $L$ of the companion according to the best fit of the SED. Some of these systems are long period binaries for which the $Gaia$ single-star astrometric solution is not necessarily accurate (when \texttt{ruwe} $\gtrsim$ 1.4), thus we also indicate the value of this parameter in the Table.
In Fig.~\ref{fig:app:HR_comp}, we show the \teff\ and luminosity of the MS companions as well as the position of the zero-age MS track for two different metallicities.
Not all of the stars for which we detected IR-excess were already known to be binaries, and for some of the known binaries, it was not previously clear from the RVs whether the companion was a WD or a low-mass MS star. We briefly discuss a few cases that are of particular interest or that have not been previously reported as composite systems.

\object{HD149382} is the brightest known hot subdwarf. It has been claimed to have a substellar companion on the basis of radial velocity measurements, 
%(Geier et al. 2009), 
however no RV variations were detected in follow-up studies \citep{2011ApJ...743...88N,2020arXiv200204545S}. The star has a visual companion 1$\arcsec$ (75 AU) away that is probably the source of the IR-excess \citep{2005ASPC..334..435O}, and likely constitute a background object \citep{schneider_thesis}.

\object{PB7032} is a slow pulsator \citep{2011MNRAS.415.3042K} that has not been reported to be part of a binary system previously. Our SED fit indicates the presence of a M-dwarf companion (\teff\ $\sim$3400~K). The quality of the TESS data for this star is poor and we find no hint of low-frequency variability in the Lomb-Scargle periodogram (LSP) that could be associated with a reflection effect.

\object{PG0014+068} is a rapid pulsator that has been extensively observed with time-series photometry and well-studied from an asteroseismic point of view \citep{2001ApJ...563.1013B,2005ASPC..334..619C}. It has always been considered a single star and the RVs of the MMT spectra do not show large variations ($\sigma$=2.2 kms$^{-1}$ for 4 measurements taken within 12 days). However, the SED of the star shows a strong IR excess that can be reproduced with a K-type companion (\teff\ $\sim$4700 K). 
The TESS light curve of PG0014+068 does not show any sign of a reflection effect. 
The system possibly has a period of the order of tens of days, or longer.

\object{PG0250+189} shows an IR-excess that corresponds to an early M or late K star (\teff\ $\sim$3800~K). This star is among the few objects located below the ZAEHB (see Fig.~\ref{fig:below_ehb}, but also \citealt{1994ApJ...432..351S}). Unfortunately, neither a TESS light curve nor RV measurements were found in the literature.

\object{PG0940+068} is a slowly pulsating sdB and a known binary with a period of 8.33 d \citep{2000MNRAS.311..877M}, however it was initially not clear whether the companion is a WD or dM star. Based on the absence of variations in the TESS light curve, \citet{2022A&A...666A.182S} classified the companion as a WD. The star's SED shows a mild IR-excess that can be reproduced with a late M-type star (\teff\ $\sim$3250~K).

\object{PG1340+607} is a slowly pulsating sdB that has not previously been reported to be part of a binary system. However, the radial velocities measured from the two MMT spectra showed a difference of 10 km s$^{-1}$, thus suggesting a possible binary nature.
The SED of the star shows a mild IR-excess that can be reproduced with a late M-type star (\teff\ $\sim$3300~K). The LSP of the TESS light curve shows a peak (SNR=8.7) at 3.8h that is separated from the $g$-modes found at shorter periods. This peak could be due to a reflection effect, although a long-period $g$-mode can not be ruled out. Additional RV measurements are needed to establish the orbital period.

\object{PG2151+100} has been reported as an RV variable star with a faint MS companion but unknown period \citep{2005ASPC..334..363G,2003PhDT........48E}. Our SED fit suggests the presence of a late M-dwarf star (\teff\ $\sim$3300~K).
In addition, the LSP of the TESS data shows a strong signal at 40.06h in an otherwise featureless light curve. The TESS signal is most likely due to the stellar spots on the companion's surface.
%\ml{I think it's worth showing the folded LC at the 40.06h period}.  
%\ml{40 hours is larger than the longest reflection effect reported in \citet{2022A&A...666A.182S} (which is 33hr, but we don't know anything about the parameters of the star (GAIA DR2 5709912046530451840). Actually, I don't know why this star wasn't picked up by Veronika, no LC in 2022 ??}

\object{PG1101+249} (Feige\,36) is a known binary with a period of 8.5 h \citep{1998ApJ...502..394S} and a companion that was believed to be a WD. However, the SED shows a mild IR-excess that we reproduce with a stellar companion at \teff\ $\sim$3300K. The TESS light curve does not show any variability at the orbital period.

\object{PG2317+046} (PB5333) is a known binary for which \citet{2004Ap&SS.291..315E} reported a period of 22.2 h. The SED shows a strong IR excess that we reproduce with an early-M type star (\teff\ $\sim$3500~K). The periodogram of the TESS data shows a clear peak at 19.87 h and the presence of the first harmonic. However recent investigations by Schaffenroth et al. (priv. comm. 2024) found the system to have a much longer period ($\sim$92 days) and a somewhat hotter companion (K8V).

\object{PG1154-070} is not a known binary, and did not show large RV variations in the study of \citet{1998ApJ...502..394S}. Its IR-excess indicates a late K-type companion (\teff\ $\sim$4000~K).

\object{HS2151+0857} (a $p$-mode pulsator according to \citealt{2001A&A...378..466O}), \object{HS1824+5745} and \object{PG2303+019} (also known as HS2303+0152, and a $p$-mode pulsator according to \citealt{2002A&A...383..239S}) have  
IR-excesses that suggest the presence of late K-type companions. The three stars are also part of the sample analyzed by Heber et al. (in prep.) who found similar properties for the companions.

\object{PG0823+546} and \object{PG2158+082} are two hot He-sdO composites with a visible IR-excess that corresponds to early K or late G companions (see also \citealt{2001PASP..113..490W}). These are interesting objects given that the fraction of composite systems among He-sdOs is relatively small ($\sim$9~\%, \citealt{2024PhDT........36D}) compared to that of sdBs ($\sim$30~\%, \citealt{2003AJ....126.1455S}).

\begin{table*}[h]
\small
\caption{Properties of the hot subdwarf systems with IR-excess}
\label{table_composites}      
\centering                    
\begin{tabular}{l c c c c c c c}        
\toprule\toprule
\noalign{\vskip4bp}
Star & \teff\ & log~$g$ & \heh\ & \teff$_{\rm ,~comp}$ & R$_{\rm comp}$ & log(L$_{\rm comp}$) & $Gaia$ \texttt{ruwe} \\ 
 & (K) & (cm s${-2}$) &  & (K) & (R$_{\odot}$) & (L$_{\odot}$) & \\
 \noalign{\vskip3bp}
 \midrule
HD149382 & 35519 & 5.80 & $-1.48$ & $2879${\raisebox{0.5ex}{\tiny$^{+233}_{-350}$}} & $0.54${\raisebox{0.5ex}{\tiny$^{+0.08}_{-0.05}$}} & $-1.75${\raisebox{0.5ex}{\tiny$^{+0.19}_{-0.23}$}} & 1.20 \\ 
PG0940+068 & 26953 & 5.41 & $-2.79$ & $3247${\raisebox{0.5ex}{\tiny$^{+474}_{-433}$}} & $0.44${\raisebox{0.5ex}{\tiny$^{+0.08}_{-0.07}$}} & $-1.72${\raisebox{0.5ex}{\tiny$^{+0.28}_{-0.30}$}} & 1.02 \\ 
PG1340+607 & 25606 & 5.30 & $-2.62$ & $3299${\raisebox{0.5ex}{\tiny$^{+347}_{-336}$}} & $0.49${\raisebox{0.5ex}{\tiny$^{+0.06}_{-0.06}$}} & $-1.59${\raisebox{0.5ex}{\tiny$^{+0.20}_{-0.22}$}} & 1.07 \\ 
PG2151+100 & 33946 & 5.61 & $-3.66$ & $3348${\raisebox{0.5ex}{\tiny$^{+186}_{-178}$}} & $0.93${\raisebox{0.5ex}{\tiny$^{+0.11}_{-0.09}$}} & $-1.01${\raisebox{0.5ex}{\tiny$^{+0.13}_{-0.13}$}} & 2.17 \\ 
PG1101+249 & 28841 & 5.67 & $-2.01$ & $3399${\raisebox{0.5ex}{\tiny$^{+255}_{-274}$}} & $0.39${\raisebox{0.5ex}{\tiny$^{+0.04}_{-0.03}$}} & $-1.75${\raisebox{0.5ex}{\tiny$^{+0.15}_{-0.17}$}} & 0.95 \\ 
PB7032 & 27122 & 5.49 & $-2.82$ & $3442${\raisebox{0.5ex}{\tiny$^{+362}_{-254}$}} & $0.44${\raisebox{0.5ex}{\tiny$^{+0.05}_{-0.06}$}} & $-1.61${\raisebox{0.5ex}{\tiny$^{+0.19}_{-0.19}$}} & 0.94 \\ 
PG2317+046 & 44970 & 6.05 & $-2.82$ & $3515${\raisebox{0.5ex}{\tiny$^{+149}_{-127}$}} & $0.74${\raisebox{0.5ex}{\tiny$^{+0.04}_{-0.04}$}} & $-1.12${\raisebox{0.5ex}{\tiny$^{+0.08}_{-0.08}$}} & 0.93 \\ 
FEIGE34 & 61095 & 5.92 & $-1.89$ & $3601${\raisebox{0.5ex}{\tiny$^{+204}_{-192}$}} & $0.52${\raisebox{0.5ex}{\tiny$^{+0.05}_{-0.04}$}} & $-1.38${\raisebox{0.5ex}{\tiny$^{+0.12}_{-0.12}$}} & 2.24 \\ 
PG0250+189 & 25562 & 5.75 & $-3.84$ & $3858${\raisebox{0.5ex}{\tiny$^{+933}_{-511}$}} & $0.27${\raisebox{0.5ex}{\tiny$^{+0.04}_{-0.03}$}} & $-1.82${\raisebox{0.5ex}{\tiny$^{+0.37}_{-0.29}$}} & 1.10 \\ 
PG1154-070 & 27637 & 5.53 & $-2.40$ & $3971${\raisebox{0.5ex}{\tiny$^{+257}_{-343}$}} & $0.45${\raisebox{0.5ex}{\tiny$^{+0.05}_{-0.04}$}} & $-1.36${\raisebox{0.5ex}{\tiny$^{+0.15}_{-0.17}$}} & 3.11 \\ 
PG1618+563 & 35405 & 5.83 & $-1.69$ & $4006${\raisebox{0.5ex}{\tiny$^{+268}_{-501}$}} & $0.75${\raisebox{0.5ex}{\tiny$^{+0.08}_{-0.08}$}} & $-0.91${\raisebox{0.5ex}{\tiny$^{+0.17}_{-0.23}$}} & 1.98 \\ 
HS2151+0857 & 35883 & 5.80 & $-1.42$ & $4060${\raisebox{0.5ex}{\tiny$^{+211}_{-258}$}} & $0.80${\raisebox{0.5ex}{\tiny$^{+0.19}_{-0.13}$}} & $-0.81${\raisebox{0.5ex}{\tiny$^{+0.21}_{-0.19}$}} & 0.99 \\ 
HS1824+5745 & 35248 & 5.84 & $-1.61$ & $4345${\raisebox{0.5ex}{\tiny$^{+129}_{-154}$}} & $0.67${\raisebox{0.5ex}{\tiny$^{+0.08}_{-0.07}$}} & $-0.85${\raisebox{0.5ex}{\tiny$^{+0.11}_{-0.11}$}} & 1.20 \\ 
PG2303+019 & 36839 & 5.77 & $-1.71$ & $4574${\raisebox{0.5ex}{\tiny$^{+181}_{-199}$}} & $0.44${\raisebox{0.5ex}{\tiny$^{+0.05}_{-0.04}$}} & $-1.12${\raisebox{0.5ex}{\tiny$^{+0.12}_{-0.12}$}} & 1.03 \\ 
PG1647+253 & 36073 & 5.84 & $-2.08$ & $4633${\raisebox{0.5ex}{\tiny$^{+44}_{-46}$}} & $0.83${\raisebox{0.5ex}{\tiny$^{+0.06}_{-0.06}$}} & $-0.54${\raisebox{0.5ex}{\tiny$^{+0.06}_{-0.06}$}} & 1.42 \\ 
PG0014+068 & 35638 & 5.93 & $-1.62$ & $4684${\raisebox{0.5ex}{\tiny$^{+181}_{-277}$}} & $0.49${\raisebox{0.5ex}{\tiny$^{+0.08}_{-0.06}$}} & $-0.99${\raisebox{0.5ex}{\tiny$^{+0.15}_{-0.15}$}} & 1.15 \\ 
PHL1079 & 32910 & 5.60 & $-2.19$ & $4759${\raisebox{0.5ex}{\tiny$^{+48}_{-53}$}} & $0.92${\raisebox{0.5ex}{\tiny$^{+0.12}_{-0.11}$}} & $-0.41${\raisebox{0.5ex}{\tiny$^{+0.11}_{-0.11}$}} & 4.67 \\ 
PG1610+519 & 43897 & 5.54 & $-3.14$ & $4773${\raisebox{0.5ex}{\tiny$^{+98}_{-103}$}} & $1.07${\raisebox{0.5ex}{\tiny$^{+0.08}_{-0.07}$}} & $-0.28${\raisebox{0.5ex}{\tiny$^{+0.07}_{-0.07}$}} & 1.98 \\ 
PG1206+165 & 28833 & 5.58 & $-2.44$ & $4808${\raisebox{0.5ex}{\tiny$^{+76}_{-75}$}} & $0.87${\raisebox{0.5ex}{\tiny$^{+0.05}_{-0.05}$}} & $-0.44${\raisebox{0.5ex}{\tiny$^{+0.06}_{-0.06}$}} & 2.02 \\ 
PG1018-047 & 31448 & 5.60 & $-4.00$ & $4828${\raisebox{0.5ex}{\tiny$^{+61}_{-118}$}} & $0.87${\raisebox{0.5ex}{\tiny$^{+0.11}_{-0.09}$}} & $-0.44${\raisebox{0.5ex}{\tiny$^{+0.10}_{-0.10}$}} & 3.48 \\ 
TON357 & 65722 & 5.94 & $-1.85$ & $4870${\raisebox{0.5ex}{\tiny$^{+66}_{-86}$}} & $0.65${\raisebox{0.5ex}{\tiny$^{+0.06}_{-0.05}$}} & $-0.67${\raisebox{0.5ex}{\tiny$^{+0.08}_{-0.07}$}} & 1.51 \\ 
PG0749+658 & 25958 & 5.57 & $-3.31$ & $4893${\raisebox{0.5ex}{\tiny$^{+40}_{-85}$}} & $0.70${\raisebox{0.5ex}{\tiny$^{+0.05}_{-0.05}$}} & $-0.60${\raisebox{0.5ex}{\tiny$^{+0.07}_{-0.07}$}} & 4.49 \\ 
PG0934+553 & 45326 & 5.79 & $-0.50$ & $4957${\raisebox{0.5ex}{\tiny$^{+89}_{-107}$}} & $0.65${\raisebox{0.5ex}{\tiny$^{+0.02}_{-0.02}$}} & $-0.64${\raisebox{0.5ex}{\tiny$^{+0.05}_{-0.05}$}} & 1.15 \\ 
PG0823+546 & 75000 & 5.76 & $0.14$ & $5099${\raisebox{0.5ex}{\tiny$^{+482}_{-477}$}} & $0.53${\raisebox{0.5ex}{\tiny$^{+0.06}_{-0.05}$}} & $-0.76${\raisebox{0.5ex}{\tiny$^{+0.18}_{-0.19}$}} & 1.09 \\ 
PG2158+082 & 62500 & 5.86 & $1.75$ & $5224${\raisebox{0.5ex}{\tiny$^{+516}_{-473}$}} & $0.65${\raisebox{0.5ex}{\tiny$^{+0.06}_{-0.05}$}} & $-0.55${\raisebox{0.5ex}{\tiny$^{+0.18}_{-0.18}$}} & 0.94 \\ 
PG1701+359 & 33152 & 5.55 & $-4.00$ & $5391${\raisebox{0.5ex}{\tiny$^{+92}_{-172}$}} & $0.70${\raisebox{0.5ex}{\tiny$^{+0.07}_{-0.06}$}} & $-0.44${\raisebox{0.5ex}{\tiny$^{+0.09}_{-0.10}$}} & 3.51 \\ 
HS0252+1025 & 45138 & 5.25 & $-2.65$ & $5445${\raisebox{0.5ex}{\tiny$^{+345}_{-370}$}} & $0.78${\raisebox{0.5ex}{\tiny$^{+0.19}_{-0.15}$}} & $-0.32${\raisebox{0.5ex}{\tiny$^{+0.22}_{-0.22}$}} & 3.75 \\ 
PG0154+182 & 36755 & 5.69 & $-1.68$ & $5473${\raisebox{0.5ex}{\tiny$^{+92}_{-193}$}} & $0.86${\raisebox{0.5ex}{\tiny$^{+0.23}_{-0.19}$}} & $-0.24${\raisebox{0.5ex}{\tiny$^{+0.21}_{-0.22}$}} & 1.02 \\ 
HS0127+3146 & 39543 & 5.19 & $-3.46$ & $5871${\raisebox{0.5ex}{\tiny$^{+370}_{-473}$}} & $1.16${\raisebox{0.5ex}{\tiny$^{+0.26}_{-0.21}$}} & $0.14${\raisebox{0.5ex}{\tiny$^{+0.21}_{-0.22}$}} & 1.04 \\ 
\bottomrule
\end{tabular}
\tablefoot{The atmospheric parameters of the hot subdwarfs (\teff, \logg, and \heh\ are obtained from the fit of the Bok spectra. The \teff, $R$, and $L$ of the companions are derived from the SED fits and parallaxes. }
\end{table*}

\section{Stars removed from the Bok sample}\label{app:removed}

\begin{figure*}[hb!]
\centering
   \includegraphics[width=0.98\textwidth]{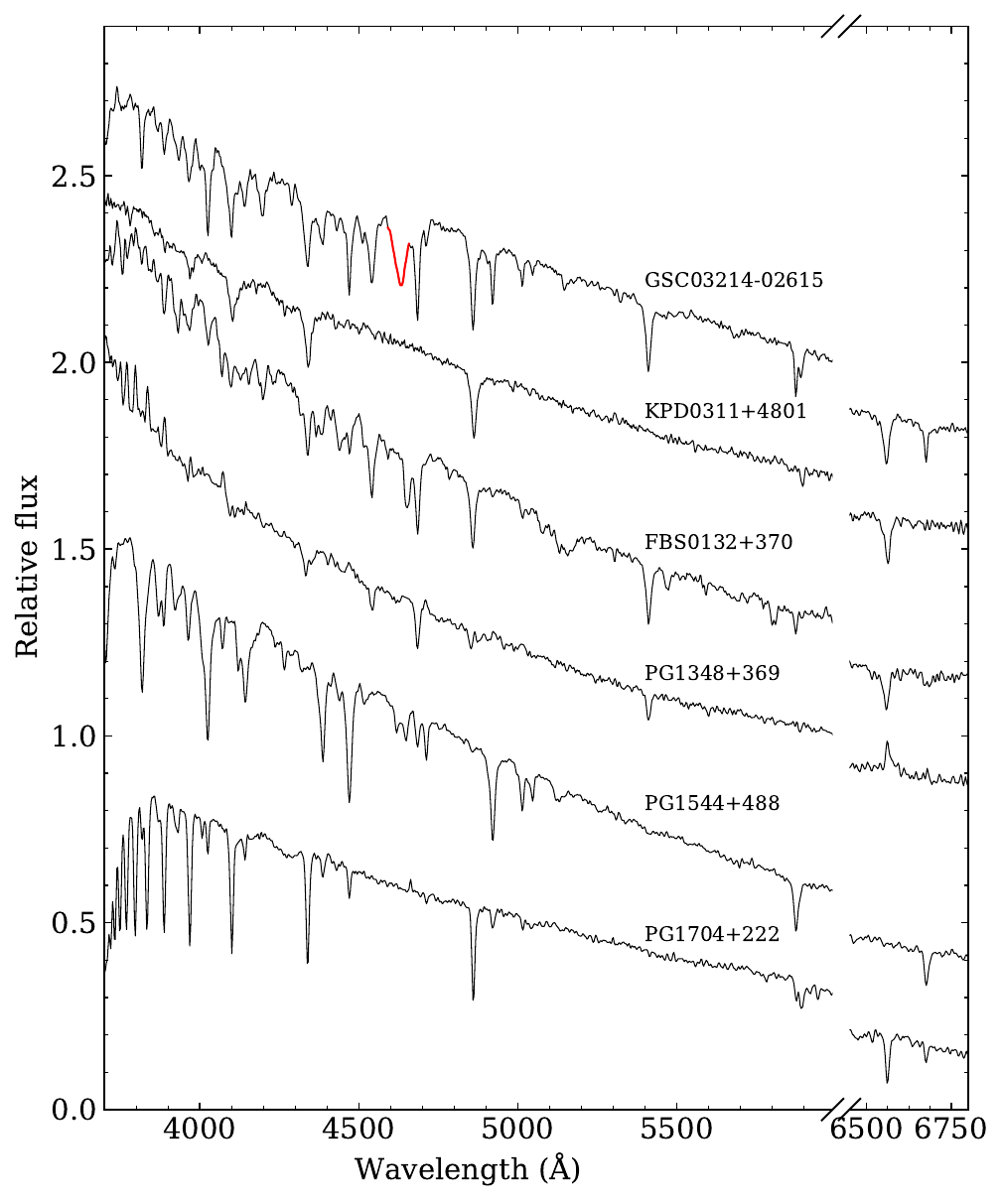}
     \caption{Bok spectra of six particular stars. The candidate magnetic star GSC03214-02615 is shown on top, with the distinctive absorption feature around 4600 \AA\ highlighted in red. The five other stars were discarded from the analysis as explained in Sect.~\ref{sec:res:bok}.}
     \label{fig:app:peculiar}
\end{figure*}

We list here the stars removed from the Bok sample, but whose spectra are still available on Vizier and shown in Fig.\ \ref{fig:app:peculiar}. 

\begin{itemize}

\item \object{PG1704+222} is a low-mass ($\sim$0.55 \msun) post-AGB giant star \citep{1998A&A...335..985M}. 
The fit resulted in \teff = 17.5~kK, log~$g$ = 2.8 and a helium abundance close to solar, in agreement with the estimates of \citet{1998A&A...335..985M}. This places the star at the lower limit of the model grid in terms of surface gravity. 

\item \object{PG1544+488} is a sdB binary composed of two similar He-rich sdB stars. This is, up to now, the only known binary system comprising two hot subdwarf stars \citep{Ahmad2004,Sener2014}. The two components cannot be disentangled in our Bok spectra. 

\item \object{KPD0311+4801} (WD0311+480) is a very hot DA white dwarf (see e.g. \citealt{Filiz2024}).

\item \object{FBS0132+370} could not be well reproduced by our model atmospheres. We estimate a relatively large \teff\ of 54 kK, a log~$g$ = 6 and some helium enrichment. The spectrum shows a few relatively strong absorption features that correspond to \ion{C}{iv} lines. It is an object similar to those recently identified by \citet{werner2022} as hot subdwarf stars with atmospheres strongly enhanced in carbon and oxygen, now identified as CO-sdO. 

\item \object{PG1348+369} shows emission lines in its spectrum. The high Balmer lines (around 3800 \AA) and H$_\alpha$ are in emission. A LAMOST spectrum was analyzed by \citet{Lei2023_mass} and the authors classified it as a hot sdO star (\teff\ = 65 kK). They did not report anything peculiar about the object.
Our SED fit of the object suggests the presence of an infrared excess, at least when fitting the SED with a stellar \teff\ of 70 kK. The star is definitely hot as indicated by the strong \ion{He}{ii} 5412 \AA\ line. The TESS light curve shows a conspicuous periodic signal at 3.316 days along with the first harmonic. This is consistent with the interpretation of \citet{2022ApJ...928...20B} that this is a reflection effect system in which the emission lines come from the heated side of the dM companion.

\end{itemize}

\clearpage

\section{PG0215+183}\label{app:pg0215}

\begin{figure}
\resizebox{\hsize}{!}{
   \includegraphics{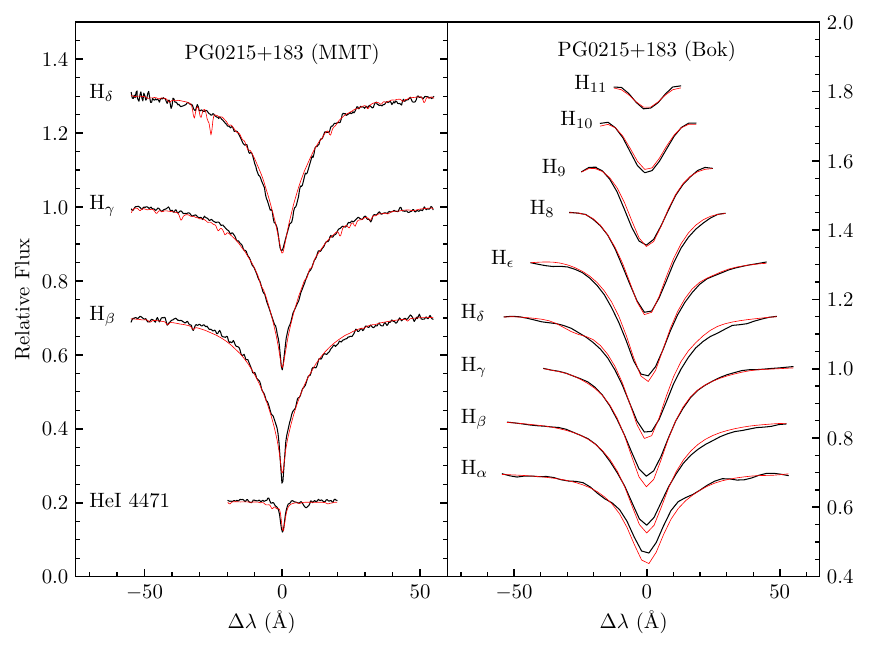}\vspace{1pt}
   }
   \resizebox{\hsize}{!}{   
    \includegraphics{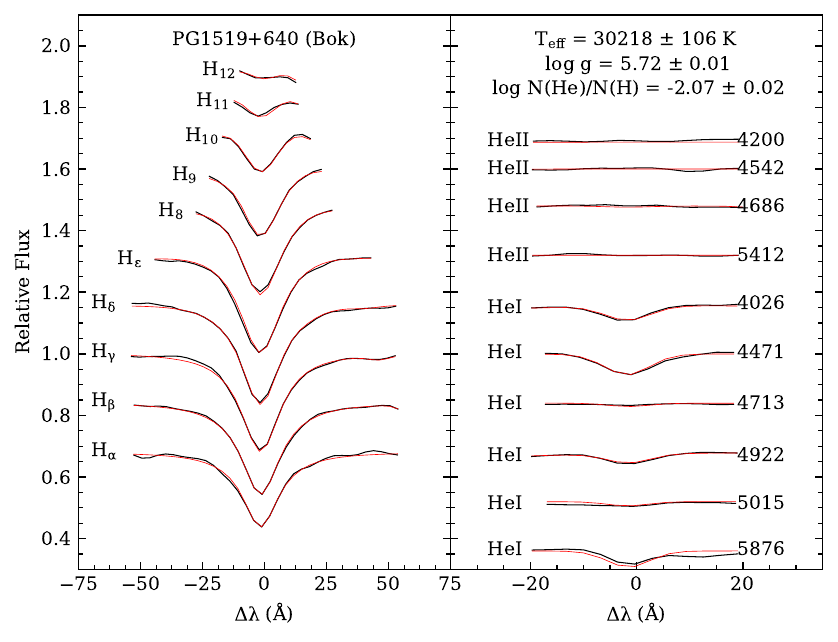}}
     \caption{Top: respective best fits to the MMT (31.2 kK, \logg\ = 6.18, \heh\ = $-$2.8) and Bok (28.9 kK, \logg\ = 5.85, \heh\ = $-$3) spectra of PG0215+183. Bottom: Best fit to the Bok spectrum of PG1519+640, a star with similar atmospheric parameters as PG0215+183 and showing hints of He stratification in its MMT spectrum.
     }
     \label{fig:PG0215}
\end{figure}

Among the stars in the MMT sample, PG0215+183 stands out because of its high surface gravity obtained from the MMT spectrum. With a log~$g$ of 6.2 and a \teff\ of 31 kK, the star lies well below the ZAEHB. Its He abundance is low and only \ion{He}{i} 4471~\AA\ is visible in the MMT spectrum. At first sight, the fit is good, although a thorough inspection shows that the cores of H$_\beta$ and H$_\gamma$ are slightly narrower than the model prediction and H$_\delta$ is slightly wider than the model (see Fig.~\ref{fig:PG0215}). The atmospheric parameters obtained from the Bok spectrum are significantly different with \teff\ = 29 kK and log~$g$ = 5.8. This is the star with a large log~$g$ difference of 0.4 dex in Fig.~\ref{fig:MMT_comp}. Although the atmospheric parameters from the Bok spectrum places the star close to the ZAEHB, and the resulting mass of 0.51 \msun\ derived from the parallax and SED fit is normal, the atmospheric fit is obviously poor, as seen in Fig.~\ref{fig:PG0215}. We verified that the difference in atmospheric parameters obtained from the two spectra comes from the wavelength range used: fitting the Bok spectrum over the 4000-4950 \AA\ range leads to the same atmospheric parameters as obtained with the MMT spectrum. Currently we do not know why the spectrum of that star cannot be properly reproduced. There are no RV variations from the eight individual MMT spectra ($\sigma_{rv}$ = 2.5 km~s$^{-1}$), the star is not pulsating and has no IR excess. The spectral fit of spectra taken with a different instrument resulted in the same mismatch between the observed and modeled Balmer lines (H.\ Dawson, priv.\ comm.\ 2025). The effective temperature of the star is in the range where \het\ and stratification are observed, but the Balmer lines in the Bok spectra of the stratified stars discussed in Sect.~\ref{sec:res:mmt:strat} are properly reproduced (see also Fig.~\ref{fig:PG0215}). 

\clearpage

\onecolumn

\section{Table of results}

\scriptsize
\begin{longtable}{l c c c c c c c c c c c} 
\caption{\label{table_res_all} Excerpt from the online table of results for all stars in the Bok sample}\\
\hline\hline
%\begin{tabular}{l c c c c c c c c c c c}        
%\toprule\toprule
%\noalign{\vskip4bp}
Star & Type & \teff\ & log~$g$ & \heh\  & $R$ & $L$ & $M$ & IR-excess & $Gaia$ \texttt{ruwe} & Pulsation & Binarity \\ 
 & & (K) & (cm s$^{-2}$) &  & ($R_\odot$) & ($L_\odot$) & ($M_\odot$) & &  &  &  \\
\hline
\endfirsthead
\caption{continued.}\\
\hline\hline
Star & Type & \teff\ & log~$g$ & \heh\  & $R$ & $L$ & $M$ & IR-excess & $Gaia$ \texttt{ruwe} & Pulsation & Binarity \\ 
 & & (K) & (cm s$^{-2}$) &  & ($R_\odot$) & ($L_\odot$) & ($M_\odot$) & &  &  &  \\ 
\hline
\endhead
\hline
\endfoot
AGK2+81$^\circ$266 & sdO & $59549${\raisebox{0.5ex}{\tiny$^{+424}_{-405}$}} & $6.03${\raisebox{0.5ex}{\tiny$^{+0.02}_{-0.02}$}} &$-3.07${\raisebox{0.5ex}{\tiny$^{+0.05}_{-0.06}$}} & $0.096${\raisebox{0.5ex}{\tiny$^{+0.003}_{-0.003}$}} &$104.7${\raisebox{0.5ex}{\tiny$^{+28.8}_{-23.7}$}} & $0.36${\raisebox{0.5ex}{\tiny$^{+0.07}_{-0.06}$}} & no & 1.08  & no & single \\ 
BD+25$^\circ$4655 & He-sdO & $38927${\raisebox{0.5ex}{\tiny$^{+115}_{-133}$}} & $5.68${\raisebox{0.5ex}{\tiny$^{+0.03}_{-0.04}$}} &$1.61${\raisebox{0.5ex}{\tiny$^{+0.04}_{-0.04}$}} & $0.141${\raisebox{0.5ex}{\tiny$^{+0.002}_{-0.002}$}} &$41.0${\raisebox{0.5ex}{\tiny$^{+3.1}_{-2.9}$}} & $0.34${\raisebox{0.5ex}{\tiny$^{+0.07}_{-0.06}$}} & no & 1.22  & no & ... \\ 
BD+28$^\circ$4211 & sdO & $70999${\raisebox{0.5ex}{\tiny$^{+517}_{-603}$}} & $6.57${\raisebox{0.5ex}{\tiny$^{+0.02}_{-0.02}$}} &$-1.28${\raisebox{0.5ex}{\tiny$^{+0.02}_{-0.02}$}} & $0.065${\raisebox{0.5ex}{\tiny$^{+0.003}_{-0.001}$}} &$99.3${\raisebox{0.5ex}{\tiny$^{+27.6}_{-22.5}$}} & $0.59${\raisebox{0.5ex}{\tiny$^{+0.12}_{-0.10}$}} & no & 1.17  & no & single \\ 
BD+37$^\circ$1977 & He-sdO & $50165${\raisebox{0.5ex}{\tiny$^{+446}_{-405}$}} & $4.25${\raisebox{0.5ex}{\tiny$^{+0.01}_{-0.00}$}} &$1.75${\raisebox{0.5ex}{\tiny$^{+0.05}_{-0.11}$}} & $1.350${\raisebox{0.5ex}{\tiny$^{+0.181}_{-0.144}$}} &$10420.4${\raisebox{0.5ex}{\tiny$^{+3640.1}_{-2575.6}$}} & $1.19${\raisebox{0.5ex}{\tiny$^{+0.42}_{-0.30}$}} & no & 1.06  & ... & ... \\ 
BD+37$^\circ$442 & He-sdO & $50284${\raisebox{0.5ex}{\tiny$^{+408}_{-457}$}} & $4.25${\raisebox{0.5ex}{\tiny$^{+0.01}_{-0.00}$}} &$1.75${\raisebox{0.5ex}{\tiny$^{+0.04}_{-0.04}$}} & $1.063${\raisebox{0.5ex}{\tiny$^{+0.088}_{-0.076}$}} &$6516.8${\raisebox{0.5ex}{\tiny$^{+1673.2}_{-1328.0}$}} & $0.74${\raisebox{0.5ex}{\tiny$^{+0.20}_{-0.15}$}} & no & 1.14  & ... & ... \\
\hline
\end{longtable}
\tablefoot{
Uncertainties on the atmospheric parameters (\teff, \logg, and, \heh) are only statisticals.
}
%\end{tabular}

%\end{appendix}

%\end{document}

\clearpage

\section{Additional Figures}\label{sec:appD}

\begin{figure}[h!]
\centering
   \includegraphics[width=0.91\textwidth]{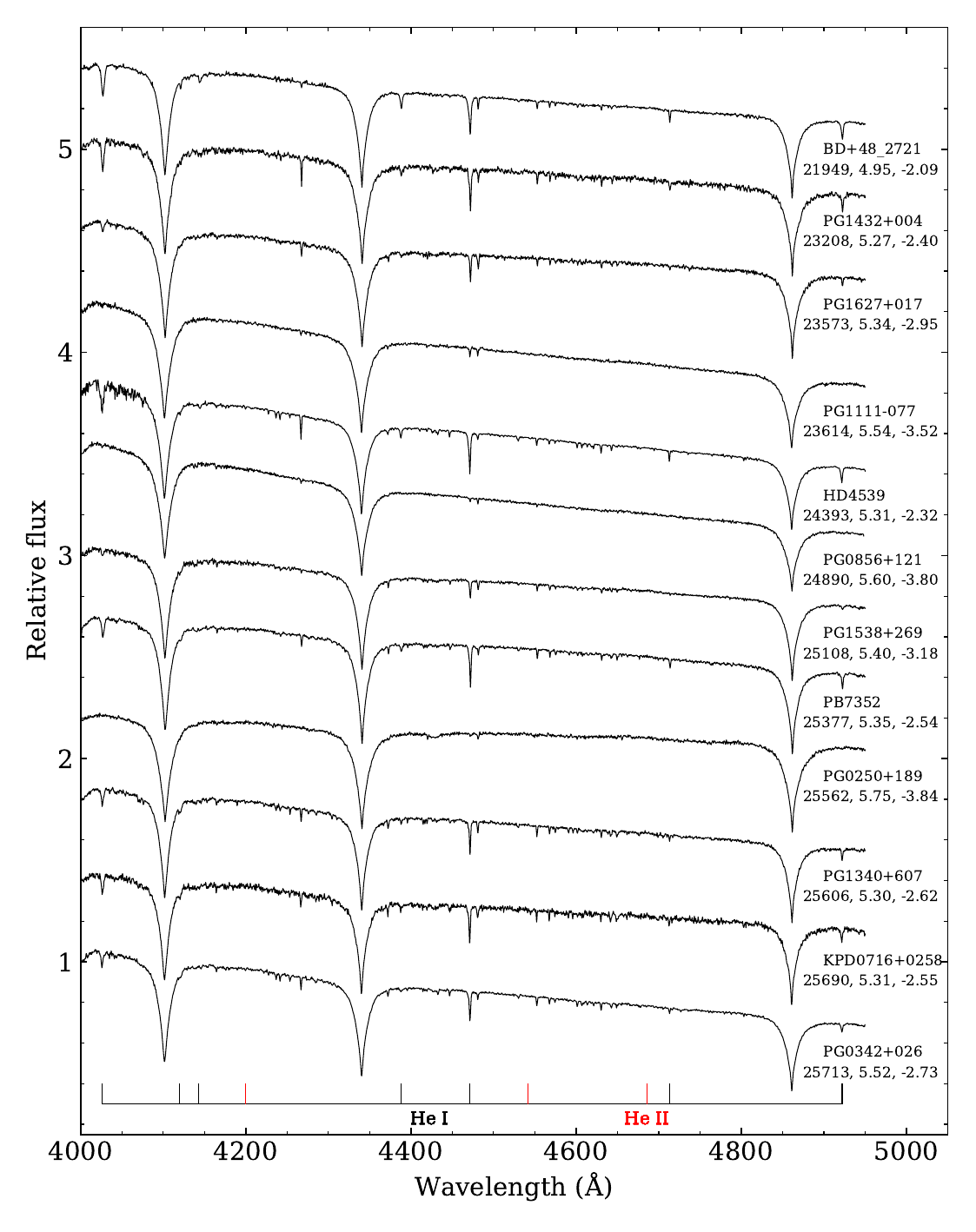}
     \caption{MMT spectra of all stars in the MMT sample ordered by increasing \teff. Under the name of each star, we indicate the \teff, \logg, and helium abundance (\heh) of the stars, obtained from the fit of their Bok spectrum. None of the stars is sufficiently hot and He-rich to display the \ion{He}{ii} lines at 4200 \AA\ and 4542 \AA. The spectra of the most reddened stars, show a small diffuse interstellar band absorption around 4450 \AA. Examples are PG0250+189, PG0215+183 and some of the KPD objects. 
     }
     \label{fig:spectra_all}
\end{figure}

\begin{figure}
\ContinuedFloat
\centering
   \includegraphics[width=0.98\textwidth]{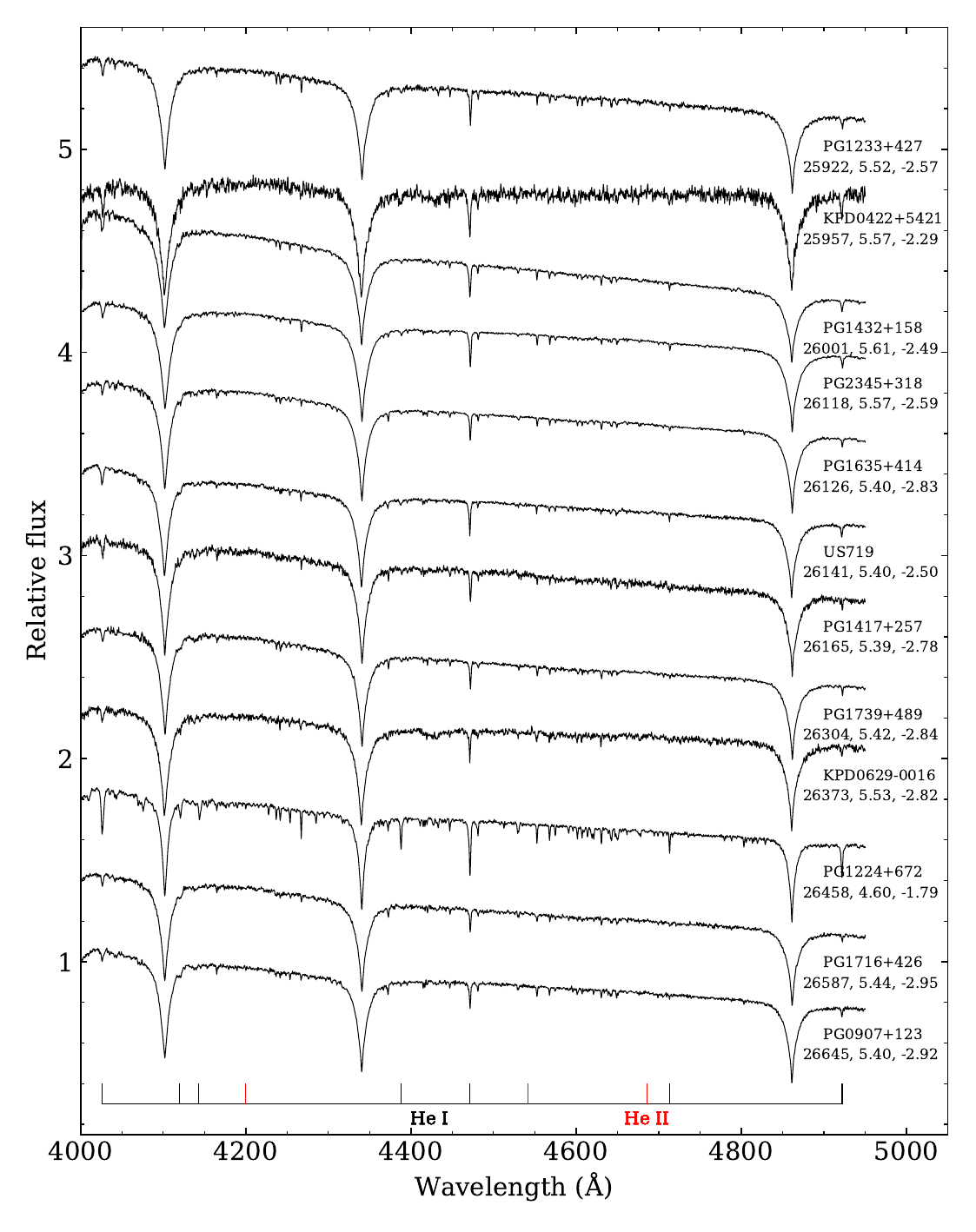}
    \caption{Continued.}
\end{figure}
\begin{figure*}
\ContinuedFloat
\centering
   \includegraphics[width=0.98\textwidth]{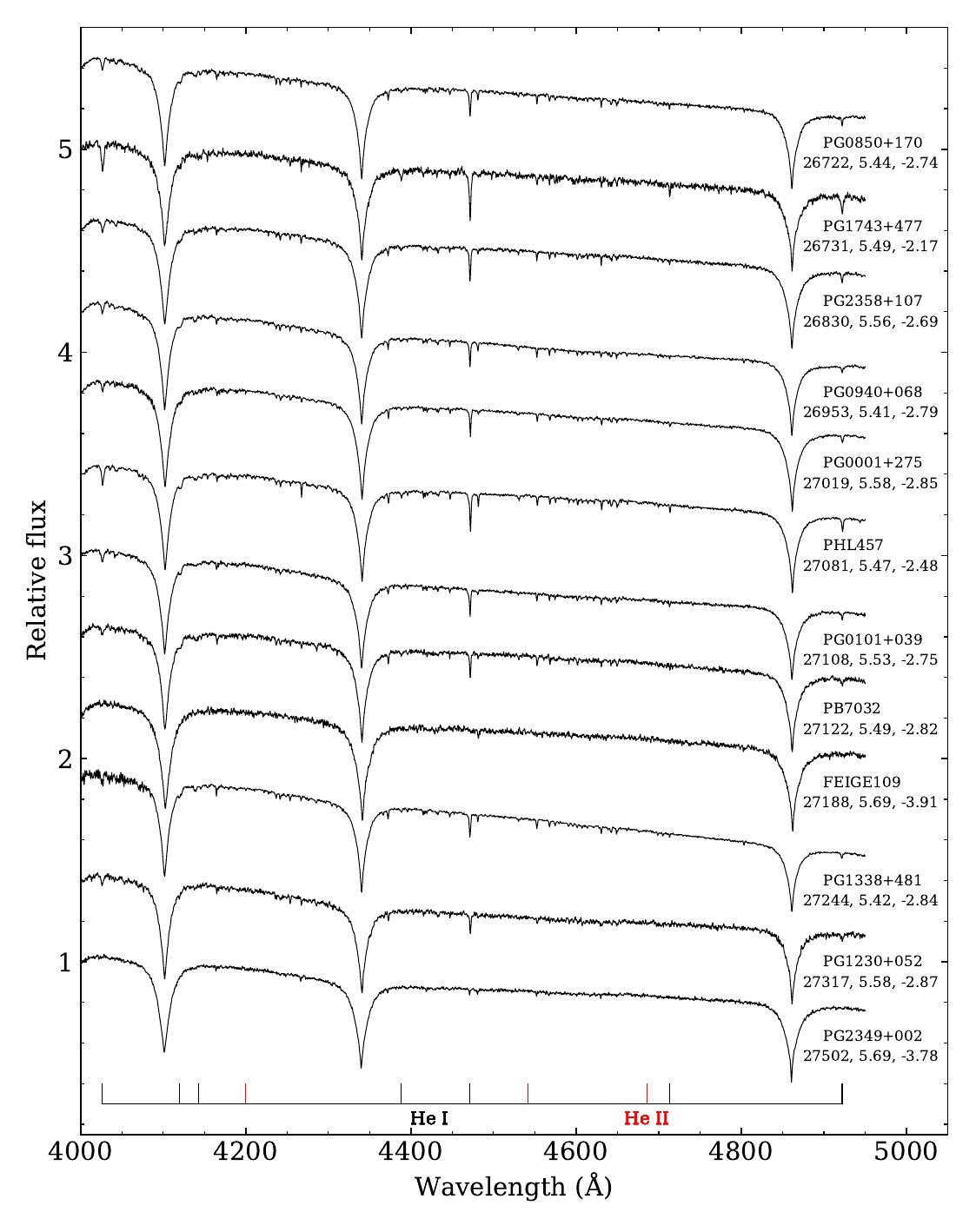}
    \caption{Continued.}
\end{figure*}
\begin{figure*}
\ContinuedFloat
\centering
   \includegraphics[width=0.98\textwidth]{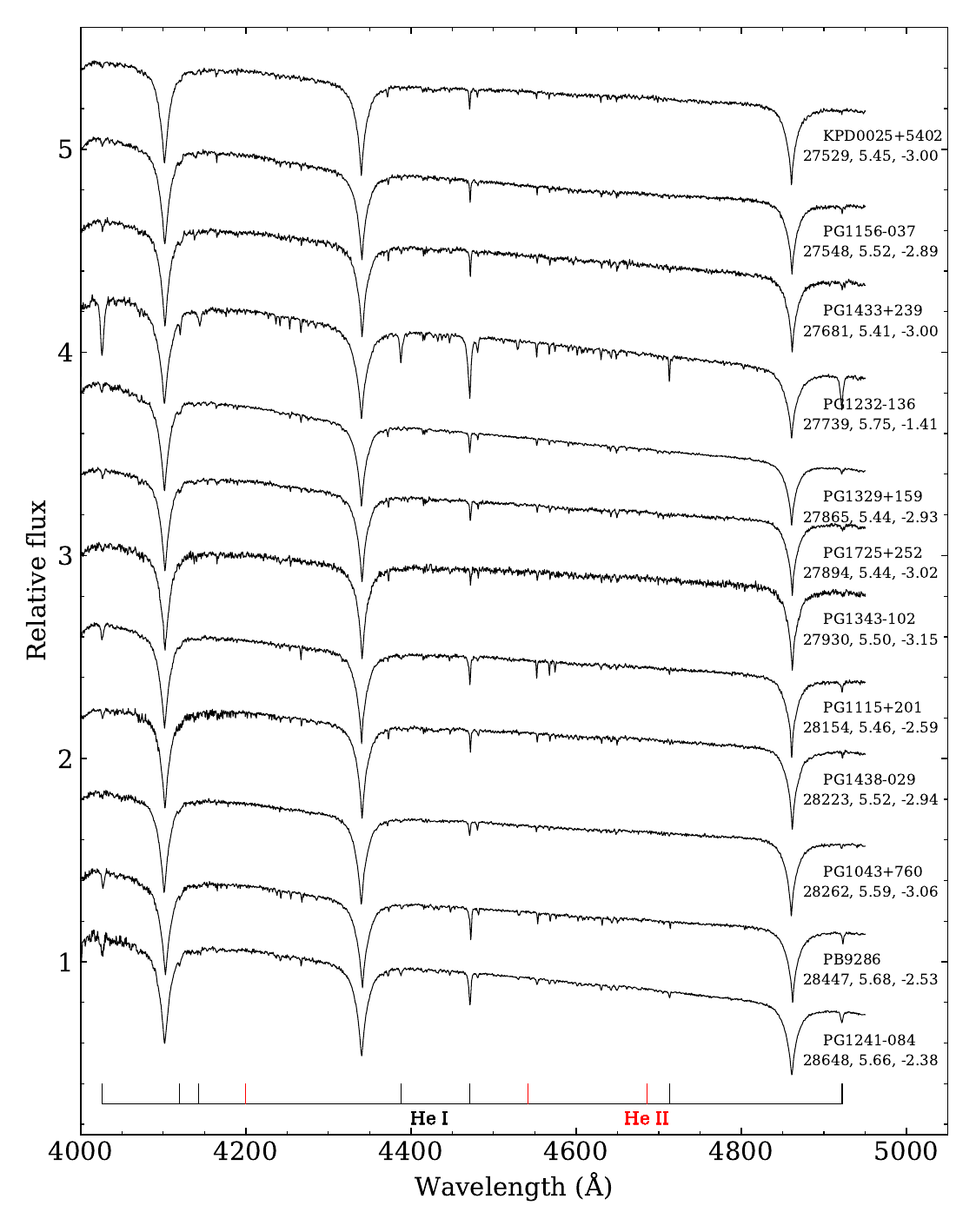}
    \caption{Continued.}
\end{figure*}
\begin{figure*}
\ContinuedFloat
\centering
   \includegraphics[width=0.98\textwidth]{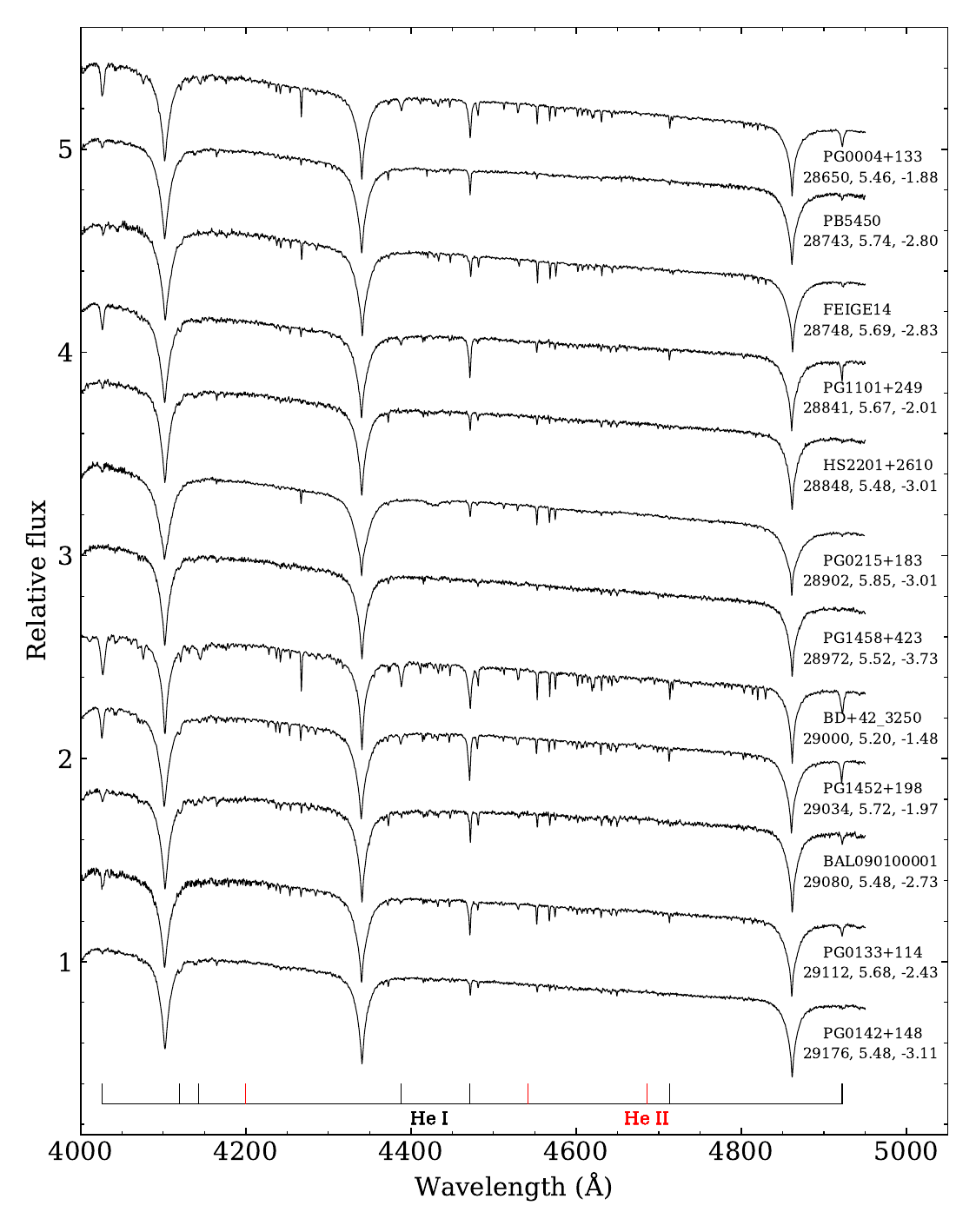}
    \caption{Continued.}
\end{figure*}
\begin{figure*}
\ContinuedFloat
\centering
   \includegraphics[width=0.98\textwidth]{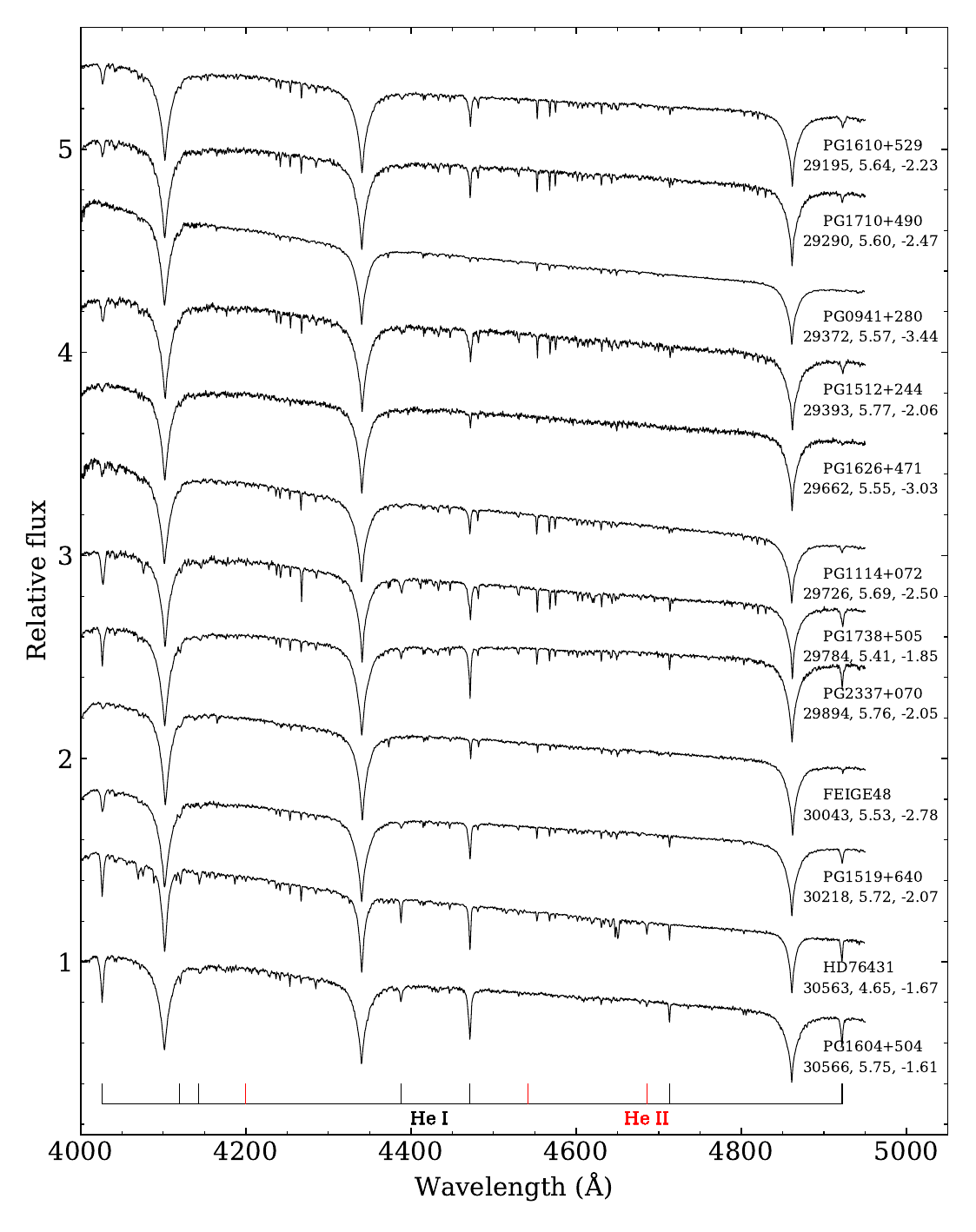}
    \caption{Continued.}
\end{figure*}
\begin{figure*}
\ContinuedFloat
\centering
   \includegraphics[width=0.98\textwidth]{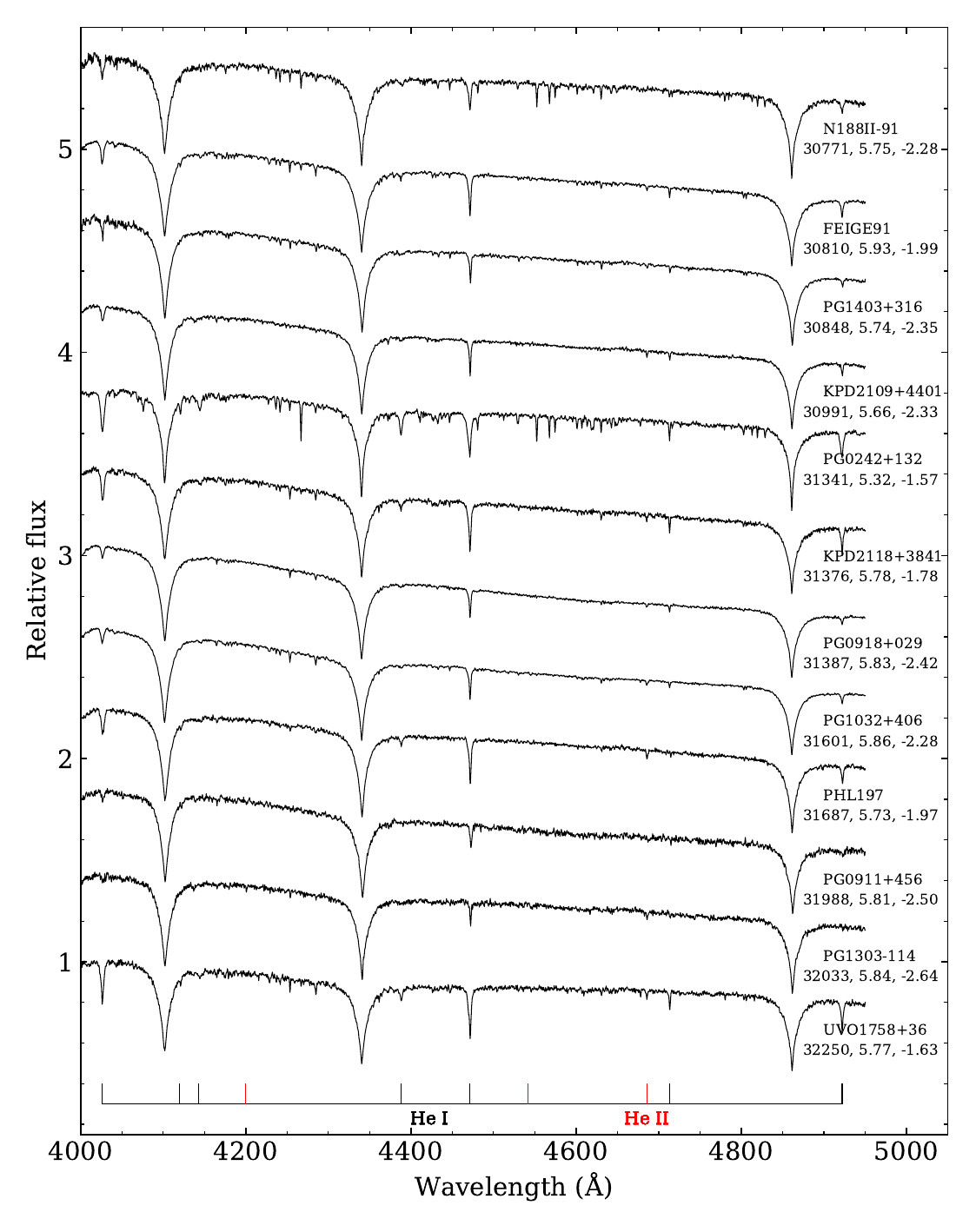}
    \caption{Continued.}
\end{figure*}
\begin{figure*}
\ContinuedFloat
\centering
   \includegraphics[width=0.98\textwidth]{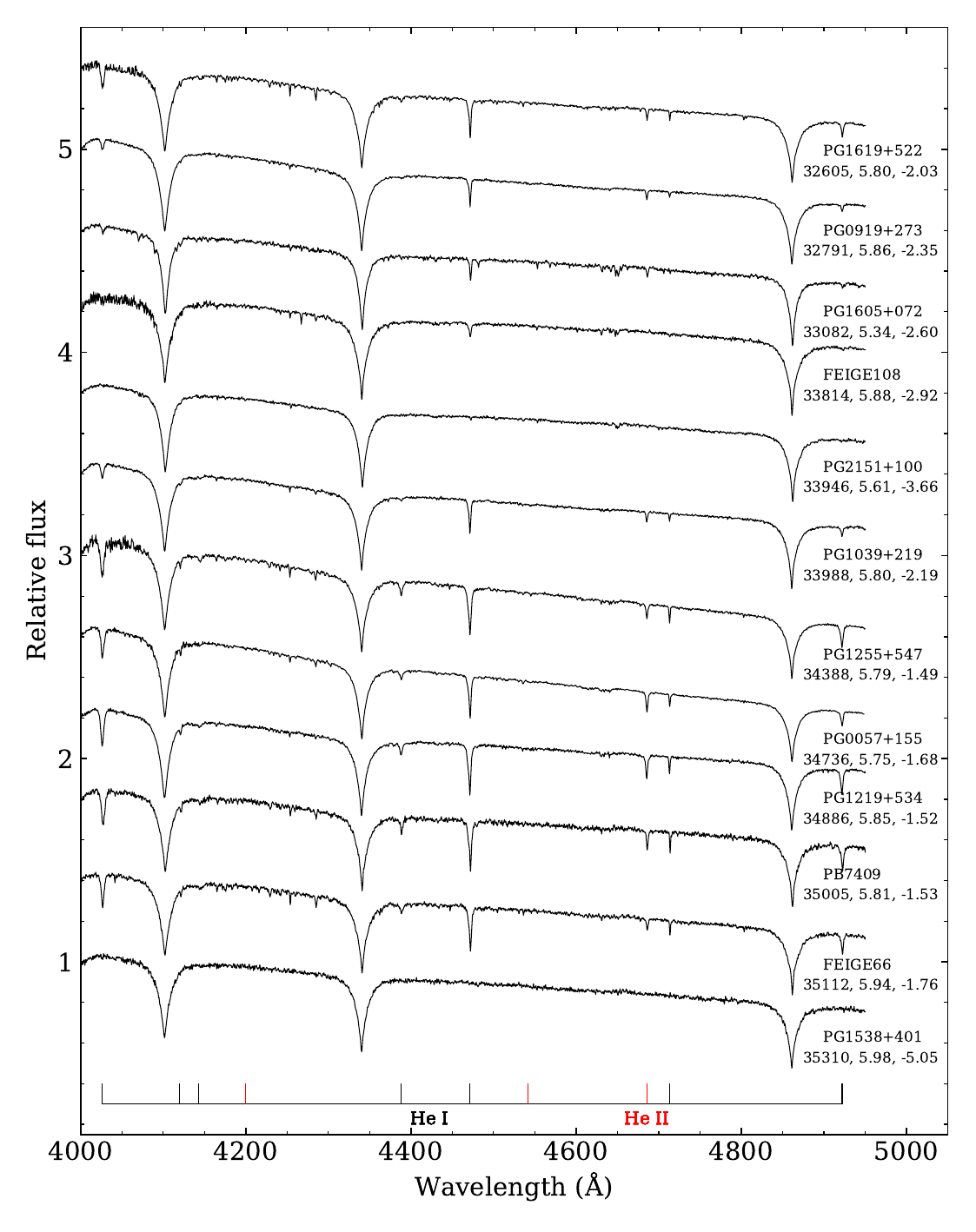}
    \caption{Continued.}
\end{figure*}
\begin{figure*}
\ContinuedFloat
\centering
   \includegraphics[width=0.98\textwidth]{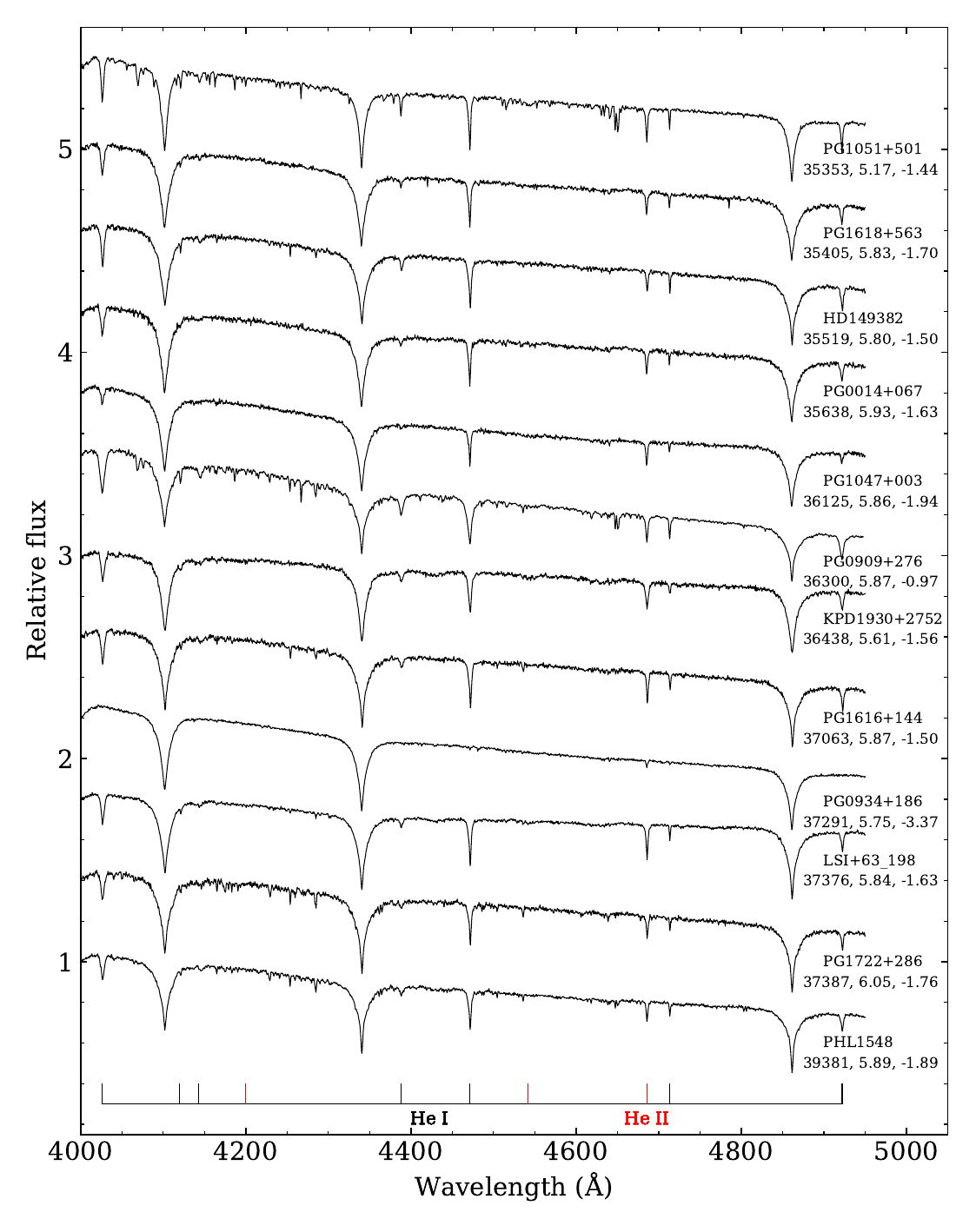}
    \caption{Continued.}
\end{figure*}
\begin{figure*}
\ContinuedFloat
\centering
   \includegraphics[width=0.98\textwidth]{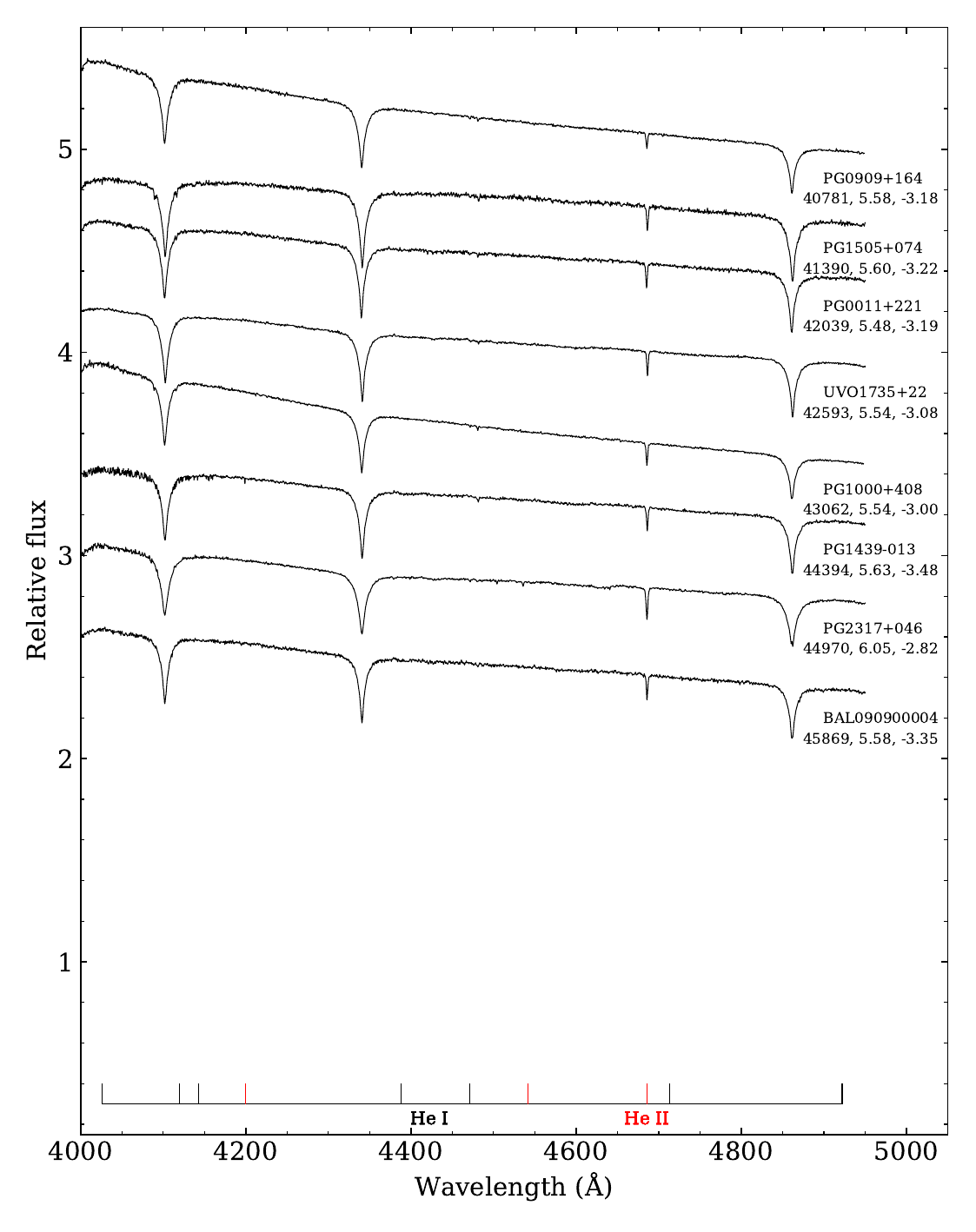}
    \caption{Continued.}
\end{figure*}

\end{appendix}

\end{document}